\documentclass[twocolumn,
 showpacs,nofootinbib,superscriptaddress,leqn]{revtex4}
\usepackage{amsmath,amssymb,setspace,
graphicx,subfigure,epsfig}
 \usepackage{amsfonts}
 \usepackage{latexsym}
 \usepackage{epsf,epsfig,subfigure,axodraw,graphicx,color}

\newcommand{\dis}[1]{\begin{equation}\begin{split}#1\end{split}\end{equation}}

\newcommand{\Dslash}{{D\hskip -0.23cm\slash}}
\newcommand{\dslash}{{\partial\hskip -0.185cm\slash}}

\newcommand{\tev}{\,\textrm{TeV}}
\newcommand{\gev}{\,\textrm{GeV}}

\newcommand{\tadec}{T_{\tilde a\rm -dcp}}
\newcommand{\tarad}{T_{\tilde a\rm =rad}}
\newcommand{\treh}{T_{\rm R}}

\newcommand{\td}{T_{\rm D}}
\newcommand{\Linst}{\Lambda_{\rm inst}}
\newcommand{\Linsteta}{\Lambda_{\eta'}}

\newcommand{\axino}{{\tilde{a}}}

\newcommand{\gluino}{{\tilde{g}}}
\newcommand{\N}{{\cal N}}
\newcommand{\Ca}{${\cal C}$}
\newcommand{\Pa}{${\cal P}$}
\newcommand{\Ti}{${\cal T}$}
\newcommand{\UPQ}{U(1)$_{\rm PQ}$}
\newcommand{\up}{\uparrow}
\newcommand{\down}{\downarrow}

\begin{document}


\title{Axions and the Strong CP Problem}%

\author{Jihn E. Kim}

 \email{jekim@ctp.snu.ac.kr}
\affiliation{%
Department of Physics and Astronomy and Center for Theoretical Physics,
Seoul National University, Seoul 151-747, Korea}

\author{Gianpaolo Carosi}%
 \email{carosi2@llnl.gov}
\affiliation{Physical Sciences Directorate, Lawrence Livermore National Laboratory, Livermore,
California, USA
}%

\begin{abstract}
Current upper bounds of the neutron electric dipole moment constrain the physically observable quantum chromodynamic (QCD) vacuum angle $|\bar\theta| \lesssim 10^{-11}$. Since QCD explains vast experimental data from the 100 MeV scale to the TeV scale, it is better to explain this smallness of $|\bar\theta|$ in the QCD framework, which is the strong \Ca\Pa~ problem. Now, there exist two plausible solutions to this problem, one of which  leads to the existence of the very light axion. The axion decay constant window, $10^9\ {\gev}\lesssim F_a\lesssim 10^{12}\ \gev$ for a ${\cal O}(1)$ initial misalignment angle $\theta_1$, has been obtained by astrophysical and cosmological data. For $F_a\gtrsim 10^{12}$ GeV with $\theta_1<{\cal O}(1)$, axions may constitute a significant fraction of dark matter of the universe. The supersymmetrized axion solution of the strong \Ca\Pa~ problem introduces its superpartner the axino which might have affected the universe evolution significantly. Here, we review the very light axion (theory, supersymmetrization, and models) with the most recent particle, astrophysical and cosmological data, and present prospects for its discovery.
\end{abstract}



\centerline{Submitted to Reviews of Modern Physics}

\maketitle

\section{Overview}

Strong interaction phenomena has revealed that the discrete symmetries of charge conjugation \Ca, parity \Pa~ and time reversal \Ti~ are separately good symmetries of nature. Therefore, quantum chromodynamics(QCD) based on the gauge group SU$(3)_c$ \citep[Han, Nambu (1965), Bardeen, Fritszch, Gell-Mann (1972)]{Han65,Bardeen72} must respect any combinations of these discrete symmetries \Ca, \Pa~ and \Ti~ to be accepted as the theory of strong interactions. Among these discrete symmetries, the \Ca\Pa~ symmetry is not necessarily respected in QCD due to the nonzero QCD vacuum angle $\theta$, an issue known as the ``strong \Ca\Pa~ problem". Since QCD is so successful phenomenologically, any possible solution to $\lq\lq$the strong  \Ca\Pa~ problem" is expected to be realized in nature. Currently the most attractive solution leads to the existence of a very light axion \citep[Kim (1979), Shifman, Vainstein, Zakharov (1980), Dine, Fischler, Srednicki (1981), Zhitnitskii(1981)]{Kim79,Shifman80,DFS81,Zhit80}. Searches for  QCD axions generated from the Sun \citep[Andriamonje {\it et. al.} (CERN Axion Search Telescope (CAST) Collaboration, 2007), Inoue {\it et. al.} (Tokyo Axion Helioscope Collaboration, 2008)]{CAST07,Inoue08} and remnant axions from the early universe \citep[Rosenberg (Axion Dark Matter Experiment(ADMX) Collaboration, 2004), Carosi (ADMX Collaboration, 2007)]{ADMX04,ADMX07} are presently ongoing.

\begin{figure}[!]
\resizebox{0.9\columnwidth}{!}
{\includegraphics{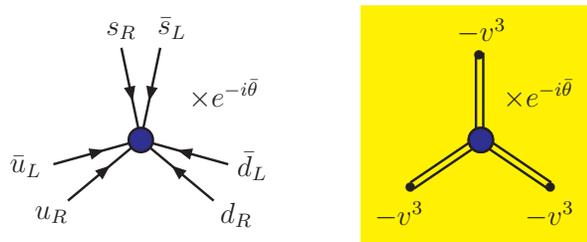}}
\caption{The determinental interaction of light quarks. Chiral symmetry breaking introduces the anomalous $\eta'$ mass term from the quark condensations.}\label{fig:U1tHooft}
\end{figure}

The story of axions started with the QCD U(1) problem \citep[Weinberg (1975)]{Wein75U1} which is now understood, having been solved by the 't Hooft determinental interaction \citep['t Hooft (1976,1986)]{tHooft76,tHooftU1}. The determinental interaction is shown pictorially as the left diagram of Fig. \ref{fig:U1tHooft} and the solution is shown as the shaded right diagram.  The strong interaction makes the quark bilinears condense with the vacuum expectation value (VEV) of order $v\simeq 260$ MeV. The phase of this interaction $\bar\theta$ originates from the QCD vacuum angle which is known to be physical \citep[Callan, Dashen, Gross (1976), Jackiw, Rebbi (1976)]{Callan76,Jackiw76} and contributes to the neutron electric dipole moment (NEDM) by an order $\bar\theta$ times the neutron size, which is absurdly large. Peccei and Quinn(PQ) observed that there exists a way to make $\bar\theta$ just a phase by introducing a symmetry, now called \UPQ, and hence physical amplitudes do not depend on  $\bar\theta$ as in the massless quark case \citep[Peccei, Quinn (1977)]{PQ77,PQ77b}. In the standard model (SM), this phase is a pseudoscalar Goldstone boson called the `axion' in the multitude of Higgs fields as noted in \citep[Weinberg (1978), Wilczek (1978)]{Wein78,Wil78}. If the PQ idea is completed with Fig. \ref{fig:U1tHooft}, this axion is exactly massless (but observable), and  $\bar\theta$ would behave  `unphysically' by the freedom of choosing an appropriate axion VEV, which was the original PQ idea. However, there exists subleading terms, proportional to one power of $m_q$, which close the quark lines with the current quark mass instead of a condensation. Then, an axion potential develops, and the axion becomes a pseudo-Goldstone boson. The axion solution of the strong \Ca\Pa~ problem is cosmological in that the axion VEV chooses $\bar\theta=0$ at the minimum of this axion potential. The currently allowed axion is very light and long lived.

The properties of the axion (denoted as $a$) are chiefly given by its decay constant $F_a$ which sets the scale of nonrenormalizable axion interactions through $a/F_a$. Initial axion searches placed $F_a$ far above the electroweak scale and additional stringent bounds on $F_a$ were obtained from the studies of stellar evolutions and cosmology \citep[Kim (1987)]{Kimrev}. Axion astrophysics started by \citep[Dicus, Kolb, Teplitz, Wagoner (1978, 1980)]{Dicus78,Dicus80}, using earlier ideas from \citep[K. Sato, H. Sato (1975), K. Sato (1978)]{Sato75,Sato78}, now gives a stringent lower bound on the decay constant, $F_a\ge 0.5\times 10^9$ GeV, from the study of SN1987A \citep[Turner (1990), Raffelt (1990)]{Turnerrev90,Raffeltrev90}. With this large decay constant, the axion flux from the Sun is a small fraction of the solar neutrino flux, but still may be detectable by the CAST experiment and by the Tokyo helioscope.

It is known that very light axions with $F_a$ in the $10^{12}$ GeV region (axion mass in the micro-eV range) can compose some part of cold dark matter(CDM) in the unverse \citep[Preskill, Wise, Wilczek (1983), Abbott, Sikivie (1983), Dine, Fischler (1983)]{Preskill83,Abbott83,DineFish83}. The exact amount of the axion CDM depends on the initial axion misalignment angle $\theta_1$ at the time of axion creation when the universe temperature was around the axion decay constant, $T\sim F_a$. This observation puts the very light axion on the list of leading CDM candidate particles. If indeed these cosmic axions compose a significant fraction of CDM in the universe, they may be detectable by collecting axion-converted photons in the cavity type detectors \citep[Sikivie (1983)]{SikDet} as tried by \citep[DePanfilis {\it et al.} (RBF Collaboration, 1987), Hagmann  {\it et al.} (Univ. of Florida, 1990)]{DePanfilis87,Hagmann90} and now continuing at the ADMX experiment.

Cosmology with CDM was the leading candidate of the early universe in the 1980s \citep[Blumenthal, Faber, Primack, Rees (1984), Kolb, Turner (1990), Weinberg (2008)]{Blum84,KolbTur90,Wein08}. Since then  this view has given way to the new cosmology with the discovery of dark energy(DE) in 1998 \citep[Perlmutter {\it et al.} (1998), Riess {\it et. al.} (1998)]{Perlmutter98,Riess98}. The current view of the dominant components of the universe is $\Omega_{\rm CDM}\simeq 0.23$ and $\Omega_\Lambda\simeq 0.73$ with only a few percent consisting of baryons \citep[Spergel {\it et al.} (Wilkinson Microwave Anisotropy Probe(WMAP) Collaboration, 2007)]{WMAPsigma}. The most plausible dark matter (DM) candidates at present are the lightest supersymmetric(SUSY) particle(LSP), the axion, the axino, and the gravitino. Here, we will review mostly on the axion and its CDM-related possibility.

The need for DM was suggested as early as the 1930s \citep[Zwicky (1933), Smith (1936)]{Zwicky33,Smith36}. Since then, evidences of nonluminous DM in the universe has been accumulating: examples of which include flat galactic rotation curves, Chandra satellite photos, and gravitational lensing effects. If the galactic bulge is the dominant mass in the galaxy, the velocity $v$ of a star located at $r$ from the center should rotate with  $v\sim r^{-1/2}$. But the observed flat rotation curve \citep[See, for example, McGaugh {\it et al.} (2007),  and references therein]{flatrot07} violates this expectation and implies an extended mass in the halo as $\rho(r)\sim 1/r^2$. Also, the Chandra observation of X-ray and gravitational lensing images implies this matter profile around the bullet cluster \citep[Clowe {\it et al.} (Chandra Collaboration, 2006)]{Chandra06}. Circular gravitational lensing images \citep[Jee {\it et al.} (2007)]{gravlens07}  also support the existence of DM. The DM density around us the Solar system is usually taken as $\rho_{\rm DM}\simeq (0.3-0.45)$ GeV/cm$^3$.

\begin{figure}[!]
\resizebox{0.9\columnwidth}{!}
{\includegraphics{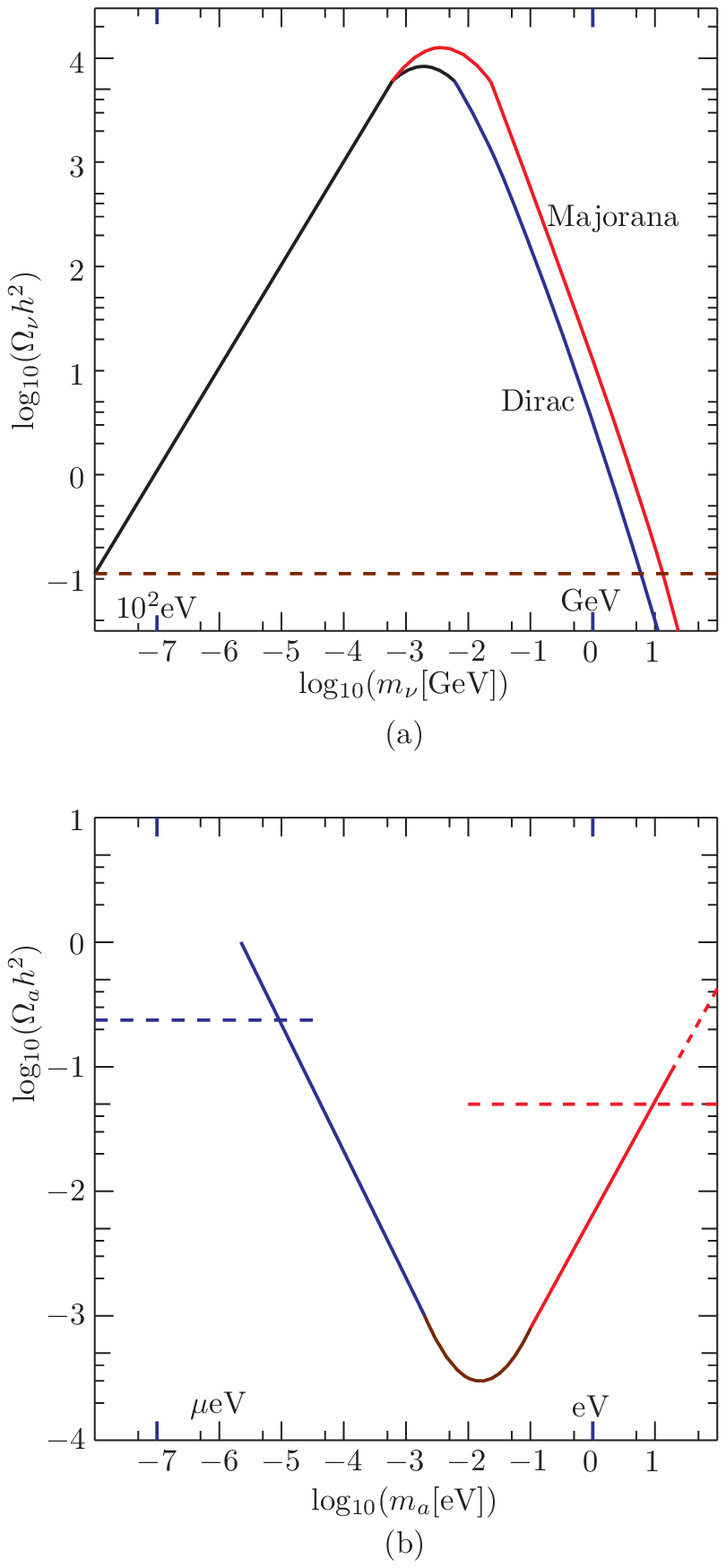}}
\caption{The Lee-Weinberg type plots for (a) the neutrino $\Omega_\nu h^2$  \cite{KolbTur90} and (b) the axion $\Omega_a h^2$, where $h$ is the present Huble constant in units of $\rm 100\, km\, s^{-1}Mpc^{-1}$. The dash line of (a) is for $\Omega_\nu h^2=0.113$. In (b), it corresponds to the hadronic axion. The blue and red dashlines correspond to the CDM and hot DM limits, respectively. }\label{fig:LeeWein}
\end{figure}

Current CDM candidates belong to either incoherent particles or coherent oscillations of spin-0 fields. From this disctinction, the bosonic collective motion such as the axion can be considered as CDM.  The popular incoherent CDM particles are the weakly interacting massive particles(WIMP) or decay products of WIMPs. A more frequently used independent distinction is thermal relics and nonthermal relics, but there is no strict correspondence relation between the incoherent--coherent particles and the thermal--nonthermal relics. WIMPs are massive particles with weak interaction cross sections, first discussed in terms of a heavy neutrino, corresponding to the RHS crossing point of Fig. \ref{fig:LeeWein}(a) \citep[Lee, Weinberg (1977)]{LeeWein77b}. The LHS crossing point corresponds to 10 eV neutrino \citep[Marx, Szalay (1972), Cowsik, McClelland (1972)]{Marx72,Cowsik72}. WIMPs, such as the LSP, are thermal relics when their number density is determined by the freezeout temperature and are called nonthermal relics if their number density is determined by the other mechanism such as by the decay of heavier relics \citep[Choi, Kim, Lee, Seto (2008)]{ChoiKY08prd}.  In Fig. \ref{fig:LeeWein}(b), we sketch the axion energy density in terms of the axion mass. The shape is a flipped one from that of Fig. \ref{fig:LeeWein}(a), because in the axion case the low mass and high mass regions contributes $\Omega_a$ from different physics, one from the vacuum misalignment and the other from the hot thermal relics.

\begin{figure}[!]
\resizebox{0.9\columnwidth}{!}
{\includegraphics{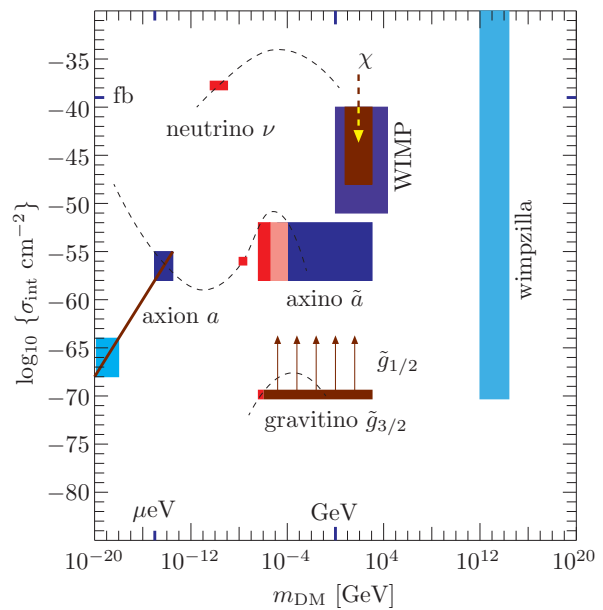}}
\caption{Some proposed particles in the interaction cross section versus the corresponding particle mass $m_i$ plane. The skeleton is taken from Ref. \cite{Roszfig04}.
The dashed curves represent schematic shapes of $\Omega_i$ versus the corresponding particle mass $m_i$. The small red square-box corresponds to the hot DM hadronic axion. Two small outside squares (in the cyan and blue colors) in the axion region are marked to show the plausible (GUT and CDM) axions, respectively. The abundances of heavy axino, gravitino and wimpzilla depend on how inflation ends. }\label{fig:sigmavsmass}
\end{figure}

In addition to the heavy neutrino, SUSY with R-parity conservation allows the LSP to be just such a WIMP particle. The LSP interaction is $\lq\lq$weak" since the interaction mediators (SUSY particles) are supposed to be in the 100 GeV range. For a WIMP to be a successful CDM candidate, usually the interaction cross section at the time of decoupling needs to be around \citep[Kolb, Turner (1990), Spergel {\it et al.} (WMAP Collaboration, 2007)]{KolbTur90,WMAPsigma}
\begin{align}
&\langle\sigma_{\rm int} v\rangle|_{\rm at\ decoupling}\approx 0.2\times
10^{-26} ~{\rm cm^3s}^{-1},\nonumber\\
&\quad\quad {\rm with}\ \Omega_m h^2\simeq 0.113\pm 0.009.\label{WIMPcond}
\end{align}
This is roughly the cross-section for the LSP from low energy SUSY, which is the reason why the DM community is so interested in the WIMP LSP. Some superweakly interacting particles such as gravitinos, axinos, and wimpzillas \citep[Chung, Kolb, Riotto (1999)]{ChungWzilla} can be CDM candidates as well, but their cross sections do not fall in the range of Eq. (\ref{WIMPcond}).  The CDM candidate particles are shown in the $\sigma_{\rm int}$ versus mass plane in Fig. \ref{fig:sigmavsmass} with minor modification from that of \citep[Roszkowski (2004)]{Roszfig04}. The incoherent fermions, such as the neutrino and left ends of the bars of the axino and gravitino correspond to the left crossing points of Fig. \ref{fig:LeeWein}(a). The rest, except the axion, corresponds more or less to the right crossing points of Fig. \ref{fig:LeeWein}(a) with the reheating after inflation considered if necessary.  Currently, there are experimental efforts to discover the LSP as predicted by SUSY models. Direct cosmological searches are also ongoing \citep[Jungman, Kamionkowski, Griest (1996), Bertone, Hooper, Silk (2005), Bernabei {\it et al.} (DAMA collaboration, 2003, 2008), Lee {\it et al.} (KIMS Collaboration, 2007), Angle {\it et al.} (XENON Collaboration, 2008), Behnke {\it et al.} (COUPP Collaboration, 2008), Ahmed {\it et al.} (CDMS Collaboration, 2008)]{Jungman96,Bertone05,Dama03,Dama08,KIMSCo07,Xenon10,
COUPP08,CDMS08}. At the LHC, the probable LSP mass ranges will be looked for by the neutralino decay producing the LSP.


It is known that density perturbations began growing much earlier than the time of recombination in order to become large enough to form galaxies in the young universe. For galaxy formation, therefore, DM is needed since proton density perturbations could not grow before the recombination time, but DM perturbations could. With DM, the equality point of radiation and matter energy densities can occur much earlier than the recombination time since DM is not prohibited in collapsing by Silk damping \citep[Silk (1968)]{Silkdamp68}.
If the WIMP mass and interaction cross section fall in the region allowed by Eq. (\ref{WIMPcond}), it can be CDM. If the LSP is the only CDM component, then the LSP mass would give one number for the DM density, which may not be the case. Thus, even if the LSP is contributing to the CDM density, we may need the axion to account for the correct amount of CDM around us. This is possible in the anthropic scenario of very light axions because it is equally probable for the initial axion misalignment angle $\theta_1$ to take any value between 0 and $\pi$ \citep[Tegmark, Aguirre, Rees, Wilczek (2006)]{Tegmark06}.

Here we review the axion, which is probably the most interesting Nambu-Goldstone boson \citep[Nambu (1960), Nambu, Jona-Lasinio (1961), Goldstone (1961)]{Nambu60, Nambu61, Goldstone61}, as well as related issues.
In Sect. \ref{sec:StrongCPsols} we discuss the strong \Ca\Pa~ problem and its plausible solutions. In Sect. \ref{sec:axions} we review the most attractive solution giving the very light axion and present the axion theory in terms of possible axion couplings defined with $c_1,c_2,$ and $c_3$ which will be used throughout this review. In Sect. \ref{sec:axionsOuterspace} we present axion astrophysics and cosmology. Here we present a new number for the cosmic axion abundance in view of the recent more accurate data on light quark masses.  In Sect. \ref{sec:AxionDet} we summarize the axion detection ideas and the ongoing axion detection experiments. In Sect. \ref{sec:AxionModels} we summarize the proposed very light axion models, including superstring axions. Finally in Sect. \ref{sec:Axinocosmology} we briefly discuss cosmology with the axino, the axion's superpartner.

If the axion was observed, it would mark one of the most profound elementary particle discoveries because it would confirm experimentally the instanton--based arguments of QCD. In addition, if it were shown to be consistent with a cosmologically significant amount of axions, the CDM idea of the bosonic collective-motion would also be confirmed experimentally. If SUSY is correct and the axion is the solution to the strong \Ca\Pa~ problem, it must affect the evolution of the universe as well.

\section{ The strong CP problem and solutions }\label{sec:StrongCPsols}

There are good reviews on the strong  \Ca\Pa~ problem \citep[Kim (1987), Cheng (1988), Peccei (1989)]{Kimrev,Chengrev88,Pecceirev89}, and here we outline a few key points. QCD with SU$(3)_c$ gluons is a confining gauge theory with three light quarks below 1 GeV and $\Lambda_{\rm QCD}=380\pm 60$ MeV \citep[Groote, K\"orner, Schilcher, Nasrallah (1998)]{QCDLambda98}. The classical gluon field equations have the instanton solution \citep[Belavin, Ployakov, Schwartz, Tyupkin (1975)]{Belavin75},
\begin{equation}
G_\mu =if(r) g^{-1}(x)\partial_\mu g(x),\quad f(r)=\frac{r^2}{r^2+\rho^2}
\end{equation}
where the gauge coupling is absorbed in the gauge field and $g(x)$ is a pure gauge form  with $G_{\mu\nu}\propto 1/r^4$ for a large $r$ and $\rho$ is the instanton size. The (anti-) instanton solution satisfies the (anti-) selfduality condition $G_{\mu\nu}=\pm \tilde G_{\mu\nu}$ which carries the integer Pontryagin index
\begin{equation}
q=\frac{1}{16\pi^2}\int d^4x~ {\rm Tr}~ G\tilde G
 =\frac{1}{32\pi^2}\int d^4x~  G_{\mu\nu}^a\tilde G^{a\mu\nu}\label{Pontryagin}
\end{equation}
where $\tilde G^{a\mu\nu}= \frac12 \epsilon^{\mu\nu\rho\sigma}G_{\rho\sigma}^a$. The classical solutions with $q=-\infty,\cdots,-1,0,+1,\cdots,+\infty$, introduces a new real number $\theta$ which parametrizes the $|\theta\rangle$ vacuum,
\begin{equation}
|\theta\rangle= \sum_{n=-\infty}^\infty  e^{in\theta}|n\rangle .\label{QCDvacuum}
\end{equation}
Since $n$s are integers, in view of Eq. (\ref{Pontryagin}), $\theta$ is a periodic variable with period $2\pi$. It has been known that $\theta$ is an observable parameter  \citep[Callan, Dashen, Gross (1976), Jackiw, Rebbi (1976)]{Callan76,Jackiw76}.  In the $\theta$ vacuum, we must consider the \Pa~ and \Ti~ (or \Ca\Pa) violating interaction parametrized by $\bar\theta=\theta_0+\theta_{\rm weak}$,\footnote{With the canonical normalization of the gauge field, the RHS of Eq. (\ref{AnomalyTheta}) is multiplied by $g_c^2$.}
\begin{equation}
{\cal L}= \bar\theta  \{G \tilde G\}
\equiv\frac{\bar\theta }{64\pi^2}\epsilon^{\mu\nu\rho\sigma}
G_{\mu\nu}^aG^a_{\rho\sigma}\label{AnomalyTheta}
\end{equation}
where the curly bracket carries $1/32\pi^2$, $\theta_0$ is the angle given above the electroweak scale and $\theta_{\rm weak}$ is the value introduced by the electroweak \Ca\Pa~  violation. This observable $\bar\theta$ has led to the so-called strong \Ca\Pa~  problem from the upper bound on the NEDM. For QCD to become a correct theory, this  \Ca\Pa~ violation by QCD must be sufficiently suppressed.

\subsection{Neutron electric dipole moment}\label{subsec:neutroned}

The interaction (\ref{AnomalyTheta}) is the anomaly term \citep[Adler (1969), Bell, Jackiw (1969)]{Adler69,Bell69} which is the basis for solving \citep['t Hooft (1986)]{tHooftU1} the old U(1) problem of QCD \citep[Weinberg (1975)]{Wein75U1}. The important size of instantons for physics is near the scale where QCD becomes strong. In \citep['t Hooft (1976)]{tHooft76}, 't Hooft has shown that the determinental interaction of light quarks carries the same global symmetry as that of Eq. (\ref{AnomalyTheta}), and it is customary to use this light quark determinental interaction rather than treating the gluon interaction (\ref{AnomalyTheta}). The early estimates of NEDM proportional to $\bar\theta$ from the determinental interaction are $2.7\times 10^{-16}\bar\theta~ e$cm \citep[Baluni (1979)]{Baluni79} and $3.6\times 10^{-16}\bar\theta~ e$cm
\citep[Crewther, Di Vecchia, Veneziano, Witten (1979, 1980(E))]{Crewther79}. Other estimates in different methods are $11\times 10^{-16}\bar\theta~ e$cm \citep[Cea, Nardulli (1984)]{Cea84}, $1.2\times 10^{-16}\bar\theta~ e$cm \citep[Schnitzer (1984)]{Schnitzer84}, $3\times 10^{-16}\bar\theta~ e$cm \citep[Musakhanov, Israilov (1984)]{Musakhanov84}, and $5.5\times 10^{-16}\bar\theta~ e$cm \citep[Kanaya, Kobayashi (1981)]{Kanaya81}. Comprehensive reviews on the NEDM exist \citep[Dar (2000), Pospelov, Ritz (2005)]{Dar00,Pospelov05}. Recently, the NEDM has been estimated in the hard wall AdS/QCD model with one extra dimension, $1.08\times 10^{-16}\bar\theta~ e$cm  \citep[Hong, Kim, Siwach, Yee (2007)]{Hong07}.

The diagrams contributing to the NEDM are restricted. The neutron magnetic dipole moment arises at one loop in the chiral perturbation theory. If we treat this neutron magnetic dipole moment operator $\mu_{\rm anom}\bar n \sigma^{\mu\nu}n F_{\mu\nu}^{\rm em}$ as a vertex, tree diagrams do not contribute to the NEDM, because the magnetic moment term has the same chiral transformation property as that of the mass term and hence by redefining an external neutron field one can remove the phases in the neutron mass and in the dipole moment operator together.

\begin{figure}[!]
\resizebox{0.9\columnwidth}{!}
{\includegraphics{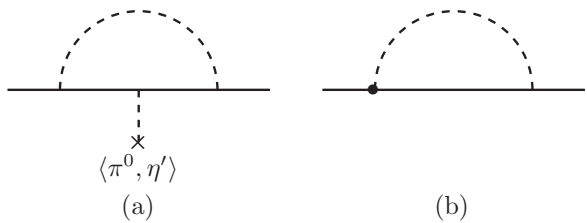}}
\caption{The insertion of the \Ca\Pa~ violation effect by VEVs of $\pi^0$ and $\eta'$ in (a). They can be transferred to one vertex shown as a bullet in (b). With this bullet, the \Ca\Pa~ violation is present by a mismatch between the \Ca\Pa~ conserving RHS vertex and \Ca\Pa~ violating LHS vertex.}\label{fig:CPVvertex}
\end{figure}

\begin{figure}[!]
\resizebox{0.9\columnwidth}{!}
{\includegraphics{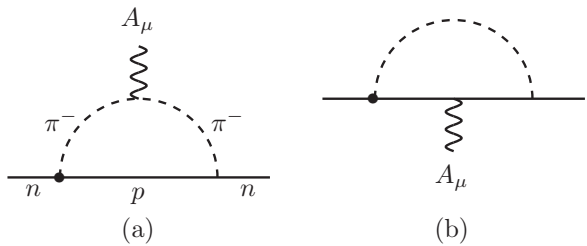}}
\caption{Diagrams contributing to the NEDM with the bullet representing the \Ca\Pa~ violation effect. Diagram (a) is the physically observable contribution.}\label{fig:NEDMcal}
\end{figure}

Let the U(1) chiral transformation of quarks in the broken phase be encoded in the neutron mass term as $m_n \bar n_L e^{i(\alpha'_1\eta'/f_{\eta'} -\alpha'_8\pi^0/f_\pi +\bar\theta/2)} n_R$  + h.c. ($e^{i\alpha'\bar\theta}$ instead of $e^{3i\alpha'\bar\theta}$ because the baryon octet is spin-$\frac12$). The vacuum expectation values(VEVs) of $\pi^0$ and $\eta'$ are calculated in Subsec. \ref{subsec:axionmass}. The \Ca\Pa~ violation is present by a mismatch between the \Ca\Pa~ conserving right-hand side(RHS) vertex and \Ca\Pa~ violating left-hand side(LHS) vertex as shown in Fig. \ref{fig:CPVvertex}(b). The mass term of  Fig. \ref{fig:CPVvertex}(b) and the neutron magnetic dipole moment term of Fig. \ref{fig:NEDMcal}(b) have the same chiral transformation property and the phases appearing there can be simultaneously removed by redefining $n_R$, for example.
However, the phase appearing in Fig. \ref{fig:NEDMcal}(a) cannot be removed by this phase redefinition and this contribution is physically observable. Since Fig. \ref{fig:NEDMcal}(a) is the physically observable NEDM,  for proton a similar argument leads to the same magnitude and opposite sign for the proton electrin dipole moment, i.e. $d_n+d_p=0$.  Now, we estimate the NEDM as
\begin{equation}
\frac{d_n}{e}=\frac{g_{\pi NN}\overline{g_{\pi NN}}}{4\pi^2 m_N}\ln\left(\frac{m_N}{m_\pi}\right)
\end{equation}
where the \Ca\Pa~ violating scalar coupling $\overline{g_{\pi NN}}$ (the bullet of  Fig. \ref{fig:NEDMcal}(a)) is estimated in Ref. \citep[Crewther, Di Vecchia, Veneziano, Witten (1980(E))]{Crewther79} as
\begin{equation}
\overline{g_{\pi NN}}=-\bar\theta \frac{2(m_{\Xi}-m_\Sigma)m_um_d}{f_\pi (m_u+m_d)
(2m_s-m_u-m_d)}\approx -0.023\bar\theta\label{CDVW}
\end{equation}
where we take $Z=m_u/m_d\approx 0.48$, $m_d\approx 4.9$ MeV and $m_s/m_d\simeq 20.1$.
From Eq. (\ref{VEVetap}) of Subsec. \ref{subsec:axionmass}, we estimate the \Ca\Pa~ violating scalar coupling as
\begin{equation}
\overline{g_{\pi NN}}=-\bar\theta\frac{Z}{(1+Z)}\simeq -\frac{\bar\theta}{3}~.\label{piVEV}
\end{equation}
Note that Eqs. (\ref{CDVW}) and (\ref{piVEV}) give a factor of $\sim$10 difference. The existing calculations vary within a factor of 10 \cite{Hong07,Cea84}. These old calculations depend on the various approximation methods used, but non of these estimated the VEV of $\pi^0$. For example, for Eq. (\ref{CDVW}) Eq. (11) of Ref. \citep[Crewther, Di Vecchia, Veneziano, Witten (1979)]{Crewther79} uses the SU(3) symmetric baryon octet coupling due to the \Ca\Pa\ violating interaction. On the other hand, for Eq. (\ref{piVEV}) the ground state vacuum of the mesonic fields has been used. After integrating out baryons, we look for the vacuum below the chiral symmetry scale. Then, the correct vacuum choice adds the value (\ref{piVEV}) to the value (7). But here we choose the one-order-larger value from the mesonic vacuum shift value (\ref{piVEV}) for an order of magnitude estimate, not to worry about the signs of the contributions. So, we estimate the NEDM  as $4.5\times 10^{-15}\bar\theta~ e$cm from Eq. (\ref{piVEV}).

Since the recent upper bound on the NEDM is $|d_n|<2.9\times 10^{-26}e$cm \citep[Baker {\it et. al.} (2006)]{Baker06}, we must require
 \begin{equation}
|\bar\theta|<0.7\times 10^{-11}.\label{thetabound}
\end{equation}
This extremely small upper bound on $\bar\theta$ has led to the so-called `strong \Ca\Pa~ problem'. $|\bar\theta| \lesssim 10^{-11}$ is perfectly allowed but its tiny value  is not explained given that it could have chosen a value anywhere between 0 and $\sim\pi$. The strong \Ca\Pa~ problem is the quest to understand more satisfactorily, $\lq\lq$Why $\bar\theta$ is so unnaturally small?".

\subsection{Possible solutions}\label{subsec:sgCPsolutions}

In the remainder of this paper, we simplify the notation, replacing $\bar\theta$ by $\theta$ since there will not be much confusion. There are three explanations for the smallness of $\theta$ in the naturalness framework:
\begin{eqnarray}
&&\rm  Case~ 1.~ Calculable~ \theta,\nonumber\\
 &&\rm   Case~ 2.~ Massless~ up~ quark,\nonumber\\
 &&\rm  Case~ 3.~Axion.\nonumber
\end{eqnarray}
In this subsection, we discuss Cases 1 and 2 briefly, and concentrate on Case 3 in the subsequent sections.

\subsubsection{Calculable $\theta$}
\label{subsub:calculable}

Naturalness of a theory with a parameter $\beta$ is elegantly defined by 't Hooft in \citep['t Hooft (1979)]{tHooftNatur79}: The theory is natural if the symmetry of the theory increases in the limit of the vanishing  $\beta$. A frequently quoted example is the Dirac fermion mass, $m\bar\psi_L\psi_R+{\rm h.c.}$, where $m\to 0$ introduces a chiral symmetry $\psi\to e^{i\beta\gamma_5}\psi$ in the theory.

Regarding the strong  \Ca\Pa~ problem, the appropriate symmetry is parity \Pa~ or  \Ca\Pa~ since the interaction (\ref{AnomalyTheta}) violates  parity \Pa, time reversal \Ti~ and  \Ca\Pa, but conserves charge conjugation \Ca. Requiring  \Ca\Pa~ invariance in the Lagrangian is equivalent to setting $\theta_0$ at zero. However, the observed weak interaction phenomena exhibit the weak \Ca\Pa~ symmetry violations in the neutral $K$ meson system and $B\to K^+\pi^-$ decay \citep[Amsler {\it et al.} (2008)]{PData08}, and hence the needed introduction of  \Ca\Pa~ violation in weak interactions with $\theta_0=0$ must be achieved spontaneously. In this process one necessarily introduces $\theta_{\rm weak}$ part in $\theta$ which can be calculated and required to be sufficiently small within the bound given in (\ref{thetabound}). Along this line, many ideas were proposed \citep[B\`eg, Tsao (1978), Mohapatra, Senjanovic (1978), Segre, Weldon (1979), Barr, Langacker (1979)]{Beg78,Moha78,SegWel79,Barr79}. This naturalness idea may be extended so as to only effect renormalizable couplings \citep[Georgi (1978)]{Georgi78}.  In any case, the introduction of weak \Ca\Pa~ violation by spontaneous mechanisms \citep[Lee (1973)]{LeeTD73} or by soft scalar masses \citep[Georgi (1978)]{Georgi78} must be checked with various weak phenomena. The current weak \Ca\Pa~ violation data fits nicely with the Kobayashi-Maskawa type \Ca\Pa~ violation  \citep[Kobayashi, Maskawa (1973)]{KobMask73}, and these drastically different spontaneous weak \Ca\Pa~ violation ideas are probably difficult to fit the data but are not considered ruled out yet \citep[He (2007)]{He07} even though the spontaneous \Ca\Pa~ violation scheme \citep[Branco (1980)]{Branco80} in the Weinberg model \citep[Weinberg (1976)]{Wein76CP} is ruled out \citep[Chang, He, McKellar (2001)]{Chang01}. It should be noted, though, that the above proposed models have the difficulty in satisfying the bounds (\ref{thetabound}).

However, the Nelson-Barr type weak \Ca\Pa~ violation is mimics the Kobayashi-Maskawa type \Ca\Pa~ violation  even though the fundamental reason for \Ca\Pa~ violation is spontaneous \citep[Nelson (1984), Barr (1984)]{Nelson84,Barr84}. The scheme is designed such that the Yukawa couplings are real, i.e. $\theta_0=0$ from the \Ca\Pa~ invariance. Next, the spontaneous \Ca\Pa~ violation is introduced through the singlet VEVs, which is the key difference from the previous calculable models. Thus, the spontaneous \Ca\Pa~ violation is required to occur much above the weak scale through the singlet VEVs, mediating it to light quarks through mixing with vector-like heavy quarks. In modern terms, the heavy quarks can be considered as the mediation sector. Then, integrating out heavy fields, we obtain the SM quarks with the Kobayashi-Maskawa type weak \Ca\Pa~ violation. To ensure Arg.Det.$M_q=0$ at tree level, the specific forms for the Higgs couplings to the SM quarks and the superheavy vectorlike quarks are needed. Beyond the tree level, however $\theta$ is generated at one loop, typically with the form \citep[Bento, Branco, Parada (1991), Goffin, Segr\`e, Welson (1980)]{Bento91,Goffin80},
\begin{equation}
\theta_{\rm weak}\approx \frac{1}{16\pi^2}\Delta f^2 \sum (\rm loop~ integrals)
\end{equation}
where $\Delta f^2$ is the product of couplings and the Feynman loop integral is of ${\cal O}(1)$. To satisfy the bound (\ref{thetabound}), the small coupling $\Delta f^2$ is needed. Some mechanism such as the family symmetry may be needed to forbid $\theta_{\rm weak}$ at one loop \citep[Chang, Keung (2004), Nelson (1984)]{ChangD04,Nelson84}.

This kind of Nelson-Barr type calculable $\theta_{\rm weak}$ can be mimicked in many extra dimensional models including superstring. Recently, for example $\theta_{\rm weak}$ is calculated to be ${\cal O}(10^{-12})$ from two loop level in a sequestered flavor and \Ca\Pa~ model \citep[Cheung, Fitzpatrick, Randall (2008)]{Cheung08}.

Strictly speaking, the axion models also belong to the calculable models but we separate it from the models with spontaneous \Ca\Pa~ violation because there it is not needed to set $\theta_0=0$.

\subsubsection{Massless up quark}\label{subsub:masslessu}

\begin{figure}[!]
\resizebox{0.9\columnwidth}{!}
{\includegraphics{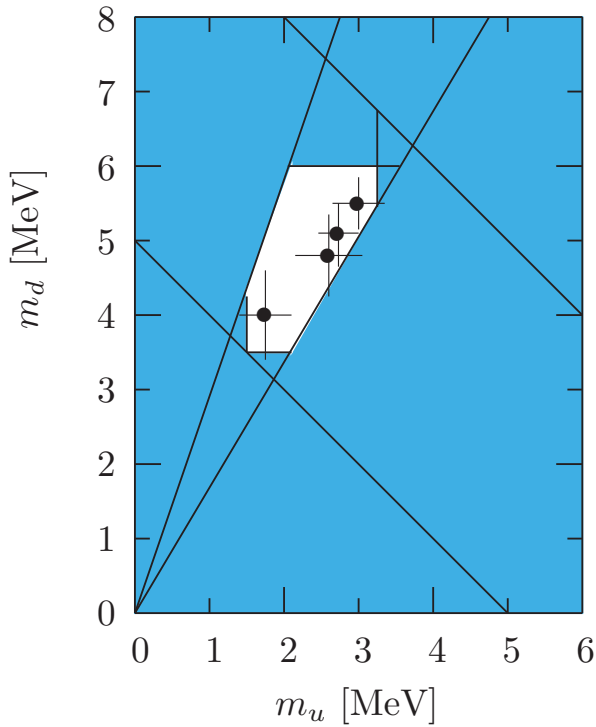}}
\caption{The allowed $m_u-m_d$ region \cite{Manohar08}. Two downward sloping lines are from the bound on $(m_u+m_d)/2$ and two rising lines are from the bound on $m_u/m_d$, which are determined by the masses of the meson octet. Two vertical and horizontal boundaries are from the Particle Data Book bounds on $m_u= [1.5,3.3]~\rm MeV $ and $m_d=[3.5,6.0]~\rm MeV $ \cite{PData08}.}
\label{fig:uqmass}
\end{figure}

  Suppose that we chiral-transform a quark as $q\to
e^{i\gamma_5\alpha}q$. Then, the QCD Lagrangian changes as
\begin{align}
&\int d^4x [-m_q \bar qq-\theta  \{g_c^2G \tilde G\} ]\to\nonumber\\
&\quad\int d^4x [-m_q \bar qe^{2i\gamma_5\alpha}q
-(\theta-2\alpha) \{g_c^2G \tilde G\}]\label{chiraltr}
\end{align}
where $\{G \tilde G\}=(1/64\pi^2)\epsilon^{\mu\nu\rho\sigma} G^a_{\mu\nu}G^a_{\rho\sigma}$.
If $m_q=0$, it is equivalent to changing $\theta\to
\theta-2\alpha$. Thus, there exists a shift symmetry
$\theta\to \theta-2\alpha$. It is known that the tunneling amplitude due to instanton solutions with a zero mass quark vanishes \citep['t Hooft (1976)]{tHooft76}, which implies that the shift symmetry is an exact symmetry. In this case, $\theta$ is not physical, and hence there is no strong  \Ca\Pa~  problem if the lightest quark (i. e. the up quark) is massless. The question for the massless up quark solution is, $\lq\lq$Is the massless up quark phenomenologically viable?" Weinberg's famous up/down quark mass ratio, $Z=m_u/m_d$, gave $Z=5/9$ \citep[Weinberg (1979)]{Wein79up}. It is very similar to the recent compilation of the light quark masses, $m_u=2.6^{+0.9}_{-1.1}~{\rm MeV}, m_d=4.9^{+1.1}_{-1.4}$ MeV, and $Z=0.48^{+1.2}_{-1.3}$ which is shown in Fig. \ref{fig:uqmass}. This compilation is convincing enough to rule out the massless up quark possibility \citep[Kaplan, Manohar (1986)]{KapMan86}. In this review, we will use $Z=0.48$ when a number is needed though the appropriate bound may be $0.35 < Z < 0.60 $ \citep[Buckley, Murayama (2007), Manohar, Sachrajda (2008)]{Buckley07,Manohar08}.

For some time the massless up quark possibility was taken seriously \citep[Kaplan, Manohar (1986)]{KapMan86}. The reason is that even if the Lagrangian mass for the up quark is zero, the 't Hooft determinental interaction may generate a useful up quark mass for the chiral perturbation. There was a confusion on this issue  for some time \citep[Leutwyler (1990), Choi (1992)]{Leutwyler90,ChoiK92}. Now, it is clear that the massless up quark possibility is ruled out, even without using the lattice calculation of the ratio, $m_u/m_d=0.410\pm 0.036$ \citep[Nelson, Fleming, Kilcup (2002)]{lattice03}.

\section{Axions}\label{sec:axions}

The axion solution seems to be the most attractive one among three possible strong \Ca\Pa~ solutions, in particular at present when the massless up quark possibility is excluded and calculable solutions need one loop suppressions.

Peccei and Quinn tried to mimic the symmetry  $\theta\to\theta-2\alpha$ of the massless quark case of (\ref{chiraltr}), by considering the full electroweak theory Lagrangian \citep[Peccei, Quinn (1977)]{PQ77,PQ77b}.
They found such a symmetry if $H_u$ is coupled only to up-type quarks and $H_d$ couples only to down-type quarks,
\begin{equation}
{\cal L}=-\bar q_L u_R H_u-\bar q_L d_RH_d-V(H_u,H_d)
+({\rm h.c.})-\theta  \{G \tilde G\}.
\end{equation}
Certainly, if we assign the same global charge under the $\gamma_5$ transformation to $H_u$ and $H_d$, $q\to e^{i\gamma_5\alpha}q, H_u\to e^{i\beta}H_u, H_d\to e^{i\beta}H_d$, the flavor independent
part changes to
\begin{align}
{\cal L}\to &-\bar q_Le^{-i\gamma_5\alpha} u_R e^{i\beta}H_u
-\bar q_Le^{-i\gamma_5\alpha} d_R e^{i\beta}H_d\nonumber\\
&-V(e^{i\beta}H_u,e^{i\beta}H_d)+({\rm h.c.})-(\theta-2\alpha)
\{G \tilde G\}.\label{PQ}
\end{align}
Since the full Lagrangian must possess the global symmetry, the potential $V$ should not allow the $(H_uH_d)^2$ term. Choosing $\beta=\alpha$ achieves the same kind of $\theta$ shift of the massless quark case, which is called the PQ global symmetry U(1)$_{\rm PQ}$. Unlike the massless up quark case, here $\theta$ is physical. Even though the coefficient of $\{G\tilde G\}$ changes in the same way in Eqs. (\ref{chiraltr}) and (\ref{PQ}), these two cases differ in that the tunneling amplitude vanishes with a massless quark (for which a detailed discussion will be presented in Subsec. \ref{subsec:axionmass}) and does not vanish without a massless quark. The reason is that the Higgs fields transform under \UPQ, and one of the Higgs fields, called `axion' $a$, has the shift symmetry $a\to a+(\rm constant)$ and corresponds to the Goldstone boson of the spontaneously broken \UPQ~ \citep[Weinberg (1978), Wilczek (1978)]{Wein78,Wil78}. So we call the resulting axion from Eq. (\ref{PQ}) the Peccei-Quinn-Weinberg-Wilczek (PQWW) axion. If the consequence of the determinental interaction is only Fig. \ref{fig:U1tHooft}, then out of two bosons $\eta'$ and $a$, only $\eta'$ obtains mass by the RHS diagram of Fig. \ref{fig:U1tHooft} and $a$ remains massless. If $a$ remains massless, the strong CP problem is solved as envisioned in \citep[Peccei, Quinn (1977)]{PQ77} since for any $\theta$ we can choose the VEV  $\langle a\rangle$ such that the final $\theta$ is zero. This was Peccei and Quinn's idea that $\langle a\rangle$ has a shift symmetry mimicking the massless quark case. However, $a$ has interactions and can be produced in the stars and $K$ meson decays, which differs from the massless quark case.

At the classical Lagrangian level, there seems to be no strong  \Ca\Pa~ problem. But, the axion coupling to $\{G\tilde G\}$ is generated at one loop level, which is the U(1)$_{\rm PQ}$-QCD-QCD anomaly. The 't Hooft determinental interaction we mentioned above is exactly this anomalous coupling. With this one loop term, the Lagrangian is not invariant under the phase shift symmetry $\beta$, or $a\to a+(\rm constant)$. Since it is explicitly broken at one loop level, the phase field $\beta$ of the Higgs fields, or axion $a$ does not have a flat potential, i.e. Fig. \ref{fig:U1tHooft} is not complete. Weinberg and Wilczek interpreted this phenomenon using the spontaneous symmetry breaking of the global symmetry U(1)$_{\rm PQ}$. It is said that $\theta$ is made dynamical where $\theta\equiv a/ F_a$, but in the PQWW axion case the component was there from the beginning in the phases of the Higgs doublet fields. The free energy depending on $-\cos\theta$ is the potential for the axion. Since it is proportional to $-\cos\theta$, the minimum of the potential is at $\theta=0$ in \Ca\Pa~ conserving theories \citep[Vafa, Witten (1984)]{VW84}, and thus the vacuum chooses $\theta=0$. We must discuss this effect below the chiral symmetry breaking scale as will be discussed in Subsec. \ref{subsec:axionmass}. Thus, the axion solution of the strong \Ca\Pa~ problem is a kind of cosmological solution. Note however that the weak \Ca\Pa~ violation shifts $\theta$ a little bit, leading to $\theta\sim O(10^{-17})$ \citep[Georgi, Randall (1986)]{Randall86}.

The PQWW axion is ruled out quickly \citep[Donnely {\it et. al.} (1978), Peccei (1978)]{Donnelly78,PQWWrule}, which was the reason for the popularity of calculable models in 1978 as discussed in Subsubsec. \ref{subsub:calculable}.  Nowadays, cosmologically considered axions are very light, which arises from the phase of SU(2)$\times$U(1) singlet scalar field $\sigma$. The simplest case is the Kim-Shifman-Vainstein-Zakharov (KSVZ) axion model \citep[Kim (1979), Shifman, Vainstein, Zhakharov (1980)]{Kim79,Shifman80} which incorporates a heavy quark $Q$ with the following coupling and the resulting chiral symmetry
\begin{align}
{\cal L}=&-\bar Q_LQ_R\sigma+({\rm h.c.})-V(|\sigma|^2)-\theta
\{F \tilde F\},\label{KSVZlag}\\
{\cal L}\to &-\bar Q_Le^{i\gamma_5\alpha} Q_R e^{i\beta}\sigma
+({\rm h.c.})-V(|\sigma|^2)\nonumber\\
&\quad\quad-(\theta-2\alpha)  \{G \tilde G\}.
\end{align}
Here, Higgs doublets are neutral under U(1)$_{\rm PQ}$. By coupling $\sigma$ to $H_u$ and $H_d$, one can introduce a PQ symmetry also, not introducing heavy quarks necessarily, and the resulting axion is called the Dine-Fischler-Srednicki-Zhitnitskii (DFSZ) axion
\citep[Dine, Fischler, Srednicki (1981), Zhitnitskii (1980)]{DFS81,Zhit80}. In string models, most probably both heavy quarks and Higgs doublets contribute to the $\sigma$ field couplings. The VEV of $\sigma$ is much above the electroweak scale and the axion is a {\it very light axion}.\footnote{Once it was called an invisible axion \citep[Wise, Georgi, Glashow (1981), Nilles, Raby (1981)]{Wise81,NillesRaby82} but is better to be called a very light axion due to the possibility of its detection.}
The SU(2)$\times$U(1) singlet $\sigma$ field may mix with the Higgs doublet component by a tiny amount, whence practically we can consider the axion as the phase of a singlet field $\sigma$, $\sigma=[(v+\rho)/\sqrt2]e^{ia/f_S} $ with the identification $a\equiv a+2\pi N_{DW}F_a$ with the axion period $2\pi N_{DW}F_a$. Note that we use $f_S$ for the VEV of $\sigma$ or the value relevant in the field space and $F_a$ defined from the coefficient of the anomaly term. Namely, the coefficient of the anomaly $\{G\tilde G\}$ defines $F_a$ as $\theta=a/F_a$ while the VEV($v$) of $\sigma, \sigma\propto e^{ia/v},$  defines $f_S$. The periodicity $2\pi$ of $\theta$ implies that $F_a$ cannot be larger than $v\equiv f_S$, and we have  $F_a=f_S/N_{DW}$. It has been shown that models with $N_{DW}\ne 1$ has an energy crisis problem in the standard Big Bang cosmology \citep[Sikivie (1982)]{Sikivie82}. But models with $N_{DW}= 1$ do not have such a problem due to the mechanism to convert the two dimensional axionic domain wall disks surrounded by the axionic strings into radiation \citep[Barr, Choi, Kim (1987)]{Barr87DW}.

\subsection{Axion shift symmetry and reparametrization invariance}\label{subsec:shiftsymm}

In the original PQWW axion model, the Lagrangian in the effective field theory language has been extensively discussed \citep[Donnelly, Freedman, Lytel, Peccei, Schwartz (1978), Peccei (1989)]{Donnelly78,Pecceirev89}. Here, due to the simplicity in the formulae, we present the variant-type axion models where the PQ charges are assigned only to the right-handed quark fields \citep[Bardeen, Peccei, Yanagida (1987)]{Bardeen87var}. This discussion will make it easier in introducing our general formulae below. The PQ current is \citep[Bardeen, Peccei, Yanagida (1987)]{Bardeen87var},
\begin{align}
J_\mu^{\rm PQ}=& F_a\partial_\mu a+x\sum_{i=1}^{N_g} \bar d_{Ri}\gamma_\mu d_{Ri}+(1/x)\sum_{i=1}^{N}\bar u_{Ri}\gamma_\mu u_{Ri}\nonumber\\
 &+(-x)\sum_{i=N+1}^{N_g}\bar u_{Ri}\gamma_\mu u_{Ri}
\end{align}
where $N_g$ is the number of families, $N$ is the number of up-type quarks coupled to $H_u$, and $x=\langle H_u\rangle /\langle H_d\rangle$. The color anomaly is non-vanishing, i.e. the divergence of $J_\mu^{\rm PQ}$ is
\begin{align}
\partial^\mu J_\mu^{\rm PQ}&= \frac12 N(x+\frac{1}{x})\frac{\alpha_c}{4\pi}G_{\mu\nu}^a \tilde G^{a\hskip 0.02cm \mu\nu}\nonumber\\
+ &m_u \bar u[i\gamma_5 e^{ia\gamma_5/F_a x}]u +m_d \bar d[i\gamma_5 e^{ia\gamma_5 x/F_a }]d\label{curdiv}
\end{align}
where we considered the one family model of $u$ and $d$ with $N=1$. If $N$ were zero, there is no color anomaly. For a nonvanishing $N$, we have to pick up the component orthogonal to the longitudinal $Z_\mu$. Since the axial-vector part of $Z_\mu$ current is proportional to $J_{\mu 3}^5$, any axial U(1) current orthogonal to the longitudinal $Z_\mu$ is an SU(2)$_{\rm flavor}$ singlet current constructed in terms of right-handed quark fields. These include the currents corresponding to both $\eta'$ and the PQ phase. Since $\eta'$ is known to be heavy, we integrate out $\eta'$ to obtain light fields below the chiral symmetry breaking scale. This corresponds to picking up an anomaly-free piece, orthogonal to the longitudinal $Z_\mu$. It is
\begin{align}
J_\mu^{a}=& J_\mu^{\rm PQ}-\frac12 N(x+\frac{1}{x})\frac{1}{1+Z}\left( \bar u\gamma_\mu\gamma_5 u+Z\bar d\gamma_\mu\gamma_5 d\right)
\label{PQvarcurrent}
\end{align}
where $Z=m_u/m_d$.  The divergence of (\ref{PQvarcurrent}) is proportional to $m_um_d$ which must be the case for the particle orthogonal to $\eta'$.

Below we use the typical axion model (\ref{KSVZlag})  because of its simplicity in assigning the PQ charges whenever an explicit example is needed. It has the following \UPQ~ charges, $\Gamma$,
\begin{table}[!h]
\begin{tabular}{c|lll}
{\rm Field}\ \ &\quad $\sigma$ &\quad $Q_L$ &\quad $Q_R$\\
 \hline
 $\Gamma$\ \ & \quad 1 &\quad $+\frac12$ &\quad$-\frac12$
\end{tabular}
\end{table}

\noindent In this example, the axial-vector current for \UPQ~ is $J_\mu^5=\bar Q\gamma_\mu\gamma_5 Q +v\partial_\mu a$ where $a$ is the phase field of $\sigma=(v/\sqrt2 )e^{ia/v}$. The current corresponds to the charge flow which satisfies the current conservation equation if the symmetry is exact. But the axial-vector current is in general violated at one loop by the anomaly \citep[Adler (1969), Bell, Jackiw (1969)]{Adler69,Bell69}, $\partial^\mu J^5_\mu=(N_Q g_c^2/32 \pi^2)G_{\mu\nu}^a \tilde G^{a\hskip 0.02cm \mu\nu}$, or $\partial^2 a=(N_Q g_c^2/32 \pi^2 v)G_{\mu\nu}^a \tilde G^{a\hskip 0.02cm \mu\nu}+(m_Q/v) \bar Q i\gamma_5 Q $ with the $Q$ number $N_Q$, which shows that the axion interaction with the SM fields is only the anomaly term (plus the anomalous coupling with the SM gauge fields). Here and in Eq. (\ref{curdiv}), we explicitly write the QCD coupling $g_c^2$, but in the remainder of the paper we absorb the gauge coupling in the gauge fields except in the experimental section, Sec. \ref{sec:AxionDet}. This axion is the one settling $\theta$ at zero; thus one needs the axion-gluon-gluon anomalous coupling for which the color anomaly of $J_\mu^5$ should exist. This kind of symmetry $\Gamma$ is called the PQ symmetry.

Axions are introduced as the Goldstone boson degree of a spontaneously broken global \UPQ~ symmetry in renormalizable gauge models \citep[Peccei, Quinn (1977), Kim (1985)]{PQ77,Kimcomp} and/or as a pseudoscalar degree in a more fundamental theory where the axion interaction arises as a nonrenormalizable anomalous interaction in the effective low energy theory. The most compelling nonrenormalizable interaction was observed in the compactification of 10-dimensional superstring models \citep[Witten (1984)]{Witten84}. Below, we will treat that the axion is present as a dynamical degree at the electroweak scale whether it arises from the spontaneously broken PQ symmetry or from a more fundamental theory with a nonrenormalizable anomalous coupling, and focus on QCD interactions containing the axion degree, $a=\theta F_a$. Then, let us collectively write the most general form of its interactions: the $c_1$ term is the derivative coupling respecting the PQ shift symmetry, the $c_2$ term is the phase in the quark mass matrix, and the $c_3$ term is the anomalous coupling or the determinental interaction ${\cal L}_{\rm det}$,
\begin{align}
{\cal L}_{\theta}&=\frac12 f_{S}^2\partial^\mu \theta\partial_\mu\theta-\frac1{4g_c^2} G_{\mu\nu}^a G^{a\hskip 0.02cm \mu\nu}+(\bar q_L i\Dslash q_L+\bar q_R i\Dslash q_R)\nonumber\\
&+ c_1(\partial_\mu \theta)\bar q\gamma^\mu\gamma_5 q
-\left(\bar q_L~ m~q_R e^{ic_2\theta}+{\rm h.c.}\right)\nonumber\\
&+c_{3} \frac{\theta}{32\pi^2}G^{a}_{\mu\nu}\tilde G^{a\hskip 0.02cm\mu\nu}\ ({\rm or}\ {\cal L}_{\rm det} ) \label{Axionint}\\
&+c_{\theta\gamma\gamma}\frac{\theta}{32\pi^2}F
^{i}_{\rm em,\mu\nu}\tilde F_{\rm em}^{i\hskip 0.02cm\mu\nu}+{\cal L}_{\rm leptons,\theta}\nonumber
\end{align}
where $\theta=a/f_S$ with the axion decay constant $f_S$ up to the domain wall number ($f_S=N_{DW}F_a$), $q$ is the fermion matrix composed of SU(3)$_c$ charge carrying fields. When the singlet scalar fields are easier to discuss, we use $f_S$, and when the anomaly term is easier to discuss, we use $F_a$. ${\cal L}_{\rm leptons,\theta}$ is the axion interaction with leptons. $c_1,c_2,$ and $c_3$ are pre-given coupling constants below the axion scale $f_S$  with the mass parameter $m$ defined to be real and positive below the electroweak scale. Then, the determinental interaction can be used instead of the $c_3$ term,
\begin{equation}
{\cal L}_{\rm det}=-2^{-1}ic_3\theta(-1)^{N_f}
\frac{e^{-ic_3\theta}}{K^{3N_f-4}} {\rm Det}(q_R\bar q_L)+{\rm h.c.}
\label{detint}
\end{equation}
where we multiplied the overall interaction by $\theta$ in the small $\theta$ region and require the periodicity condition, $c_3\theta=c_3\theta+2\pi$. The periodicity can be accomodated automatically if we replace $-2^{-1}ic_3\theta$ by 1, but then we must add a constant so that it vanishes at $\theta=0$. The sign is chosen such that the potential is a minimum at $\theta=0$ \citep[Vafa, Witten (1984)]{VW84}.
With the fixed phases, the $c_3$ term is given from the QCD vacuum structure (\ref{QCDvacuum}) which does not have any dimensional coupling. But the instanton physics necessarily introduces the instanton sizes and hence a kind of QCD scale $K$ for the interaction respecting the chiral transformation property for a flavor singlet operator ${\cal L}_{\rm det}$. We will use either the anomaly term or ${\cal L}_{\rm det}$. The $\theta$ dependence of the form (\ref{detint}) is $-c_3\theta\sin(c_3\theta)$ which has the parity symmetry $\theta\to -\theta$. The Fourier expansion satisfying these is
$$
 -2^{-1}c_3\theta\sin(c_3\theta)=-2^{-1}
 [1-\cos(c_3\theta)]+\sum_{n=2}
 a_n\cos(nc_3\theta)
 $$
where the Fourier coefficients satisfy, $\sum_{n=1}^\infty n^{2i}a_n =\delta_{i0}$. Neglecting the $n\ge 2$ terms, we will use just the $\cos(c_3\theta)$ dependence.

In the defining phase Eq. (\ref{Axionint}), the PQWW axion is given by $c_1=0, c_2\ne 0$, and $c_3=0$, the KSVZ axion by $c_1=0, c_2= 0$, and $c_3\ne 0$, the model-independent axion \citep[Witten (1984)]{Witten84} in superstring models  by $c_1=0, c_2= 0$, and $c_3\ne 0$, and the DFSZ axion by $c_1=0, c_2\ne 0$, and $c_3=0$. In general, axion models from high energy will have $c_2\ne 0$, and $c_3\ne 0$, and the shift symmetry allows $c_1\ne 0$ in a different basis. For simplicity, we discuss Eq. (\ref{Axionint}) for one flavor QCD first. For $N_f$ flavors, both $c_i$ and $\theta$ are defined from $N_f\times N_f$ matrices in addition to the anomalous coupling and hence the axion is included in Tr$\theta$ which also contains the $\eta'$ meson part of QCD. For $N_f$ flavors, $c_i\theta$ must be replaced by Tr$c_i\theta$. For the following illustrative discussion, we refer to one flavor QCD, but in Subsec. \ref{subsec:axionmass} in the axion mass estimation we present the full $N_f$ flavor QCD result with the chiral symmetry breaking taken into account.

For the case of axion mass, $c_1, c_2$ and $c_3$ terms may be relevant, but only the combination $c_2+c_3$ appears. This Lagrangian has a shift symmetry $a\to a+$ (constant), which reparametrizes the couplings between $c_1,c_2,$ and $c_3$. Explicitly, the axion field dependent changes of the quark fields $q_L\to e^{i\alpha a(x)}q_L$ and $q_R\to e^{-i\alpha a(x)}q_R$ give
$
c_1\to c_1-\alpha,
c_2\to c_2-2\alpha,
c_3\to c_3+2\alpha,
$
and it must give the same physics, i.e. \citep[Kim (1987), Georgi, Tomaras, Pais (1981)]{Kimrev,GeorgiTom81},
\begin{widetext}
\begin{eqnarray}
\Gamma_{1PI}[a(x), A_\mu^a(x); c_1, c_2, c_3,  m, \Lambda_{\rm QCD}]= \Gamma_{1PI}[a(x), A_\mu^a(x); c_1 -\alpha, c_2-2\alpha, c_3+2\alpha, m, \Lambda_{\rm QCD}].
\label{Gammaone}
\end{eqnarray}
\end{widetext}

\begin{figure}[!]
\resizebox{0.95\columnwidth}{!}
{\includegraphics{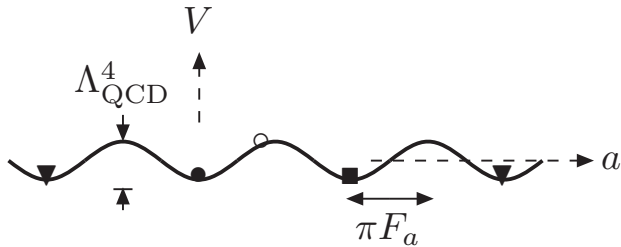}}
\caption{The case with $N_{DW}=3$ where three vacua
are distinguished.}\label{fig:axionDW}
\end{figure}

The reparametrization symmetry dictates the non-derivative couplings satisfying $c_2+c_3=$ (constant), which is one of the reasons that we use $\theta=\theta_{\rm QFD}+\theta_{\rm QCD} =\theta_0+\theta_{\rm weak}$ as a physical parameter in axion models. Usually, transferring all couplings of axion to the coefficient of $G\tilde G$, the axion decay constant $F_a$ and $\theta$ are defined. Instead, if we use $f_S$ (defined to be the VEV of the singlet Higgs field $\sigma$), there exists the coefficient $c_3$ defined in Eq. (\ref{Axionint}). The triangle diagrams may give an integer times $\theta$ and the instanton potential comes to the original value by a $\theta$ shift of $2\pi/(c_2+c_3)$, with $c_2+c_3=N_{DW}$ not necessarily 1 in the pseudoscalar field space. Thus, this integer is called the domain wall number $N_{DW}$ \citep[Sikivie (1982)]{Sikivie82}
\begin{equation}
N_{DW}=|c_2+c_3|={\rm Tr} \Gamma (f_{\rm colored}) \ell(f_{\rm colored})\label{DWnumber}
\end{equation}
where the trace is taken over all heavy and light quarks and $\ell$ is the index of $SU(3)_c$ representation of colored fermions and the PQ charge is given for the left-handed chiral representations.
The height of the potential is $O(\Lambda^4_{\rm QCD})$ of the nonabelian gauge interaction, which is shown in Fig. \ref{fig:axionDW} with the domain wall number $N_{DW}=3$: the bullet, the square and the triangle denote different vacua.
Two important properties of axions in \Ca\Pa~ conserving theories are:
\begin{itemize}
\item[(i)] periodic potential with the period $2\pi F_a$ where $F_a$ is defined in (\ref{Axionint}) with $F_a\equiv f_S/N_{DW}$, and
\item[(ii)] the minima are at $a=0,\ 2\pi F_a,\ 4\pi F_a, \cdots$.
\end{itemize}
This determines the cosine form of the potential. There exist the  axion mixing with quark condensations as we will discuss it in more detail later.

The derivative coupling, i.e. the $c_1$ term, can never contribute to the PQ symmetry breaking effect, especially to the axion mass. This axion gets its mass from the $\theta$ anomaly term which breaks the PQ symmetry. The global symmetry is not broken by the derivative term which therefore cannot contribute to the axion mass. From the reparametrization invariance (\ref{Gammaone}), the combination $c_2+c_3$ is the correct combination for the axion mass, which is shown explicitly below.  This derivation is included with a more complicated expression in the SUSY extension, but we show the $c_2+c_3$ dependence in this supergravity framework because it is the underlying symmetry in many axion models. Some of the following discussion is derived from \cite[Choi, Kim, Nilles (2007)]{ChoiKNilles07}.

\subsubsection*{Supersymmetrization}

We will discuss the reparametrization invariance with the SUSY generalization. In the $\N=1$ SUSY models with chiral fields $z$, there are the superpotential $W(z)$ and the gauge kinetic function $f(z)$ both of which are the holomorphic functions of $z$. The superpotential gives the $c_2$ term and the gauge kinetic function gives the anomaly term $c_3$. The PQ invariant Lagrangian, the $c_1$ part, has the shift symmetry under the shift of the axion supermultiplet: $A\to A+i({\rm constant})$. This derivative coupling must appear from the D terms in SUSY models, i.e. through the K\"ahler potential. The real K\"ahler potential $K(z,z^*)$ must respect the PQ symmetry in the form of $A+\overline{A}$,
\begin{align}
K=K_0[A+\overline{A}]+ \{Z_q[A+\overline{A}]\bar q_1q_2 +{\rm h.c.}\}\label{Kaehlerp}
\end{align}
where the $\vartheta^0$ components of the fields are implied and `$q$'s denotes quark supermultiplets,
\begin{equation}
q=\varphi_q+i\vartheta \psi_q
\end{equation}
with the anticommuting variable $\vartheta$. Here, we used $\vartheta$ for the anticommuting Grassmann number since $\theta$  in this review is reserved for the axion $\theta=a/F_a$.

\subsection{Axion mass}\label{subsec:axionmass}

The axion mass arises from the anomaly coupling $\theta G\tilde G$. In this subsection, first we show that only the $c_2$ and $c_3$ couplings are the relevant ones for the axion mass, and then we present the axion mass in the broken phase of the chiral symmetry.
With SUSY, the discussion is a bit tricky, because the axion remains massless due to the {\it massless gluino} (as in the massless up quark case with a spontaneously broken PQ symmetry). For the axion mass, therefore SUSY must be broken and here one has to look at all supergravity terms how they contribute to the axion mass. Nevertheless, we have the master formula (\ref{Gammaone}) for the axion which must be valid even when SUSY is broken. In this regard,  SUSY is not special for the axion mass but only the anomaly consideration is the chief constraint. Thus, the following discussion applies even without SUSY, but let us discuss the axion mass in detail with the SUSY generalization to include the gluino effects and hence the $c_1$ type derivative couplings to matter (quarks) and gauginos (gluinos).

We have noted that there exists the famous anomaly coupling of the $\eta'$ meson which is the mechanism solving the old U(1) problem of QCD. In addition to $\eta'$, the axion $a$ is introduced in the anomaly coupling and hence one must consider the mixing of $\eta'$ and axion \cite[Baluni (1979), Bardeen, Tye (1978), Kim, Kim (2006)]{Baluni79,Bardeen78,IWKim06}.

The $c_3$ term is the anomaly coupling of the axion, and we normalize the anomaly as the coefficient of $\epsilon^{\alpha\beta\gamma\delta} \varepsilon_{1\alpha} \varepsilon_{2\beta} k_{1\gamma} k_{2\beta}$. With this normalization, from $\epsilon_{\alpha\beta\gamma\delta}\partial^\alpha A^\beta \partial^\gamma A^\delta$ leading to $-\epsilon^{\alpha\beta\gamma\delta} k_{1\alpha} \varepsilon_{1\beta} k_{2\gamma} \varepsilon_{2\delta}$, the $c_3$ term anomaly is defined with ${\cal A}_3=1$.

It can be shown that, using the K\"ahler potential (\ref{Kaehlerp}),  the kinetic energy terms of fermions contain \citep[Cremmer, Ferrara, Girardello, van Pr\"oyen (1983), Nilles (1984)]{Cremmer83,Nilles84}
\begin{widetext}
\begin{equation}
\sum_\psi Z_q \left(\bar\psi i\dslash\psi+\frac16 B_\mu\bar\psi \gamma^\mu\gamma_5\psi+\frac12 Y_{q,\mu}\bar\psi\gamma^\mu\gamma_5\psi\right) +\sum_\lambda \left(\bar\lambda i\dslash\lambda-\frac12 B_\mu \bar\lambda\gamma^\mu\lambda
\right)
\end{equation}
where $B_\mu$ and $Y_{q,\mu}$ come from the auxilliary components of real $K_0$ and $Z_q$, respectively. In terms of the real parts $\cal R$ and $\cal Y$ of $K_0$ and ${\cal Z}$ (redefined from $Z_q$ of (\ref{Kaehlerp})), we obtain
\begin{align}
&B_\mu= \frac{i}{2}\left( \frac{\partial K_0}{\partial A} \partial_\mu A -\frac{\partial K_0}{\partial\overline{A}} \partial_\mu\overline{A} \right)=-\left( \frac{\partial {\cal R}}{\partial A}\partial_\mu a\right)\\
&Y_{q,\mu}=\frac{i}{2}\left( \frac{\partial { K}_0}{\partial A}\partial_\mu A -\frac{\partial { K}_0}{\partial \overline{A}}\partial_\mu \overline{A}\right)
-i\left( \frac{\partial\ln {Z}_q}{\partial A}\partial_\mu A - \frac{\partial\ln {Z}_q}{\partial \overline{A}}\partial_\mu \overline{A}\right)=2\left( \frac{\partial\ln {\cal Y}}{\partial A} \right)\partial_\mu a
\end{align}
\end{widetext}
where
\begin{align}
&G=-3\ln\left(\frac{-K}{3}\right)+\ln |W|^2,\ K=-e^{-K_0/3},\  Z_q=e^{{\cal Z}},\nonumber\\
&K_0={\cal R}+iK_{0I}={\cal R},\  Z_q={\cal Y}+i{\cal I}={\cal Y}\ .
\end{align}

\begin{figure}[!h]
\resizebox{.7\columnwidth}{!}
{\includegraphics{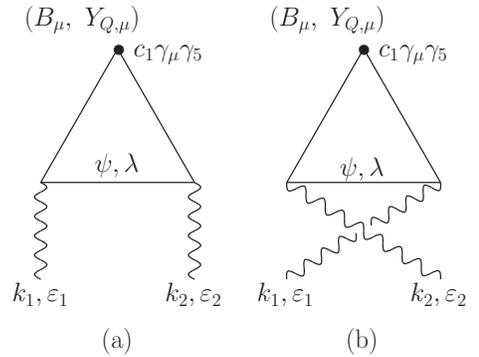}}
\caption{The Feynman diagrams for generating anomalous $\theta G\tilde G$ couplings from $c_1$ for a fermion with mass $m$. For $c_2$, we replace $c_1\gamma_\mu\gamma_5$ by $c_2m\gamma_5$.}\label{fig:Anomaly}
\end{figure}

The $c_3$ term is an anomaly term. In addition to the $c_3$ term, the $c_1$ and $c_2$ couplings via loops of Fig. \ref{fig:Anomaly} will also generate the anomaly terms. The derivative coupling, if it has ever to contribute to the axion mass, should contribute to the axion mass by the anomaly through loops.
In Fig. \ref{fig:Anomaly}, the couplings for the triangle diagrams are represented in terms of $c_1$ and $c_2$. In supergravity models, we consider $B_\mu$ and $Y_{q,\mu}$ couplings which are nothing but $c_1$. Consider a fermion with mass $m$. The derivative coupling through Fig. \ref{fig:Anomaly} has the anomaly coupling by picking up the coefficient of $\epsilon^{\alpha\beta\gamma\delta} \varepsilon_{1\alpha} \varepsilon_{2\beta} k_{1\gamma} k_{2\beta}$ \citep[See, for example, Georgi,Tomaras, Pais(1981)]{GeorgiTom81},
\begin{equation}
{\cal A}_1=\int_0^1 dx_1\int_0^{1-x_1} dx_2
\frac{-4f(x_1,x_2;q,k_1,k_2)
}{m^2-f(x_1,x_2;q,k_1,k_2)}\label{A1anom}
\end{equation}
where
$$
f=(x_1+x_2)(1-x_1-x_2) q^2 +2x_1 (1-x_1-x_2)q\cdot k_1
+x_1^2 k_1^2.
$$
Also, the quark mass term of Fig. \ref{fig:Anomaly} gives
\begin{equation}
{\cal A}_2=\int_0^1 dx_1\int_0^{1-x_1} dx_2
\frac{2m^2}{m^2-f(x_1,x_2;q,k_1,k_2)}.\label{A2anom}
\end{equation}
From Eqs. (\ref{A1anom}) and (\ref{A2anom}), we construct
\begin{eqnarray}
\frac12{\cal A}_1+{\cal A}_2=\int_0^1 dx_1\int_0^{1-x_1}2 ~dx_2 =1.\label{feynfunc}
\end{eqnarray}

When we calculate the axion mass in the real and positive quark mass basis as usual, the anomaly coupling $(a/F_a)\{G\tilde G\}$ coupling (including the loop effect) is the sole source to the axion mass. In this basis, and also in any basis due to the reparametrization invariant combination $c_2+c_3$, we do not have to discuss the derivative couplings toward the axion mass. But, even though the derivative coupling generates the anomaly, it being derivative does not contribute to the axion mass.

For {\it one flavor QCD}, we can check the above statement explicitly using Eqs. (\ref{A1anom},\ref{A2anom},\ref{feynfunc}).
In the following two limiting cases, the integrals are easily computed as, using Eqs. (\ref{A1anom}) and (\ref{A2anom}),
\begin{widetext}
\begin{eqnarray}
&&{\rm Case~ (i)}:\ m\ll\Lambda_{QCD}: \Gamma_{1PI}=\frac{1}{16\pi^2} \epsilon^{\alpha\beta\gamma\delta}k_{1\gamma}k_{2\beta}
\left(c_3+2c_1+{\cal O}(\frac{m^2}{k^2}) \right)\label{anomalylight}\\
&&{\rm Case~ (ii)}:\ m\gg\Lambda_{QCD}: \Gamma_{1PI}=\frac{1}{16\pi^2} \epsilon^{\alpha\beta\gamma\delta}k_{1\gamma}k_{2\beta}
\left(c_3+c_2+{\cal O}(\frac{k^2}{m^2}) \right) \label{anomalyheavy}
\end{eqnarray}
\end{widetext}
Consider the quark mass term and the one flavor determinental interaction with the quark condensation, $\langle\bar q_Lq_R \rangle\sim \Lambda_{\rm QCD}^3e^{i\eta'/f}$. Then, the potential takes the form
\begin{equation}
V=m\langle\bar q_L q_R\rangle e^{ic_2\theta}+{\rm h.c.} +(c_3+c_1{\cal A}_1+c_2{\cal A}_2)\left\{G\tilde G\right\}.
\end{equation}
For the anomaly combination $c_3+c_1{\cal A}_1+c_2{\cal A}_2$, the reparametrization invariance Eq. (\ref{Gammaone}) transforms $c_3+c_1{\cal A}_1+c_2{\cal A}_2$ to $c_3+2\alpha+(c_1-\alpha){\cal A}_1+(c_2-2\alpha){\cal A}_2= c_3+c_1{\cal A}_1+c_2{\cal A}_2$ where (\ref{feynfunc}) is used, i.e. it is reparametrization invariant.

For {Case (i)}, we consider the light quark below the scale $\Lambda_{\rm QCD}^4 $. Thus, we have
$$
V=mv^3\cos\left(\frac{\eta'}{f}-c_2\theta\right) +\Lambda_{\rm QCD}^4 \cos\left((c_3+2c_1)\theta+\frac{\eta'}{f}\right)
$$
for which we choose $c_1=0$. [If we kept $c_1$, we must consider the kinetic mixing of $a$ and $\eta'$.]
Integrating out the heavy $\eta'$ field as $\frac{\eta'}{f}=-c_3\frac{a}{f_S}$ from the $\Lambda_{\rm QCD}^4$ term which is the larger one, we obtain
$$
V= mv^3\cos\left((c_2+c_3) \frac{a}{f_S}\right)
$$
from which we obtain
\begin{equation}
 m_a\sim
{\sqrt{m\Lambda_{\rm QCD}^3}}~\frac{|c_2+c_3|}{f_S}.
\label{axionmass1}
\end{equation}
Quarks $u,d,$ and $s$ belong to this category.

For {Case (ii)}, the heavy quark does not condense and integrating out the heavy quark gives
$$
V=\Lambda_{\rm QCD}^4 \cos\left((c_3+c_2)\frac{a}{f_S}\right)
$$
from which the axion mass is given by
$$
m_a\sim {\Lambda_{\rm QCD}^2}~\frac{|c_2+c_3|}{f_S}.
$$
Again the axion mass depends only on the combination $c_2+c_3$. Heavy quarks above the chiral symmetry breaking scale $c, b,$  and $t$ give the $c_2$ term and vectorlike heavy quarks above the electroweak scale give the $c_3$ term when we wrote Eq. (\ref{Axionint}) just below the electroweak scale.
\vskip 0.3cm

\subsubsection*{Axion mass with light quarks}

\begin{figure}[!]
\resizebox{0.85\columnwidth}{!}
{\includegraphics{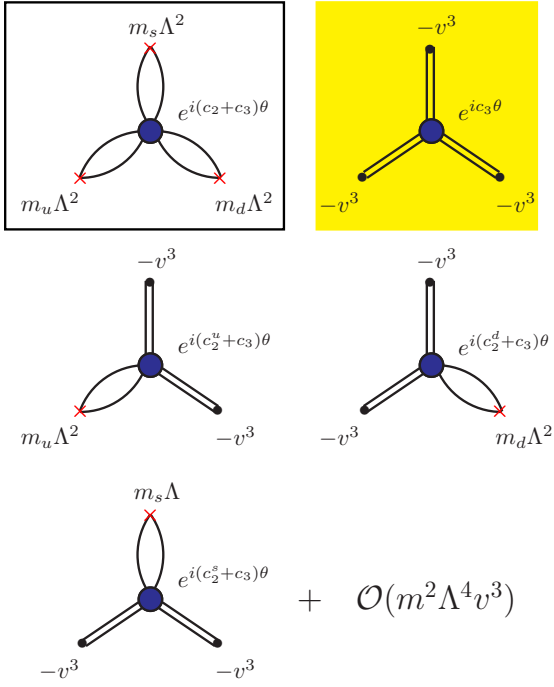}}
\caption{The 't Hooft determinental interaction. $\bullet$ denotes the quark condensation and $\times$ denotes the insertion of the current quark mass. The diagram highlighted with yellow predominantly contributes to $\eta'$ mass, and ${\cal O}(m_um_d)$ is neglected.}\label{fig:tHooft}
\end{figure}

In the real world, there exist three light quarks whose masses are much smaller than the QCD scale $\Lambda_{\rm QCD}$, and therefore the axion mass has the form anticipated in Eq. (\ref{axionmass1}). Even though there are two light quarks the axion mass dependence has the form $\sqrt{m}$ because of $F_a\gg f_\pi$. This is because of the way in picking up the leading term from the anomalous determinental interaction \citep[Kim, Kim (2006)]{IWKim06} as depicted in Fig. \ref{fig:tHooft}.

In fact, this is simply obtained just by noting that the instanton interaction is a U(1) singlet \citep[Kim (1987)]{Kimrev}. Suppose we integrate out quark fields; then the quark mass parameters appear in the effective interaction as shown in the first diagram of Fig. \ref{fig:tHooft}. In this vacuum in a theory with a massless quark, the tunneling amplitude vanishes so that the strength of the first diagram must be proportional to $m_q$. With three quarks, we can generalize it as $1/(m_u^{-1}+m_d^{-1}+m_s^{-1})$. Suppose that there are only gluons and a very light axion $a$ at low energy. Integrating out heavy fields, we are left with the flavor independent coupling $aG\tilde G$. Here we are not considering $\eta'$ even below the quark condensation scale. If quarks are added, the flavor singlet coupling $aG\tilde G$ can be split to quark mass terms with
$\alpha_u\propto x/m_u,~ \alpha_d\propto x/m_d,~ \alpha_s\propto x/m_s $, etc., as if quarks are not integrated over, $m_u\bar u_L u_Re^{i\alpha_u \theta}+m_d\bar d_L d_Re^{i\alpha_d \theta}+\cdots$, which shows that the flavor singlet coupling is of order ${\cal O}(a/F_a)$. Then, even below the chiral symmetry breaking scale, we have the PQ charges proportional to $1/m_q$. With this definition of quark charges, the axion mass comes from integrating out $G\tilde G$, and is proportional to $\alpha_u+\alpha_d+\alpha_s$ which is $\sim m_um_dm_s/ [m_s(m_u+m_d)+m_um_d]$, which was shown first for the PQWW axion \citep[Baluni (1979)]{Baluni79}. This is true even in the heavy quark KSVZ type axion models.
Even if the light quarks do not have the same PQ charge as in some variant axion models
\cite[Krauss, Wilczek (1986), Peccei, Wu, Yanagida (1986), Bardeen, Peccei, Yanagida (1987), Kim, Lee (1989), Hindmarsh, Moulatsiotis (1997)]{varKrausWil86,Peccei86,Bardeen87var,KimLee89,varHind97}, the axion mass has the same final form due to the reparametrization invariance, which will be shown below. So, the axion mass formula we write below is quite general.

However, there was an assumption in this statement: $\eta'$ was integrated out. So, it is necessary to include $\eta'$ to obtain a more accurate axion mass. The light mesons and axion interactions must appear from those of Fig. \ref{fig:tHooft}. In this framework, however, {\it the flavor singlet condition  must be invoked as a constraint}. [This flavor singlet condition is the anomaly matching without $\eta'$.] Along the way, we would like to see how the $\sqrt{m}$ dependence arises below the chiral symmetry breaking scale in the KSVZ model.

In the presence of vectorlike heavy quarks, the heavy fields are integrated out, whose sole effect is encoded in the low energy effective theory as nonrenormalizable couplings suppressed by $F_a$, e.g. in the anomalous $c_3$ type couplings with the SM gauge bosons. It is assumed that the heavy quark does not condense, since the QCD coupling is very small due to the asymptotic freedom at the heavy-quark Compton wave length scale and there does not exist a strong force to attract heavy quarks. Below the heavy quark scale, there is no massless mesons composed of heavy quarks. Therefore, the general form of the axion interaction Eq. (\ref{Axionint}) is valid at low energy. Firstly, the determinental interaction has the same chiral symmetry behavior as that of the anomaly term, and  the anomaly term is removed in favor of the determinental interaction to include $\eta'$ explicitly. Second, let us choose the basis where the $u$ and $d$ quark masses are real. Since the strange quark mass is known to be below the QCD scale, we must include the strange quark with real and positive mass also in the instanton interaction. For simplicity, $\pi^0$ and $\eta'$, arising from quark condensations $\bar uu$ and $\bar dd$ with decay constants $f_\pi$ and $f_{\eta'}(\approx f_\pi)$ \citep[Gell-Mann, Oakes, Renner (1968)]{GellMann68}, are considered explicitly but with the $\eta$ meson frozen. The effects of heavy quarks are included in the $c_3$ term. If we keep $c_1$, the kinetic mixing of mesons and axion are present, due to the PCAC relation $\langle 0|J^\mu_{5i}(x)| {\rm meson}_j(k) \rangle=-ik^\mu f_i^2 e^{-ik\cdot x}\delta_{ij}$ where $J^\mu_{5i}\sim\bar q\gamma^\mu\gamma_5 T_i q$. This would modify the axion mass, and hence it is easiest to calculate the axion mass by choosing the reparametrization parameter $\alpha$ such that $c_1=0$. In this basis, denoting $\pi^0,\eta'$ and $a$ in terms of dimensionless fields, $\theta_\pi=\pi^0/f_\pi, \theta_{\eta'}=\eta'/f_{\eta'}, \theta=a/F_a$, we obtain the following effective interaction below the chiral symmetry breaking scale, \begin{widetext}
\begin{equation}
{\cal L}=-m_u\langle \bar u_Lu_R\rangle e^{i[(\theta_\pi+\theta_{\eta'}) +c_2^{u}\theta]}-m_d \langle \bar d_Ld_R\rangle e^{i[(-\theta_\pi+\theta_{\eta'}) +c_2^{d}\theta]}+{\rm h.c.}+{\cal L}_{\rm det}
\end{equation}
where ${\cal L}_{\rm det}$ is given in Eq. (\ref{detint})
\begin{equation}
{\cal L}_{\rm det}= (-1)^{N_f} K^{-5}\left(\langle\bar u_Lu_R\rangle\langle
\bar d_Ld_R\rangle\langle \bar s_Ls_R\rangle e^{i(2\theta_{\eta'}-c_3\theta)}+\cdots+{\rm flavor~singlet~ constraint}\right)+ {\rm h.c.},
\label{DetCondInt}
\end{equation}
\end{widetext}
where $K$ has the mass dimension arising from QCD instanton physics. The above form is consistent with the anomaly (\ref{anomalylight}) with $c_1=0$. Note that the log~det form in the effective Lagrangian was used in \citep[Veneziano (1979), Di Vecchia, Veneziano (1980), Witten (1979, 1980), Di Vecchia, Nocodemi, Pettorino, Veneziano (1981)]{Veneziano79,DiVecchia80,Witten79,
Witten80Ann,DiVecchia81} from the $1/N_c$ expansion consideration, but we use (\ref{DetCondInt}) because of its simplicity in the diagramatic expansion. The sign of the first diagram inside the box in Fig. \ref{fig:tHooft} is determined to be minus without the weak \Ca\Pa\ violation \citep[Vafa, Witten (1984)]{VW84}.  The QCD vacuum with the flavor independence of light quarks without the determinental interaction chooses $m_q\langle \bar qq\rangle=-|m_q|v^3$ and we choose the sign of all quark masses be positive so that $\langle \bar qq\rangle= \langle \bar uu\rangle= \langle \bar dd\rangle=-v^3$ \citep[Dashen (1971), Langacker, Pagels (1973, 1979), Gasser, Leutwyler (1982)]{Dashen71,Langacker73,Langacker79,Gasser82}. Eq. (\ref{DetCondInt}) is the instanton interaction of Fig. \ref{fig:tHooft}, which gives $\Lambda^4 , m_u\Lambda^3, m_d\Lambda^3,\cdots$ by many ways of closing quark lines, shown in Fig. \ref{fig:tHooft}, but here one must invoke the {\it flavor singlet constraint}. The dominant term is the second diagram highlighted as yellow, which is flavor singlet and is the main source for the $\eta'$ mass.

Now let us restrict to the two flavor case. For the axion, the key diagrams are those in the second line of Fig. \ref{fig:tHooft}. If there are more than one QCD axion, then ${\cal O}(m_um_d)$ diagram will be important at the next level axion mass. Integration over the instanton size includes very large instantons which covers the chiral symmetry breaking range where mesons appear as dynamical degrees, where we will invoke the flavor singlet constraint. The effective interaction Hamiltonian of $\theta_\pi,\theta_{\eta'}$ and $\theta=a/f_S$ can be written, using the reparametrization invariance (\ref{Gammaone}) with $N_f=3$ and $\eta$ fixed, as \citep[Huang (1993), Kim, Kim (2006)]{Huang93,IWKim06},
\begin{widetext}
\begin{eqnarray}
-V&=& m_uv^3\cos\left({\theta_\pi+\theta_{\eta'}} \right) +m_dv^3\cos\left({-\theta_\pi+\theta_{\eta'}} \right)+\frac{v^9}{K^5}\cos\left(2\theta_{\eta'}
-(c_2^u+c_2^d+c_3)\theta \right)\nonumber\\
&& +m_u\frac{\Lambda_u^2 v^6}{K^5} \cos \left(-\theta_\pi+{\theta_{\eta'}} -(c_2^u+c_2^d+c_3)\theta \right)+m_d\frac{\Lambda_d^2 v^6}{K^5} \cos \left(\theta_\pi+{\theta_{\eta'}} -(c_2^u+c_2^d+c_3)\theta \right)\label{axpotentbasic},
\end{eqnarray}
\end{widetext}
where $\Lambda_u$ and $\Lambda_d$ are parameters describing the result of the Feynman and instanton size integrations. The $(-1)^{N_f}$ is cancelled by the fermion loop or $(-v)$ factors. If $m_u=m_d$, $\Lambda_u$ and $\Lambda_d$ are equal. For $m_u\ne m_d$,  $\Lambda_u$ and $\Lambda_d$ must be different. The instanton interaction is flavor independent, which should be respected in the interaction (\ref{axpotentbasic}).  The $m_u$ and $m_d$ linear terms from the determinental interaction should be flavor independent, i.e. $m_u\Lambda_u^2+ m_d\Lambda_d^2=\textrm{flavor independent}$.  Since it vanishes if one quark is massless, it must be a function of $m_u m_d$. Thus, the instanton size integration with current quark masses must give $m_u\Lambda_u^2+m_d\Lambda_d^2=2m_u m_d\tilde L^2/(m_u+m_d)$, which vanishes if any quark is massless. This is because the original gluon anomaly term $\{G\tilde G\}$ does not distinguish flavors, and the smallness of the current quark masses enable us to expand the 't Hooft determinental interaction in terms of powers of the current quark masses. Then, the $3\times 3$ mass matrix $M^2$ of $a,\eta'$ and $\pi^0$, taking into account of the chiral symmetry breaking and the solution of the U(1) problem is given as
\begin{widetext}
\begin{eqnarray}
M^2_{a,\eta',\pi^0}=\left(\begin{array}{ccc}
c^2[\Linsteta^4+2\mu\Linst^3]/F^2 \ \
&-2c[\Linsteta^4+\mu\Linst^3]/f'F& 0\\ &&\\
-2c[\Linsteta^4+\mu\Linst^3]/f'F\ \  & [4\Linsteta^4
 +2\mu\Linst^3+{m_+}v^3  ]/f'^2  & -m_-v^3/ff'\\
&& \\
0 &-m_-v^3/ff' &(m_+v^3+2\mu\Linst^3)/f^2
\end{array}\right)\label{massmatrix}
\end{eqnarray}
\end{widetext}
where $ c=c_2^u+c_2^d+c_3, F=f_S, f=f_\pi,f'=f_{\eta'},\Linsteta^4=v^6/K'^2$, and $\Linst^3=\tilde L^2v^3/K^2, m_+=m_u+m_d, m_-=m_d-m_u$, and
\begin{equation}
  \mu=\frac{m_um_d}{(m_u+m_d)}.
\end{equation}
Certainly, Eq. (\ref{massmatrix}) realizes the solution of the U(1) problem due to the $\Linsteta^4$ term in the (22) component.  In the limit $f/F, f'/F\ll 1$, we obtain
\begin{eqnarray}
m^2_{\pi^0} &\simeq& \frac{m_+v^3+2\mu \Linst^3}{f_{\pi}^2}\label{pimass}\\
m^2_{\eta'} &\simeq& \frac{4\Linsteta^4+m_+v^3+2\mu \Linst^3}{f_{\eta'}^2}\label{etapmass}\\
m^2_a &\simeq&
\frac{c^2}{F^2}\frac{Z}{(1+Z)^2}f_{\pi}^2 m_{\pi^0}^2\left(1+\Delta \right)
\label{manomatch}
\end{eqnarray}
where
\begin{equation}
\Delta=\frac{m_-^2}{m_+}~\frac{\Linst^3(m_+v^3
+\mu\Linst^3)}{m_{\pi^0}^4 f_\pi^4}.\label{AxmassDelta}
\end{equation}
In this form, the $\pi$ mass has the standard $m_+v^3$ plus the instanton contribution to light quark masses \citep[Kaplan, Manohar (1986), Choi, Kim, Sze (1988)]{KapMan86,ChoiK88b}.  From (\ref{pimass},\ref{etapmass}), we estimate the parameter $\Linsteta^4$ which is the source of the solution of the U(1) problem: $\Linsteta^4=(f^2_{\eta'}m^2_{\eta'}-f^2_{\pi}m_\pi^2)/4
\approx (202~\rm MeV)^4$ with $f_{\eta'}\simeq 86$ MeV and $f_\pi\simeq$ 93 MeV.
In any axion model, this form is valid with $|c|=N_{DW}$.
Using the standard definition on the axion decay constant $F_a=F/c$, we obtain
\begin{eqnarray}
m^2_a &\simeq&
\frac{Z}{(1+Z)^2}\frac{f_\pi^2 m_\pi^2}{F_a^2}\left(1+\Delta \right).
\label{axionmass}
\end{eqnarray}
So, even though the instanton diagrams of Fig. \ref{fig:tHooft} contain the linear quark mass diagrams summed over, the diagonalization process with mesons signals the predominant contribution of the lightest quark. The flavor singlet condition we discussed before chooses the following linear quark mass dependence
\begin{equation}
\mu=\left(\frac{1}{m_u}+\frac{1}{m_d}+\cdots\right)^{-1}.
\label{Axphase}
\end{equation}
Neglecting instanton contribution to the current quark masses, we obtain $m_a\approx$ 0.60~eV~$[10^7{\rm GeV}/F_a]$, for the mass ratio $Z\simeq 0.48$ as summarized in \citep[Manohar, Sachrajda (2008)]{Manohar08}. An earlier frequently cited $Z$ is $5/9$ \citep[Weinberg (1979), Gasser, Leutwyler (1982)]{Wein79up,Gasser82}. The correct axion mass has to include the current quark mass change due to instantons. However, the resulting estimate of $\Delta$ turns out to be small.

\subsubsection*{Comparison with old calculation}

Now, let us comment on the old anomaly matching condition. If any quark mass is zero, there exists an exact symmetry $a\to a+{\rm (constant)}$, i.e. axion is massless, above the chiral symmetry breaking scale. Below the chiral symmetry breaking scale, it is likely that this condition is satisfied. Let us denote the original current as $J_{\rm PQ}^\mu$. This current is anomalous above the chiral symmetry breaking scale, $\partial_\mu J_{\rm PQ}^\mu = (N_Q/32\pi^2)G_{\mu\nu}^a \tilde G^{a\hskip 0.02cm \mu\nu}$ where $N_Q$ is the number of heavy quarks with $\Gamma=1/2$. Below the chiral symmetry breaking scale, we considered two pseudoscalar mesons which have anomalous couplings: $\eta'$ and $a$. The global anomaly matching condition will work if there were no chiral symmetry breaking \citep['t Hooft (1979)]{tHooftNatur79}. For chiral symmetry breaking, there is no massless fermions and we consider only color singlet mesons below the chiral symmetry breaking scale. Thus, the bosonic current must be anomaly free after integrating out all heavy fields including $\eta'$, i.e. we consider an anomaly free current $J^\mu_a$ instead of $J_{\rm PQ}^\mu$ below the chiral symmetry breaking scale \citep[Kim (1987)]{Kimrev},
\begin{equation}
J^\mu_a= J_{\rm PQ}^\mu-\frac{N_Q}{2(1+Z)} (\bar u\gamma_\mu\gamma_5 u +Z\bar d\gamma_\mu\gamma_5 d) \label{axioncurrent}
\end{equation}
where the divergence of the second current gives a singlet pseudoscalar density so that axion does not mix with $\pi^0$. The expression (\ref{axionmass}) with $\Delta$ shows that the finite $\eta'$ mass enters into the $a-\eta'$ mixing.

\subsubsection*{Mesons without axion}

Even if there is no axion, we can diagonalize the mass matrix. If $m_u=0$, one starts with an exact up quark chiral transformation, which would lead to a Goldstone boson $\theta$ in the vacuum, $\langle \bar uu\rangle\ne 0$. This Goldstone boson couples to neutron through $(c_1\partial_\mu\theta) \bar n\gamma^\mu\gamma_5 n$. In reality, $\eta'$ obtains mass by the anomaly, and the symmetry remains unbroken: it is the phase symmetry of $\langle \bar uu\rangle$. Therefore, any violation of the shift symmetry must be invoked such that it goes away in the limit $\mu\to 0$, which is Dashen's theorem \citep[Dashen (1971)]{Dashen71}.  Thus, from Eq. (\ref{axpotentbasic}) we obtain the VEVs of $\eta'$ and $\pi^0$ for a small $\bar\theta$
\begin{eqnarray}
&&\frac{\langle\eta'\rangle}{f'}\simeq -\frac{\bar\theta}{2} (1+Z) \frac{\mu v^3}{\Linsteta^4}\nonumber\\
&&\frac{\langle\pi^0\rangle}{f}\simeq {\bar\theta} (1+Z) \frac{\mu}{m_+}\ .\label{VEVetap}
\end{eqnarray}
The VEVs of $\eta'$ and $\pi^0$ are vanishing if $\bar\theta=0$ or any quark mass is zero.
In addition, we can estimate the $\eta'$ properties from the interaction $(v^9/K^5)\cos (2\eta'/f_{\eta'})$ where $f_{\eta'}$ is the $\eta'$ decay constant and $K$ has a mass dimension. It comes from the yellowed diagram of Fig. \ref{fig:tHooft}. Comparing $\pi^0\to 2\gamma$ and $\eta'\to 2\gamma$ decay widths, 7.74 eV and 4.3 keV, respectively \citep[Amsler {\it et. al.} (Particle Data Group, 2008)]{PData08}, we obtain $f_{\eta'}^2=(4/3)(m_{\eta'}^3/m_\pi^3)(\Gamma(\pi^0\to 2\gamma/\Gamma(\eta'\to 2\gamma))f_\pi^2$, or $f_{\eta'}\approx 86$ MeV.  Fitting to the $\eta'$ mass, we obtain $K=(v^9/f_{\eta'}^2 m_\eta'^2)^{1/5}=240$ MeV.

\subsubsection*{The $\theta=0$ vacuum with axion}

We have shown above that the Lagrangian (\ref{axpotentbasic}) chooses $\theta=0$ in \Ca\Pa~ conserving theories if $\theta_\pi=0$ and $\theta_{\eta'}=0$, which is determined by QCD dynamics. However, if \Ca\Pa~ is broken,  the vacuum value of $\theta$ is shifted from the $\theta=0$ value, by the presence of any linear term of $\theta_\pi, \theta_{\eta'}$ or/and $\theta$. The meson potential is invariant under \Ca\Pa~ with ${\cal CP}(\pi)={\cal CP}(\eta')= {\cal CP}(a)=-1$. Such linear terms are generated by considering \Ca\Pa~ violating phases and chirality flipping (L$\leftrightarrow$R) insertions. So, linear terms of $a$ are generated by combining the 't Hooft determinental interaction and \Ca\Pa~ violating weak interactions. Linear terms of $\pi$ and $\eta'$ can also be generated by considering the weak interactions alone without the determinental interaction, but the conditions of the flavor singlet, chirality flipping (L$\leftrightarrow$R) and \Ca\Pa~ violating effects do not occur at one loop level. In the SM with the Kobayashi-Maskawa \Ca\Pa~ violation, Ref. \citep[Ellis, Gaillard (1979)]{EllisGai79} shows that a finite correction occurs at the fourth order, ${\cal O}(\alpha^2)$, leading to a very small NEDM, but infinite corrections occur from ${\cal O}(\alpha^7)$. These can give rise to a linear term of $\pi$. In the SM, the pioneering calculation with axion has been performed in chiral perturbation theory to obtain $\theta\le 10^{-17}$  \cite[Georgi, Kaplan, Randall (1986), Georgi, Randall (1986)]{Georgi86,Randall86}. The estimated $\theta$ however is far below the current experimental limit of $10^{-11}$.

\subsection{Axion couplings} \label{subsec:axioncouplings}

\begin{figure}[!h]
\resizebox{0.9\columnwidth}{!}
{\includegraphics{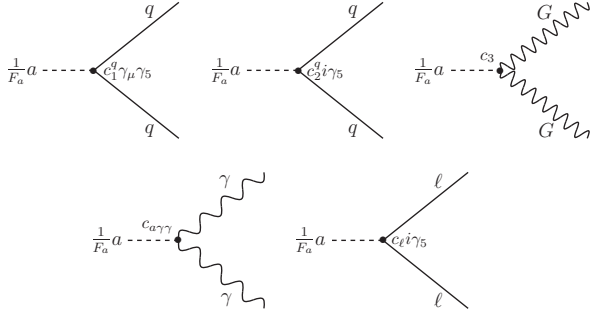}}
\caption{The Feynman diagrams of axion couplings. $G$ and $\gamma$ are gluon and photon, respectively. $c_3$ and $c_{a\gamma\gamma}$ couplings are anomalous.} \label{fig:axInt}
\end{figure}

The axion interactions are given in Eq. (\ref{Axionint}) which are depicted in Fig. \ref{fig:axInt} where we have not drawn $aW\tilde W$ and $aZ\tilde Z$ diagrams which are orthogonal to the $a\gamma\tilde\gamma$. The diagrams of Fig. \ref{fig:axInt} are complete for the low energy axion phenomenology, where the suppression factor, $1/F_a$, by the axion decay constant is explicitly shown.

\subsubsection{Axion hadron coupling} \label{subsub:ahadron}

When we discuss axion--hadron interactions, which are the relevant ones for low energy laboratory experiments and physics at the core of supernovae, we must integrate out gluon fields. Technically, it is achieved by using the reparametrization invariance to remove the $c_3\theta G\tilde G$ coupling. If we keep the $c_3$ coupling, we must consider the axion--gluon--gluon interactions also, which is hard to be treated accurately with its face value but must be the same as the one in the $c_3=0$ basis. In this way, the quark interactions are changed from the original value as
\begin{eqnarray}
\begin{array}{l}
c_1\to \bar c_1=c_1+\frac12 c_3\\
 c_2\to \bar c_2=c_2+c_3\\
c_3\to \bar c_3=c_3-c_3=0.
\end{array}\label{axhadint1}
\end{eqnarray}
In the barred notation, there exist only $\bar c_1$ and $\bar c_2$.

We will discuss one family without separating $c_{1,2}$ into $c_{1,2}^{u,d}$ first for an illustration, and then we will discuss the cases with $c_{1,2}^{u,d}$ and write down formulae for three families. Let us define the initial parameters $c_1, c_2$ and $c_3$ together with the definition of the vacuum angle $\theta_0\equiv \theta_{\rm QCD}$. In principle, the initial vacuum angle can be a free parameter. Here, the vacuum angle $\theta_{\rm QCD}$ is defined such that $c_1=0$.
Picking up the axion dependent chiral rotation charge defined below the chiral symmetry breaking scale Eq. (\ref{axioncurrent}), the chiral quarks in the chiral perturbation theory are transforming as $q_L\to \exp(iQ_A\theta)q_L, q_R\to \exp(-iQ_A\theta)q_R$ where
\begin{equation}
Q_A=\frac12 \frac{M^{-1}}{{\rm Tr}M^{-1}},\quad M^{-1}={\rm diag.}(\frac{1}{m_u},\frac{1}{m_d}).
\label{anomalytoqk}
\end{equation}
The derivative interactions of axion is obtained in this way \cite[Kaplan (1985), Georgi, Kaplan, Randall (1986)]{Kaplan85,Georgi86}.

For the KSVZ axion, we have $c_1=c_2=0$ and $c_3=1$, and the coefficient of the gluon anomaly term is $(a/F_a)+\theta_{\rm QCD}$. Hence, redefining the axion as $a+F_a\theta_{\rm QCD}$, we obtain\footnote{The sign convention is stated below.}
\begin{eqnarray}
\begin{array}{cc}
{\rm KSVZ\ axion\ }(c_1=0,c_2=0):\\
  \bar c_1=\frac12 c_3=\frac12,\\
   \bar c_2=c_2+c_3=1.
\end{array}\label{axhadint2}
\end{eqnarray}
Here, $\bar c_2$ must be split according to the flavor singlet condition to $\bar c_2^u+\bar c_2^d$, (\ref{axioncurrent}) or (\ref{anomalytoqk}).

For the DFSZ and PQWW axions, $c_1=0, c_2\ne 0$ and $c_3=0$. If a non-vanishing  $\theta_{\rm QCD}$ is introduced here, we have, using the reparametrization invariance (\ref{Gammaone}), $c'_1=-c_2/2, c'_2=0$, and $c_3'=c_2$. Then, the coefficient of the gluon anomaly term is $c_2(a/f_S)+\theta_{\rm QCD}$, and hence redefining axion as $a+(f_S/c_2)\theta_{\rm QCD}$ and going back to the $\bar c_3=0$ basis,
we obtain for one family
\begin{eqnarray}
\begin{array}{c}
{\rm DFSZ\ and\ PQWW\ axions\ }:\\
\bar c_1=\frac12(-c_2+\bar c_2),\\
 \bar c_2\ne0,\ \ \bar c_3=0 .
\end{array}\label{axhaintdfsz}
\end{eqnarray}
Again, $c_3'$ must be split according to the flavor singlet condition to $\bar c_2^u+\bar c_2^d$ according to the anomaly matching condition, Eq. (\ref{axioncurrent}).

Integrating out the heavy $\sigma$ field and heavy quark fields, the massless (at this level) degree $a=F_a\theta$ which appears from the phase of the singlet field $\sigma=(\langle \sigma\rangle+\frac{\rho}{\sqrt2} )e^{i\theta}$ appears in the effective low energy Lagrangian. If there are multiple SM singlets $S_i$ carrying PQ charges and VEVs, then the axion component is proportional to
\begin{equation}
a=\frac{1}{V_a}\sum_i \Gamma_i V_ia(S_i),\ \ V_a=(\sum_i \Gamma_i^2 V_i^2)^{1/2}\label{axcompo}
\end{equation}
where $a(S_i)$ is the phase field of $S_i$. The PQ charges are defined such that the smallest nonzero absolute value(s) of the PQ charges is 1 so that every scalar field returns to its original value after $2\pi$ shift of its phase. Let us discuss axion couplings after integrating out the heavy fields.

In the KSVZ model $c_3$ is calculated by the triangle diagram of heavy quarks for the global anomaly. The domain wall number $N_{DW}$ is
$
 N_{DW}={\rm Tr}\Gamma(Q_L)l({Q_L}),\quad {\rm with\ }F_a={V_a}/{N_{DW}}
$
where $\Gamma(Q_L)$ (defined as $Q_L\to e^{i\Gamma(Q_L)\theta} Q_L$ under $a\to a+F_a\theta$) is the PQ charge and $l({Q_L})$ is the index of $SU(3)_c$ representation. Every field is represented in terms of left-handed fields, and the PQ charges are defined such that the SM singlet $\sigma$ coupling to heavy quarks carries one unit of the PQ charge. If the light quarks also carry the PQ charge, then (\ref{DWnumber}) gives $N_{DW}$, which belongs to the generic very light axion model discussed below. The anomaly calculation gives the one loop coupling $(N_{DW}a/V_a)\{ G\tilde G\}$, but since the vacuum angle $\theta$ or axion is given by the coefficient of $\{ G\tilde G\}$, $F_a$ is defined by dividing $V_a$ of (\ref{axcompo}) by $N_{DW}$ and hence $ c_3=\pm 1$
where the sign coincides with that of Tr$\Gamma(Q_L)l({Q_L})$. As a convention, let us choose it to be $c_3=+1$, which is choosing the effective PQ charges of heavy quarks to be positive. Transferring $c_3$ to $c_2$, we split $c_3=c_2^u+c_2^d$ using the PQ charges of (\ref{anomalytoqk}),
\begin{equation}
\begin{array}{ll}
{\rm KSVZ}&{\rm axion}:\\[0.3em]
&\hskip -0.7cm \bar c_1^{u,d} = \frac12\bar c_2^{u,d}\\[0.3em]
&\hskip -0.7cm  \bar c_2^u=\frac{1}{1+Z},\quad \bar c_2^d=\frac{Z}{1+Z},\ \end{array}
\end{equation}

In the DFSZ model, $c_2^u$ and $c_2^d$ are calculated by transferring the phase of $\sigma$ to $H_u$ and $H_d$ by the PQ symmetry such that $\langle H_u^0\rangle=\sqrt2 v_2 e^{i\Gamma_ua/V_\sigma}$ and $\langle H_d^0\rangle=\sqrt2 v_1 e^{i\Gamma_da/V_\sigma}$ if $H_u^*H_d^*\sigma^2$ defines the PQ charge of $\sigma$ in terms of PQ charges $\Gamma_u$ and $\Gamma_d$ of $H_u$ and $H_d$. Here, $a=V_\sigma\theta$ is not the mass eigenstate and instead of $V_\sigma$ the mass eigenstate $\tilde a$ uses the decay constant $F_a=[{(\Gamma_u+\Gamma_d)^2V_\sigma^2+
\Gamma_u^2v_u^2+\Gamma_d^{2}v_d^2}]^{1/2}\simeq (\Gamma_u+\Gamma_d)V_\sigma$ for $V_\sigma\gg v_u,v_d$, and the axion component $\tilde a=[(\Gamma_u +\Gamma_d)V_\sigma a+\Gamma_u v_{EW}a(H_u)+\Gamma_d v_{EW}a(H_d)]/F_a\simeq a$, and $F_a=V_\sigma/N_{DW}$.  So, in the DFSZ model they are given by \citep[Carena, Peccei (1989)]{Carena89}:
$ c_2^u= \frac{|v_d|^2}{v_{EW}^2},\  c_2^d=  \frac{|v_u|^2}{v_{EW}^2},$ and $ c_1^{u,d}=c_3=0$. By the reparametrization invariance, Eq. (\ref{Gammaone}), we can use $c_1'^u=-c_2^u/2,~ c_1'^u=-c_2^d/2,c_2'^u=c_2'^d=0$, and $c_3'=c_2'^u+c_2'^d=1$. Removing $c_3'$ according to the flavor singlet condition, we obtain for one family
\begin{equation}
\begin{array}{ll}
{\rm DFSZ}& {\rm axion\ for\ one\ family}:\\[0.3em]
 &\hskip -0.7cm \bar c_1^u=-\frac{|v_d|^2}{2v_{EW}^2}+ \frac12\bar c_2^u,\quad  \bar c_1^d=  -\frac{|v_u|^2}{2v_{EW}^2}+\frac12\bar c_2^d,\\[0.5em]
&\bar c_2^u=\frac{1}{1+Z},\quad \bar c_2^d=\frac{Z}{1+Z},
\end{array}
\end{equation}
where $v_u=|\langle \sqrt2 H_u^0\rangle|, v_d=|\langle \sqrt2 H_d^0\rangle|, v_{EW}=(v_u^2+v_d^2)^{1/2}$. The PQ charges $ c_2^u= \frac{|v_d|^2}{v_{EW}^2}$ and $ c_2^d= \frac{|v_u|^2}{v_{EW}^2}$ of $H_u$ and $H_d$ are obtained by considering the orthogonal component to the longitudinal mode of $Z$ boson. Remember that the signs of $c_2^{u,d}$  are chosen from the convention that the PQ charges of $H_{u,d}$ are positive. This result is for one family.

If we have $N_g$ families, we can calculate the couplings just below the electroweak scale where all quarks obtain masses. Then, we obtain for three families, $ c_2^u=c_2^c=c_2^t = \frac{|v_d|^2}{v_{EW}^2}$ and $ c_2^d=c_2^s=c_2^b= \frac{|v_u|^2}{v_{EW}^2}$. Using the reparametrization invariance, we can calculate $c_1', c_2'$ and $c_3'$, just above 1 GeV: $c_2'=0, c_1^{i\prime} =-\frac12 c_2^i$ and $c_3'=N_g(\sum_i c_2^i)=N_g$. Then, we integrate out heavy quarks $c,b,$ and $t$ to obtain the effective couplings just above 1 GeV, which does not introduce any new $c_2$ terms. Now, there are three light quarks $u, d,$ and $s$ for which we use the reparametrization invariance to remove the $c_3'$ term such that the isosinglet condition is satisfied, $\partial^\mu J_\mu^a$ is anomaly free where $J^a_\mu= J_\mu^{\rm PQ}-\alpha_u \bar u\gamma_\mu\gamma_5 u-\alpha_d \bar d\gamma_\mu \gamma_5 d-\alpha_s \bar s\gamma_\mu\gamma_5  s$ and $\partial^\mu J_\mu^{\rm PQ}=N_g\{G\tilde G\} $. Thus, $\alpha_u+\alpha_d+\alpha_s=N_g$ is satisfied and the SU(3)$_{\rm flavor}$ singlet condition of $\partial^\mu J_\mu ^a$  determines
\begin{eqnarray}
\alpha_u=\frac{m\alpha}{m_u},\ \alpha_d=\frac{m\alpha}{m_d},\ \alpha_s=\frac{m\alpha}{m_s},
\end{eqnarray}
with
\begin{eqnarray}
m\alpha=\frac{N_gm_u m_d m_s}{m_u m_d+ m_u m_s+ m_d m_s}\simeq \frac{N_g(m_u+m_d)Z}{(1+Z)^2} .\nonumber
\end{eqnarray}
Therefore, removing $c_3'$ by the reparametrization invariance, we obtain
\begin{eqnarray}
&&\hskip -0.5cm{\rm DFSZ}~ {\rm axion\ for}\ N_g\ {\rm families}:\nonumber\\[0.3em]
&&\bar c_2^u=\frac{1}{1+Z}N_g,\ \
\bar c_2^d=\frac{Z}{1+Z}N_g,\label{DFSZc2}\\
&&\bar c_1^u=\frac12 \bar c_2^u-\frac{v_d^2}{v^2_{EW}},\ \ \bar c_1^d=\frac12 \bar c_2^d-\frac{v_u^2}{v^2_{EW}}. \label{DFSZc1}
\end{eqnarray}

If heavy quarks and also $H_{u,d}$ carry PQ charges, we must consider all these. If one SM singlet $\sigma$ houses the axion, then we obtain
\begin{eqnarray}
{\rm General}&&\hskip -0.2cm {\rm very}\ {\rm light\ axion}:\nonumber\\
 &&\hskip -1.2cm \bar c_2^u=\frac{1}{1+Z}(1\pm N_g)\\
 &&\hskip -1.2cm  \bar c_2^d=\frac{Z}{1+Z}(1\pm N_g)\\
  &&\hskip -1.2cm \bar c_1^{u}= \frac{1}{2(1+Z)} (1\pm N_g) \mp\frac{|v_d|^2}{2v_{EW}^2}
  \delta_{H_u}  \label{coneu}\\
  &&\hskip -1.2cm \bar c_1^{d}= \frac{Z}{2(1+Z)} (1\pm N_g) \mp\frac{|v_u|^2}{2v_{EW}^2}
  \delta_{H_d}
  \label{coned}
\end{eqnarray}
where the PQ charges (+ or --) of $H_u$ and $H_d$ determine the sign (-- or +) in front of DFSZ component and $\delta_H=1$ or 0 if the corresponding Higgs doublets carry the PQ charges or not. For the MI axion from superstring which is a hadronic axion, in principle there can exist an additional contribution to $c_1$ as will be pointed out in \ref{subsubsec:MIa}.

If there is no heavy degrees carrying the PQ charges above the electroweak scale, then $c_2$ in the so-called PQWW model is given by the PQ charges of $H_u$ and $H_d$,
\begin{equation}
{\rm PQWW}~ {\rm axion}:\
 {\rm Same ~as~ Eqs.~} (\ref{DFSZc2},\ref{DFSZc1}).
\label{aquarkPQWW}
\end{equation}

All models have $\bar c_1$ and $\bar c_2$.  For the original $c_2$ term, different models give different values, for example some variant axion models \citep[Krauss, Wilczek (1986), Bardeen, Peccei, Yanagida (1987), Kim, Lee (1989), Hindmarsh (1997)]{varKrausWil86,Bardeen87var,KimLee89,varHind97} have different $c_2$s from those of the PQWW axion. For astrophysical application, we must keep both $\bar c_1$ and $\bar c_2$. The $\bar c_1$ and $\bar c_2$ terms respectively give the axial vector and pseudoscalar couplings. The axion operator in the flavor SU(3) space can be written as
\begin{equation}
(\bar c_{1,2}^u-\bar c_{1,2}^d)F_3+\frac{\bar c_{1,2}^u + \bar c_{1,2}^d}{\sqrt3}F_8 +\frac{\bar c_{1,2}^u+ \bar c_{1,2}^d}{6}{\bf 1}
\end{equation}
where $F_3$ and $F_8/\sqrt3$ are the third component of the isospin and the hypercharge operators, respectively, and {\bf 1} is the identity operator.

\begin{widetext}
Then, the derivative couplings with  nucleons and mesons below the chiral symmetry breaking are defined as
\begin{eqnarray}
&&{\cal L}_{AV}^{\bar c_1} = \frac{\partial^\mu a}{F_a}\left[ C_{app}\bar p\gamma_\mu\gamma_5 p +C_{ann}\bar n\gamma_\mu\gamma_5 n+ iC_{a\pi NN}\left(\frac{\pi^+}{f_\pi}\bar p\gamma_\mu n -\frac{\pi^-}{f_\pi}\bar n\gamma_\mu p\right)
\right]\label{nucleoncoupling}\\
&&{\cal L}_{a\pi\pi\pi}^{\bar c_1} =C_{a\pi\pi\pi}\frac{\partial^\mu a}{F_a f_\pi}(\pi^0\pi^+\partial_\mu\pi^- +\pi^0\pi^-\partial_\mu\pi^+ -2\pi^+\pi^-\partial_\mu\pi^0),
\end{eqnarray}
where
\begin{eqnarray}
C_{app}=\bar c_1^u F+\frac{\bar c_1^u-2\bar c_1^d}{3}D+\frac{\bar c_1^u+\bar c_1^d}{6} S&,&\ C_{ann}=\bar c_1^dF+ \frac{\bar c_1^d-2\bar c_1^u}{3}D+ \frac{\bar c_1^u+\bar c_1^d}{6} S\label{CaNN}\\
C_{a\pi NN}=\frac{\bar c_1^u-\bar c_1^d}{\sqrt2}&,&\ C_{a\pi\pi\pi}=\frac{2(\bar c_1^u-\bar c_1^d)}{3}.
\end{eqnarray}
Here the axial vector coupling parameters of the nucleon octet are given by $F=0.47, D=0.81, S\simeq 0.13\pm 0.2$ \citep[Amsler {\it et al.} (Particle Data Group, 2008)]{PData08}. For example, for the hadronic axion couplings we obtain the results given in  \citep[Kaplan (1985), Chang, Choi (1994)]{Kaplan85,Chang93},
\begin{eqnarray}
C_{app}=\frac{1}{2(1+Z)}F+\frac{1-2Z}{6(1+Z)}D+\frac16 S&,&\ C_{ann}=\frac{Z}{2(1+Z)}F+ \frac{Z-2}{6(1+Z)}D+\frac16 S \nonumber\\
C_{a\pi NN}=\frac{1-Z}{2\sqrt2 (1+Z)}&,&\ C_{a\pi\pi\pi}=\frac{1-Z}{3(1+Z)}.\label{aNNcoupling}
\end{eqnarray}
\end{widetext}
For the DFSZ axion, there exist additional contributions from the extra terms in Eqs. (\ref{DFSZc2},\ref{DFSZc1}).

Similar expressions might be attempted for the pseudoscalar couplings in terms of $\bar c_{2}^{u,d}$ and the pseudoscalar coefficients $F', D'$ and $S'$. But,
for the axion current, corresponding to $J_\mu ^a$, there does not exist an anomaly as discussed in Eq. (\ref{axioncurrent}) and we do not write down the axion pseudoscalar couplings. The anomaly carried by axion above the chiral symmetry breaking scale is left over to $\eta'$ below the chiral symmetry breaking scale and hence these pseudoscalar couplings are for the $\eta'$ meson. The axial vector current of $\eta'$ to the nucleon octet $N=q\otimes q\otimes q$ is,
\begin{equation}
J_\mu^{\eta'}=f_{\eta'}\partial_\mu \eta' + g^5_N\bar N\gamma_\mu\gamma_5 T_0N
\end{equation}
where $T_0$ is properly normalized Tr$T_0^2=\frac12$ or $T_0={\bf 1}/\sqrt{2N_f}$, and $g^5_N$ is determined by strong interaction dynamics. The original global symmetry breaking term (\ref{detint}) is transferred to $\bar N_LN_R e^{i\alpha'_1\eta'/f_{\eta'}}$ which is actually the nucleon mass term,
\begin{eqnarray}
\Delta {\cal L}&=&-m_N \bar N_LN_R e^{i\alpha'_1\eta'/f_{\eta'}} +{\rm h.c.}\label{etapnnint}
\end{eqnarray}
For example, the SU(6) wave function of spin-up neutron is
\begin{equation}
\begin{array}{l}
|n^\up\rangle=\frac{1}{6\sqrt2}|4d^\up u^\down d^\up - 2d^\down u^\up d^\up- 2d^\up u^\up d^\down- 2u^\up  d^\down d^\up\\
\ +4u^\down d^\up d^\up- 2u^\up d^\up d^\down  - 2d^\up d^\down u^\up- 2d^\down d^\up u^\up+4d^\up d^\up u^\down\rangle
\end{array}
\end{equation}
where the quarks are now interpreted as constituent quarks below the chiral symmetry breaking. At low energy, the symmetry is the only relevant one for consideration. The octet charge $\alpha'_1$ is determined by strong interaction dynamics. The ducaplet has a different U(1) charge $\alpha''_1$. Two anomaly matching conditions, PQ-B-B and PQ-Q$_{\rm em}$-Q$_{\rm em}$, may be used but do not give a useful information because of many form factors. So, simply interpreting the PQ charges of the current quarks being transferred to the constituent quarks in the octet with a multiplcation factor $g_N^5$, we obtain the PQ charge of neutron as the PQ charge of one constituent quark. Thus, $\bar N_LN_R$ has the phase $\alpha'=2g_N^5\sqrt{2N_f}/N_f$. Guessing that $g_N^5$ is similar to the octet form factor $g_A\simeq 0.75$ \citep[Georgi (1984), p.100]{Georgi84}, $\alpha'_1$ is estimated as $1.22$.

\subsubsection{Axion-photon-photon coupling}\label{subsubsec:agamma}

As we calculated the $c_3$ coupling for the KSVZ axion, we can calculate the axion-photon-photon coupling, by simply substituting the gluon lines by photon lines and the quark triangles by charged fermion triangles. Since we are interested in low temperature experiments, we consider the energy scale below the electron mass. Therefore, considering $V_a=N_{DW}F_a$, $c_{a\gamma\gamma}^0$ calculated from the PQ charges of charged fermions becomes
\begin{equation}
 c_{a\gamma\gamma}^0=\frac{{\rm Tr}\Gamma(Q_L)Q^2_{\rm em}}{N_{DW}}.
\label{cagg0}
\end{equation}
Below the QCD chiral symmetry breaking scale, we chiral transform light quarks to obtain
\begin{equation}
{\cal L}_{a\gamma\gamma}= c_{a\gamma\gamma}\frac{e^2}{32\pi^2 F_a}~aF^{\rm em}_{\mu\nu}\tilde F_{\rm em}^{\mu\nu} \label{aggLag}
 \end{equation}
where
\begin{equation}
c_{a\gamma\gamma}\simeq c_{a\gamma\gamma}^0-c_{\chi SB} \label{aggcoupling}
 \end{equation}
where the chiral symmetry breaking effect is, including the strange quark mass effect,
\begin{equation}
c_{\chi SB} =\frac{\frac23(4+1.05 Z)}{1+1.05 Z}=[1.762,2.260]
\end{equation}
for the 20\% allowance from the tree level chiral perturbation theory estimation \citep[Kaplan, Manohar (1986)]{KapMan86}. For illustration, let us take $c_{\chi SB} \approx 1.98$ for $Z\simeq 0.5$ \citep[Manohar, Sachrajda (2008)]{Manohar08}.

In the KSVZ model, $c_{a\gamma\gamma}^0$ is determined by the PQ charge carrying heavy fermions. If there is only one neutral quark for this, then $c_{a\gamma\gamma}^0$ would be zero. If there is only one PQ charge carrying heavy quark with the electromagnetic charge $Q_{\rm em}$, then  $c_{a\gamma\gamma}^0=Q_{\rm em}^2$.  But, in realistic models from a fundamental theory it is more likely that there exist many PQ charge carrying quarks, and the coupling given for one PQ charge carrying heavy quark is presented just an illustration.

In the DFSZ model, we consider only light quarks and leptons. The PQ charges of $H_u$ and $H_d$ determine the PQ charges of $u$ and $d$ quarks. For the PQ charge of $e$, we have two possibilities: $H_d$ gives mass to $e$ and the PQ charge of $e$ is the same as that of $d$, or $H_u$ gives mass to $e$ and the PQ charge of $e$ is the opposite to that of $u$:
\begin{eqnarray}
&&c_{a\gamma\gamma}^0=-\frac{2v_d^2}{v^2_{EW}}
(\frac23)^2\cdot 3 -\frac{2v_u^2}{v^2_{EW}}\left((-\frac13)^2\cdot 3+(-1)^2\right)\nonumber\\
&&\hskip 1.5cm =-\frac83,\quad {\rm electron\ mass\ by\ }H_d\label{eduni}\\
&&c_{a\gamma\gamma}^0=-\frac{2v_d^2}{v^2_{EW}}\left( (\frac23)^2\cdot 3-(-1)^2\right)- \frac{2v_u^2}{v^2_{EW}}(-\frac13)^2\cdot 3\nonumber\\
&&\hskip 1.5cm =-\frac23,\quad {\rm electron\ mass\ by\ }H_u^\dagger\label{euuni}
\end{eqnarray}
where the PQ charges of $H_{u,d}$ were chosen to be positive before. So, in applying Eq. (\ref{aggcoupling}), we must choose the PQ charges of light quarks to be positive and hence the signs of (\ref{eduni},\ref{euuni}) must be reversed. For the PQWW axion, the coupling is the same as those of (\ref{eduni},\ref{euuni}) with positive signs.

For a general light axion, the axion-photon-photon coupling depends on the ultraviolet completion of the theory. If the axion mass is lighter than $2m_e$, its lifetime is
\begin{eqnarray}
&&\tau(a\to 2\gamma)=\frac{2^8\pi^3}{c^2_{a\gamma\gamma}\alpha_{\rm em}^2}\frac{F_a^2}{m_a^3}\simeq \frac{3.65\times 10^{24}}{c^2_{a\gamma\gamma}}\left(\frac{\rm eV}{m_a}\right)^5  {\rm s}\nonumber\\
&&\hskip 2cm
\simeq \frac{0.8\times 10^{7}t_U}{c^2_{a\gamma\gamma}}\left(\frac{\rm eV}{m_a}\right)^5  \label{axlifetime}
\end{eqnarray}
where we used $Z\simeq 0.5$ and the age of universe $t_U\approx 4.35\times 10^{17}$ s. For $c_{a\gamma\gamma}={\cal O}(1)$, the axion with 24 eV mass has the lifetime $t_U$ \citep[Moroi, Murayama (1998), Hannestad, Mirizzi, Raffelt, Wong (2008)]{Moroi98,Hannestad08}.

\subsubsection{Axion lepton couplings}\label{subsub:alepton}

The tree level axion lepton ($l$) coupling arises in the DFSZ and PQWW axions where the lepton mass term through the PQ charges of $H_d$ or $H_u$ defines the $c_l$ of Fig. \ref{fig:axInt}. The removal of $c_3$ term does not change the coupling $c_l$, and hence we obtain the following tree level couplings of axion and lepton:
\begin{eqnarray}
&&{\rm DFSZ}\ {\rm axion}:\nonumber\\ &&\quad\frac{m_lv_u^2}{N_{DW}F_av^2_{EW}}~\bar li\gamma_5 l a,\ \ {\rm lepton\ mass\ by\ }H_d\label{axionelectron}\\
&&\quad \frac{m_lv_d^2}{N_{DW}F_av^2_{EW}}~\bar li\gamma_5 l a,\ \ {\rm lepton\ mass\ by\ }H_u^\dagger\label{alephu}
\end{eqnarray}
where the PQ charges of $H_{u,d}$ are chosen to be positive. For the PQWW axion, just $F_a$ is replaced by $v_{EW}$. For the KSVZ axion, the axion lepton coupling occurs at higher order which is negligible in astrophysical applications. For the generic very light axion, the couplings given in Eqs. (\ref{axionelectron},\ref{alephu}) are applicable.

Even though, the tree level coupling of the axion with electron is absent in the KSVZ model, the axion--electron coupling is present at one loop through the $c_{a\gamma\gamma}$ coupling \citep[Srednicki (1985)]{Srednicki85}
\begin{equation}
2.2\times 10^{-15}\left(\frac{m_a}{\rm eV}\right)
\left[c_{a\gamma\gamma}^0\ln\frac{F_a}{m_e}
-\frac23\frac{4+Z}{1+Z}\ln\frac{\Lambda}{m_e}\right]
\end{equation}
where $\Lambda$ is the chiral symmetry breaking scale and $N_{DW}^{-1}$ must be multiplied in models with $N_{DW}\ne 1$. On the other hand, the DFSZ axion coupling to electron is
\begin{equation}
1.4\times 10^{-11}X_d\left(\frac{3}{N_g}\right) \left(\frac{m_a}{\rm eV}\right)
\end{equation}
where $N_g=N_{DW}/2$ is the number of families and $X_d=\sin^2\beta=v_u^2/v^2_{EW}$ for the case of Eq. (\ref{axionelectron}).

\subsection{Old laboratory bounds on $F_a$} \label{subsec:oldbounds}

With the axion couplings discussed in Subsec. \ref{subsec:axioncouplings}, one can estimate the axion production rates in various experiments. The null experimental results give the bounds on the relevant axion couplings. These are extensively discussed in earlier reviews \citep[Kim (1987), Cheng (1988), Peccei (1989)]{Kimrev,Chengrev88,Pecceirev89}. These old laboratory bounds, immediately studied after the proposal of the PQWW axion, basically rule out the PQWW axion, i.e. give the axion decay constant $F_a $ greater than $\cal O$(10 TeV),
\begin{equation}
F_a\gtrsim 10^4\ {\rm GeV}:\ {\rm old~ laboratory bound}.
\end{equation}

\section{Axions from outer space}\label{sec:axionsOuterspace}

From Eq. (\ref{axlifetime}), we note that the axion lifetime is longer than $t_U$ for $m_a\sim 24$ eV, and this kind of axion is important in cosmology. For $m_a\lesssim 23$ keV with $c_{a\gamma\gamma}=1$, the axion lifetime is longer than 10 min, allowing solar generated axions below this mass have time to reach Earth. These examples illustrated the importance of studying low mass axion effects in astrophysics and cosmology.

The window for $F_a$ obtained from the astrophysical and cosmological constraints is given by
\begin{equation}
0.5\times 10^{9}\ {\rm GeV} \lesssim
F_a \lesssim 2.5\times 10^{12}\ \rm GeV
\end{equation}
where the upper bound is understood with the initial misalignment angle of order 1.

\subsection{Axions  from stars}

In this Subsec. we present the key arguments leading to the axion constraints from astrophysical sources.
Axions have very small masses and therefore can be emitted without important threshold effects from stars, in analogy to neutrinos. The method to constrain axion models is basically the overall energy loss rate, whether using the individual stars (e.g. Sun, SN1987A) or the statistical properties of stellar populations (e.g. the stars in a globular cluster being a test population) \citep[Kolb, Turner (1990), Raffelt (1996)]{KolbTur90,Raffelt96Bk}.

We may use the axion couplings to $\gamma, p, n,$ and $ e$ to study the core evolution of a star. The simple bounds are obtained just by comparing the energy loss rates by the axion emission and by the neutrino emission. Studying the evolutionary history of a star by the axion emission may give a stronger bound than the one obtained from the energy loss rate but may not be so reliable compared to the one obtained from the energy loss rate.
Since there are good reviews on axion astrophysics \citep[Turner (1990), Raffelt (1990, 2008), Amsler {\it et al.} (Particle Data Group, 2008)]{Turnerrev90,Raffeltrev90, Raffelt08LNP,PData08}, here we briefly comment on axion physics in stars (Sun, low mass red giants, supernovae) to cite the reliable $F_a$ bound.

With the axion emission, the Sun consumes more fuel and needs an increased core temperature. From the Primakoff process $\gamma+Ze\to a+Ze$ in the hadronic axion models, Ref. \citep[Schlattl, Weiss, Raffelt (1999)]{Raffelt99} gives the axion emission rate $L_a\simeq 3.7\times 10^{-2}L_{\odot}$ with a 20\% increase of the $^8$B flux with the increased core temperature.  The $^8$B neutrino flux gives the best bound on the solar axion emission rate. The measured $^8$B neutrino flux $4.94\times 10^6 {\rm cm^{-2}s^{-1}}$ \citep[SNO Collaboration (2005a, 2005b)]{SNO05,SNO05PRC} is consistent with the axion emission if  $L_a\le 0.04L_{\odot}$ \citep[Bahcall, Serenelli, Basu (2005)]{Bahcall05}. This translates to the $F_a$ bound of $F_a/c_{a\gamma\gamma}\ge 2.6\times 10^{6}{\rm GeV}$ for $L_a\le 0.04L_{\odot}$ \citep[Schlattl, Weiss, Raffelt (1999)]{Raffelt99}.

For the axion-electron coupling such as in the DFSZ axion models, the axion emission from the globular clusters gives a useful $F_a$ bound \citep[Raffelt, Dearborn (1987)]{Raffelt87Dear}. Stars in a globular cluster are assumed to have the identical $Y$ (helium fraction) and metallicity fraction. The helium core before ignition is degenerate and the bremsstrahlung emission is very effective, whereas the Primakoff emission is suppressed by the large plasma frequency and the helium ignition does not give a useful $F_a$ bound for the KSVZ axion. However, after helium ignition the core degeneracy is lifted, the Primakoff effect becomes important, and the consumption of helium fuel is accelerated by the axion energy loss, shortening the helium-burning lifetimes. Horizontal branch stars in several globular clusters confirm the expected helium-burning lifetimes, which agrees with the standard prediction and the axion losses should not exceed  $\varepsilon_a<10~{\rm erg~ g^{-1}s^{-1}}$ in the cores of horizontal branch stars \citep[Raffelt (1990), Catelan, de Freista Pacheco, Horvath (1996)]{Raffelt90Coremass,Catelan96}, which leads to $F_a/c_{a\gamma\gamma}\ge 2\times 10^{7}{\rm GeV}$, a factor of 10 improvement over the solar bound. Note that this globular cluster bound is for models with an appreciable axion-electron coupling.

In the study of the axion emission in the small mass red giants, the processes $\gamma+Z\to a+Z, e+Z\to a+e+Z,$ and $\gamma+e\to a+e$ were considered. The early studies were the simple comparison of the axion emission and the neutrino emission \citep[Fukugita, Watamura, Yoshimura (1982a,1982b), Krauss, Moody, Wilczek (1984)]{Fukugita82a,Fukugita82b,Krauss84RG}.  In the study \citep[Dearborn, Schramm, Steigman]{Dearborn86}, it is summarized as $F_a\ge 2.1\times 10^7c_{a\gamma\gamma}$ GeV if the Primakoff process $\gamma+Z\to a+Z$ dominates and $F_a\ge 3.7\times 10^{9}\sin^2\beta$ GeV if the Compton process dominates  (which is for the DFSZ axion, viz. Eq. (\ref{axionelectron})). The Primakoff process is present in any axion model, and hence the Primakoff process bound is almost model-independent except in the region $m_a>200$ keV where $a$ is too heavy to be produced in the core of a star. But this threshold effect is irrelevant since the PQWW axion region is excluded already. Note however that there is no confirmed observation of neutrinos from the small mass red giants unlike from the Sun and SN1987A, and the possibility of dominant axion emission from red giants is not excluded by observation \citep[Raffelt (2008)]{Raffelt08}.  For the DFSZ axion, the region $m_a>10^{-2}$ is excluded due to the large axion-electron coupling. For the hadronic axion, Ref. \citep[Raffelt, Dearborn]{Raffelt87Dear} argue that the the axion mass greater than about 2eV/[($E/N-1.95$)/0.72] would reduce the helium-burning time scale and is not allowed.

For the supernovae explosion, the core temperature can go much higher than the temperature in the ignition phase of Helium in the small mass red giant cores. For supernovae, therefore, nuclear reactions are more important and the $F_a$ bound can be very strong. So we use the axion couplings to nucleons discussed in Subsec. \ref{subsub:ahadron} to study the core evolution of supernovae. In the beginning, the bounds on the axion decay constant were simply obtained by comparing the nuclear burning rates of producing axions and neutrinos \citep[Iwamoto (1984), Pantziris, Kang (1986)]{Pantziris86,Iwamoto84}. The discovery of SN1987A was important in that it propelled a great deal of interest anew in the calculation of axion production rate \citep[Raffelt, Seckel (1988), Turner (1988), Mayle, Ellis, Olive, Schramm, Steigman (1988), Hatsuda, Yoshimura  (1988)]{Raffelt88,Turner88,Mayle88,Hatsuda88}. In principle, the same kind of bound on $F_a$ could be obtained from the earlier supernovae studies. The studies after the discovery of SN1987A were performed with the derivative coupling and quartic terms of Subsec. \ref{subsub:ahadron} and obtained a bound $F_a \gtrsim10^9$ GeV. But as pointed out in \citep[Choi, Kang, Kim (1989), Kang, Pantzisis (1991), Turner, Kang, Steigman (1989), Carena, Peccei (1989)]{ChoiKang89,TurnerOPE89,
Carena89,KangK91} and Subsec. \ref{subsub:ahadron}, a proper treatment of nucleon states must be taken into account. For the axion emission from supernovae, one must constrain the energy output to
$\epsilon_a\le 1\times 10^{19}{\rm erg~g^{-1}s^{-1}}$ \citep[Raffelt (1990)]{Raffeltrev90}. The axion emission rate calculation of Ref. \citep[Raffelt (2008)]{Raffelt08LNP} is
\begin{eqnarray}
\epsilon_a=3.0\times 10^{37}\left[{\rm erg~g^{-1}s^{-1}}\right]C_N^2 F_{a,\rm GeV}^{-2} T_{a,\rm 30MeV}^{4}F \label{axemissSNova}
\end{eqnarray}
where $F_{a,\rm GeV}={F_a}/{\rm GeV} ,T_{a,\rm 30MeV}={T}/(\rm 30~MeV)$, and $F={\cal O}(1)$. In the  supernovae explosion the axion emission can be comparable to neutrino emission. Such remnant axions from all the past supernovae explosions may be around us but will be difficult to be detected because of the small $1/F_a$ \citep[Raffelt (2008)]{Raffelt08}. For the smaller $F_a$ region from supernovae explosion, axions can be trapped if the axion-nucleon interaction is strong enough. For the hadronic axion, it gives the bound on $m_a\ge 1$ eV \citep[Turner (1990), Raffelt (1990)]{Turnerrev90,Raffeltrev90}, and we have the hadronic axion window in the eV range.

For the KSVZ axion and the MI superstring axion, $\bar c_1$ terms are present. For example we can simply take $\bar c_1^u=\frac13$ and $\bar c_1^d=\frac16$, corresponding to $Z=0.5$, and hence obtain $c_{app}=\frac13 F+\frac{1}{12}S\simeq 0.17$ for the KSVZ axion. Using $c_{app}$ as $C_N$ in Eq. (\ref{axemissSNova}), we obtain an $F_a$ bound from supernovae,
\begin{equation}
F_a\ge 0.5\times 10^{9}\ {\rm GeV}.
\end{equation}

The white dwarfs in the final evolutionary stage of low mass stars ($M<10\pm 2 ~M_{\odot}$), with the theoretical model implemented in the DFSZ model, may give a stronger bound on $F_a$ \citep[Raffelt (1986)]{Raffelt86} for some region of the DFSZ parameter $\tan\beta=\langle H_u\rangle/\langle H_d\rangle$. The recent study of the bremsstrahlung process gives the bound $F_a\ge 0.6\times 10^{10}\sin^2\beta$ GeV, and even fits the cooling diagram nicely with $F_a\simeq 1.2\times 10^{10}\times\sin^2\beta$ GeV for $H_d$ giving mass to electron \citep[Isern, Garc\'ia-Berro, Torres, Catal\'an (2008)]{Isern08}.  Note that $\tan\beta$ is known to be large ($\ge 30$) in SUSY GUT models, and the white dwarfs may give the strongest $F_a$ bound for some DFSZ axion models.

The axion-nucleon coupling gets enhanced in strong
magnetic field. Magnetic fields as strong as $B>10^{18}$ Gauss in neutron stars have been assumed in the scalar virial theorem \citep[Woltjer (1964)]{neutronstarsB}.  With $B>10^{20}$ Gauss at the surface, the axion emission rate from neutron stars or white dwarfs will get enhanced by ${\cal O}(1)$ compared to the $B=0$ case \citep[Hong (1998)]{Hong98}.

In summary, axions once produced in the hot plasma of a star most probably escape the core, taking out energy. This contributes to the energy loss mechanism of a star and is used to constrain axion models. From the nucleon-nucleon-$a$ coupling, SN1987A gives the strongest astrophysical bound on the axion decay constant, $F_a>0.5\times 10^9$ GeV \citep[Turner (1990), Raffelt (1990), Raffelt (2008)]{Turnerrev90,Raffeltrev90,Raffelt08LNP}.

\subsection{Axions in the universe}\label{subsec:Axioncosmology}

Axions with  $m_a\gtrsim 24$ eV have a lifetime shorter than the age of the universe. In this case, axion decay can lead to photons that can be tested against the observed electromagnetic background of the universe, as in some spontaneously broken flavor symmetric models, $\nu_i\to\nu_ja\to\nu_j\gamma\gamma$ \citep[Berezhiani, Khlopov, Khomeriki (1990)]{Berezhiani90}. However, in this case the needed decay constant, $10^6$ GeV, is outside the current bound on $F_a$.

The axion for $m_a\lesssim 24$ eV has a longer lifetime than the age of the universe and can affect its evolution. The heavy thermal axions around the eV mass range of Fig. \ref{fig:LeeWein}(b) become the hot DM in the universe. For the 3--8 eV mass range, they accumulate in galaxy clusters where their slow decay produces a sharp line that, in principle, can be observed by telescope searches as suggested in \citep[Bershady, Ressell, Turner (1991)]{RessellTur91}. In this case, the neutrino and axion hot DM must be considered together, which now constrains the axion mass to $m_a<1.02$ eV \citep[Hannestad, Mirizzi, Raffelt, Wong (2007, 2008)]{Hannestad07,Hannestad08}, almost closing the hadronic axion window of 1--20 eV of Fig. \ref{fig:FaCartoon}.

\begin{figure}[!h]
\resizebox{0.9\columnwidth}{!}
{\includegraphics{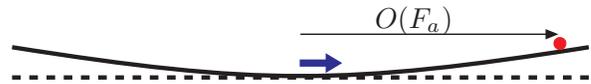}}
\caption{The almost flat axion potential. The misalignment angle is expected to be of order 1 but can also be very small as shown by the thick blue arrow.}\label{fig:flataxionp}
\end{figure}

But more attention is paid to axions behaving as the CDM candidate. The axion potential is almost flat as depicted in Fig. \ref{fig:flataxionp}. Therefore, a chosen vacuum  stays there for a long time, and starts to oscillate when the Hubble time $H^{-1}$ is comparable to the oscillation period (the inverse axion mass), $3H\approx m_a$.  This occurs when the temperature of the universe is about 1 GeV \citep[Preskill, Wise, Wilczek (1983), Abbott, Sikivie (1983), Dine, Fischler (1983)]{Preskill83,Abbott83,
DineFish83}. There exists the domain wall problem in the standard Big Bang cosmology \citep[Sikivie (1982)]{Sikivie82}. The axion strings and the domain wall problem have been elegantly summarized in \citep[Sikivie (2008)]{Sikivie08}.  The axion cosmology is correlated to the reheating temperature $T_{\rm RH}$ in the inflationary models, where one must deal with both the inflaton and the axion. The density perturbations produced by the perturbations of the inflaton field is adiabatic, $\delta\rho_{\rm matter}/ \rho_{\rm matter}=(3/4)\delta\rho_{\rm rad}/\rho_{\rm rad}$. On the other hand, the perturbations produced by the fluctuations of the axion field is isocurvature. If the reheating temperature $T_{\rm RH}$ is above the axion scale $F_a$, the limit on the isocurvature less than 30\% from the large scale structure data can be used \citep[Beltr\'an, Garc\'ia-Bellido, Lesgourgues (2007)]{Beltran07}. This will be commented more in Subsec. \ref{subsec:Anthropic} on the anthropic argument.

In supersymmetric models, the reheating temperature is constrained to $T_{\rm RH}<10^9$ GeV or $10^7$ GeV (if the gluino is lighter than the gravitino) from nucleosynthesis requirements in models with a heavy gravitino \citep[Ellis, Kim, Nanopoulos (1984), Kawasaki, Kohri, Moroi (2005)]{Ellis84,Kawasaki05}. So, with SUSY the domain wall is not so problematic. In this case, the problem of string radiated axions requiring axion mass $m_a>10^{-3}$ eV \citep[Davis (1985), Harari, Sikivie (1987), Dabholkar, Quashnock (1990)]{Davis85,Harari87,Dabholkar90} is no longer problematic.

Axions are created at $T\simeq F_a$, but the axion vacuum $\langle a\rangle$ does not begin to roll until the Hubble parameter reaches the axion mass $3H=m_a$, which occurs  at $T\simeq 1$ GeV. From then on, the classical field  $\langle a\rangle$  starts to oscillate. For a small misalignment angle, the energy density behaves like that in the harmonic oscillator $m_a^2F_a^2$ which is proportional to the axion mass times the number density. Thus, its behavior is like that of CDM, which is the reason that the axion DM is CDM even though its mass is very small and its interaction strength is much weaker than $\lq\lq$weak". Even for a large misalignment angle, an adiabatic invariant $I$ exists and one can estimate the current axion energy density. The axion field evolution with the adiabatic change of the axion mass has been considered before \citep[Chang, Hagmann, Sikivie (1999, 1998)]{Chang99,Chang98conf}.

The temperature dependent axion mass \citep[Gross, Pisarski, Yaffe (1981)]{Yaffe81} enters in the determination of cosmic temperature $T_1$ where $3H(T_1)\simeq m_a(T_1)$. The new estimate of $T_1$ for $F_a\ll 10^{16}$ GeV is a bit below 1 GeV, $T_1\simeq 0.92$ GeV \citep[Bae, Huh, Kim (2008)]{Bae08}.
QCD has two phases: the quark-gluon phase and the chiral symmetry breaking hadronic phase. Near the critical temperature $T_c$, these two phases are separated above and below $T_c$. The critical temperature is estimated as $148^{+32}_{-31}(172^{+40}_{-34})$ MeV for three (two) light quark flavors \citep[Braun, Gies (2007)]{Braun07}. So, cosmology near $T_c$ needs an information on the temperature dependent axion mass.  This region is in the boundary of weak and strong coupling regimes and it is very difficult to estimate the axion mass accurately. Early attempts in this direction are given in \citep[Steinhardt, Turner (1983), Seckel, Turner (1985), Turner (1986)]{Steinhardt83,Seckel85,Turner86}.

The 't Hooft determinental interaction is given schematically in Fig. \ref{fig:tHooft}. In the quark-gluon phase, we have the first diagram in the box, which is parametrized as $-K^{-5}(m_u m_d m_s /\bar\rho^{6})\cos[(c_2+c_3)\theta]$ where $\bar\rho$ is the effective instanton size in the instanton size integration. Ref. \citep[Gross, Pisarski, Yaffe (1981), Eq. (6.15)]{Yaffe81} expresses the result as
\begin{equation}
 n(\rho,0)e^{\left\{-\frac13\lambda^2(2N+N_f)
 -12 A(\lambda)\left[1+\frac16(N-N_f)\right]\right\}}
\label{InstantonCal}
\end{equation}
where $\lambda=\pi\rho T, A(\lambda)\simeq -\frac{1}{12}\ln (1+\lambda^2/3)+\alpha(1+\gamma\lambda^{-2/3})^{-8}$ with $\alpha=0.01289764$ and $\gamma=0.15858$, and the pre-factor $n(\rho,0)$ is the zero temperature density
\begin{eqnarray}
 n(\rho,0)&=&m_um_dm_s
 C_N(\xi\rho)^3\frac{1}{\rho^5}\nonumber\\
&&\cdot \left(\frac{4\pi^2}{g^2
(1/\rho)}\right)^{2N}e^{-8\pi^2/g^2(1/\rho)}.
\label{Instantonzero}
\end{eqnarray}
Here, the parameters are $\xi= 1.33876$ and $C_N=0.097163$ for $N=3$. For the QCD coupling constant,
we use the three loop result \citep[Amsler {\it et. al.} (2008), QCD by Hinchliffe]{PData08},
\begin{eqnarray}
\alpha_c(\mu)&= \frac{g_c^2(\mu)}{4\pi}
 \simeq  \frac{4\pi}{\beta_0\ln(\mu^2/
 \Lambda_{\rm QCD}^2)}\left[1 -\frac{2\beta_1}{\beta_0^2}
\frac{\ln[\ln(\mu^2/
 \Lambda_{\rm QCD}^2)]}{\ln(\mu^2/\Lambda_{\rm QCD}^2)}\right.\nonumber\\
 &+\frac{4\beta_1^2}{\beta_0^4 \ln^2(\mu^2/\Lambda_{\rm QCD}^2)} \times\Big(\left(\ln\Big[\ln(\mu^2/
\Lambda_{\rm QCD}^2)\Big]
-\frac12\right)^2\nonumber\\
& \left.\left.+\frac{\beta_2\beta_0}{8\beta_1^2}
-\frac54\right)\right]
\end{eqnarray}
where $\beta_0=11-\frac23 N_f,\beta_1=51-\frac{19}3 N_f$, and $\beta_2=2857-\frac{5033}9N_f+\frac{325}{27}N_f^2$.  At $T=T_{\rm GeV}$ GeV (from 700 MeV to 1.3 GeV), let us parametrize the instanton size integration of (\ref{InstantonCal}) as
\begin{equation}
 V(\theta) =-C(T) \cos(\theta),
\end{equation}
where $\theta=a/F_a$ and  $C(T)$ is
\begin{equation}
C(T)=\alpha_{\rm inst} \textrm{GeV}^4
 (T_\textrm{GeV})^{-n} .
\end{equation}
We obtain $\alpha_{\rm inst}=2.858\times10^{-12} ~(9.184\times 10^{-12}, 7.185\times10^{-13}), n=6.878~(6.789,6.967)$ for $\Lambda_{\rm QCD}=380~ (440,320)$ MeV \citep[Bae, Huh, Kim (2008)]{Bae08}.
Equating $3H(T)$ and $m(T)=\sqrt{C(T)/F_a^2}$, we obtain the following $T_1$ for $\Lambda_{\rm QCD}=380$ MeV \citep[Bae, Huh, Kim (2008)]{Bae08},
\begin{equation}
T_\textrm{1,GeV}=0.889
 \left( \frac{F_{a,\textrm  GeV}}{10^{12}}
\right)^{-0.184} .\label{T1Fa}
\end{equation}
For $F_a=10^{12}$GeV, we obtain $T_1\simeq 0.89$ GeV. This number is smaller than those given in 1980s because we used the smaller number for the product of current quark masses $m_um_dm_s$ based on the recent compilation of light quark masses \citep[Manohar, Sachrajda (2008)]{Manohar08}.
\vskip 0.3cm
\noindent {(1)\it  No sudden change of $m_a(T)$}:

Since the potential varies much more slowly than the field itself, we can use the so-called `adiabatic invariant theorem' that  if the potential is adiabatically changed, the area in the phase space swept by the periodic motion is unchanged per one axion oscillation \citep[Landau, Lifshitz (1976)]{Landau76}. In this case, for a small misalignment angle the adiabatic invariant is $\rho(t)/m(t)$ which can be interpreted as the conservation of the total axion number. For a large $\theta_1$, the invariant is not the axion number density, but the CDM energy density which can be related to the axion number density by a correction factor  \citep[Bae, Huh, Kim (2008)]{Bae08}. If we apply this until now, we obtain
\begin{widetext}
\begin{eqnarray}
&& \rho_a(T_\gamma=2.73\textrm{K})=m_a(T_\gamma) n_a(T_\gamma)f_1(\theta_2) =\frac{\sqrt{Z}}{1+Z} m_{\pi}f_{\pi} \frac{3\cdot 1.66 g_{*s}(T_{\gamma})T_\gamma^3
}{2\sqrt{g_{*}(T_1)}M_{\rm P}}\frac{F_a}{T_1}\frac{\theta_2^2f_1(\theta_2)}{\gamma}
\left(\frac{T_2}{T_1}\right)^{-3-n/2}
\label{axionenergy}
\end{eqnarray}
\end{widetext}
where $f_1(\theta_2)$ is the anharmonic correction and  we used $Z\equiv m_u/m_d\simeq 0.5$, $m_\pi=135.5\textrm{MeV}$, $f_\pi=93\textrm{MeV}$ and $g_{*s}(\textrm{present})=3.91$. $\gamma$ is the entropy increase ratio from extra particles beyond the SM. This becomes roughly $ 1.449\times 10^{-11}~\frac{\theta_1^2}{\gamma}
\left(\frac{F_{a,{\rm Gev}}}{10^{12}~T_{1,\rm GeV}}\right)F(\theta_1,n)
~{\rm eV}^4$, where $\theta_1$ is the initial misalignment angle at $T_1$ and $\theta_2$ is the angle at somewhat lower temperature $T_2$ where the adiabatic invariant $I$ is calculated. The total correction factor $F(\theta_1,n)$ takes into account the anharmonic effect and the initial overshoot of the misalignment angle, presented in \citep[Bae, Huh, Kim (2008)]{Bae08}.
For the critical density $\rho_c=3.9784\times10^{-11}({h}/{0.701}
)^2~{(\textrm{eV})}^4$ and ${\rm\Lambda_{QCD}=380\mp 60~MeV}$, the axion energy fraction, in terms of $F_a$ only, is \citep[Bae, Huh, Kim (2008)]{Bae08}
\begin{eqnarray}
\hskip-0.5cm\Omega_a&=&0.397ABC\left( \frac{\theta_1^2 F(\theta_1)}{\gamma}\right)\left( \frac{0.701}{h}\right)^2
\label{axionenergyfrac}
\end{eqnarray}
where $A= ({m_u m_d m_s}/{ 3\cdot 6\cdot 103}\ {\rm MeV}^3)^{-0.092},B=(F_a/10^{12}~{\rm GeV})^{1.184 - 0.010x} $ with $x=(\Lambda_{\rm QCD}/380~\rm MeV) - 1$, and $C= (\Lambda_{\rm QCD}/380~\rm MeV)^{-0.733}$.

\vskip 0.3cm
\noindent {(2)\it  Sudden change of $m_a(T)$}:

Let us try to calculate the misalignment angle below the critical temperature of chiral symmetry breaking where a sudden phase change is experienced near the critical temperature $T_c$. The QCD interaction for light quarks below 1 GeV can be written as
\begin{equation}
\begin{array}{l}
{\cal L}=-\left(m_u \bar u_Lu_R +m_d  \bar d_Ld_R+m_s  \bar s_Ls_R +{\rm h.c.}\right)\\
\quad\quad - K^{-5}(\bar u_Lu_R\bar d_Ld_R\bar s_Ls_R e^{-i\bar c_3\theta}+ {\rm h.c.})
\end{array}\label{quarkint}
\end{equation}
where $K$ has the mass dimension arising from QCD instanton physics. The 't Hooft determinental interaction \citep[`t Hooft (1976)]{tHooft76} written above is equivalent to the anomaly term and has the same chiral symmetry behavior. For $T_c\lesssim T\lesssim F_a$, quark bilinears are not developing VEVs, and the relevant determinental interaction for the axion is the first diagram inside the box of Fig. \ref{fig:tHooft}. Now, the importance of the determination of $T_1$ is how much the misalignment angle $\theta_1$ can be shrunk at $T_c$.

\begin{figure}[!h]
\resizebox{0.95\columnwidth}{!}
{\includegraphics{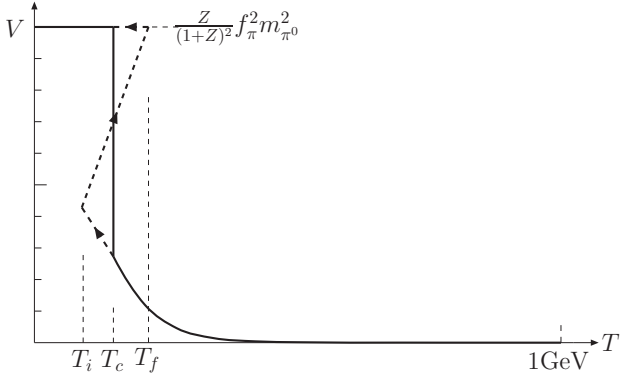}}
\caption{Phase transition near the critical temperature $T_c\approx 150$ MeV.}
\label{fig:axmassTemp}
\end{figure}

In the hadronic phase below the critical temperature $T_c$, the axion potential is schematically depicted in Fig. \ref{fig:axmassTemp}.
The value $\theta_1(T_c)$ is the boundary value at $T_c$ we use in the effective Lagrangian below $T_c$. Below $T_c$, the quark bilinears develop VEVs and we must consider the possibilities of $\bar q_Lq_R$ replaced with $\langle \bar q_Lq_R\rangle$. The effective Lagrangian from the determinental interaction is depicted in Fig. \ref{fig:tHooft}.

In the limit $F_a\gg f'$, the mass eigenstates in one flavor QCD are
\begin{eqnarray}
\eta'_{\rm mass}\simeq \left(
\begin{array}{c}
1\\ \frac{f'/F_a}{1+m/K'}
\end{array}
\right),
 \quad a_{\rm mass}\simeq \left(
\begin{array}{c}
 \frac{-f'/F_a}{1+m/K'}\\ 1
\end{array}
\right)
\end{eqnarray}
Eq. (\ref{axpotentbasic}) with $v^3=0$ has minima at $\theta=2\pi n~(n={\rm integer}$). For $v^3\ne 0$, minima are at $\theta_{\eta'}=2\pi m~(m={\rm integer})$ and  $\theta=2\pi n~(n={\rm integer})$. Therefore, the $\theta$ direction can be taken as the approximate axion direction even below $T_c$. The minimum point in the direction of the axion is not changed when one goes from $\theta_{\eta'}\ne 0$ to $\theta_{\eta'}=0$, i.e. above and below the critical temperature. [If the minimum of $\theta$ were shifted by $\pi$ from going from $\theta_{\eta'}=0$ to $\theta_{\eta'}=\pi$, the shrunk $\theta_1(T_c)$ at $T_c$ is near $\pi$, and we must start from ${\cal O}(1)$ misalignment angle at $T_c$.]
In most regions of the phase transition space, a time scale $\Delta t$ is needed for the sound wave of quark bilinears to propagate to a large distance, which releases the latent heat to keep the constant temperature during the first order phase transition in \citep[Mukhanov (2005), p.149]{Mukhanov05}. Even if one considers supercooling toward a sudden phase transition, the parameter space for a sudden phase change is almost nill and the axion energy density presented in (\ref{axionenergyfrac}) is reliable \citep[Bae, Huh, Kim (2008)]{Bae08}.

\begin{figure}[!h]
\resizebox{0.95\columnwidth}{!}
{\includegraphics{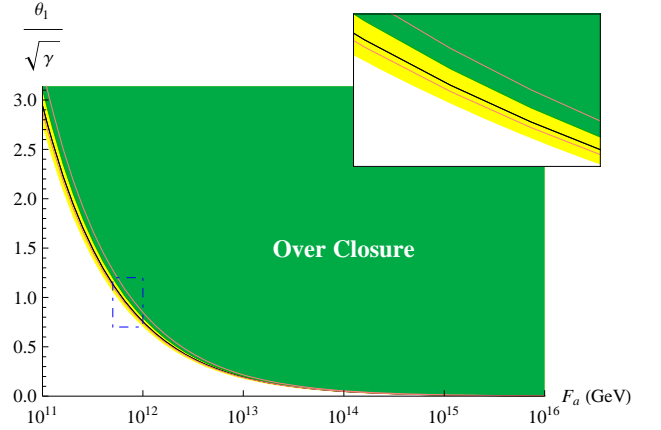}}
\caption{The plot for $F_a$ versus the misalignment angle $\theta_1/\sqrt{\gamma}$ as a function of $\Omega_a$. The overclosure portion is from the precision measurement requiring $\Omega_a<0.23$ \cite{WMAP08}. The green region is the excluded region by the condition $\Omega_a\gtrsim 0.23$. The yellow band is the error bar region of $\Lambda_{QCD}$ and two red lines are the limits from the light quark mass bounds.}
\label{fig:axionOmega}
\end{figure}
In Fig. \ref{fig:axionOmega}, we present the exclusion plot for $m_u=2.55$ MeV, $m_d=5.04$ MeV and $m_s=104$ MeV \citep[Manohar, Sachrajda (2008)]{Manohar08} in the $F_a$ versus $\theta_1/\sqrt{\gamma}$ space, including the anharmonic effect and the WMAP value \citep[Dunkley {\it et al.} (WMAP Collaboration, 2008)]{WMAP08Dun} of CDM density combined with additional data \citep[Komatsu {\it et al.}  (WMAP Collaboration, 2008)]{WMAP08} $\Omega_{\rm DM}h^2\simeq 0.1143\pm 0.0034$. Note that $F_a$ of order $10^{13}$ GeV is not very unnatural, which results from the new smaller masses for $u$ and $d$ \citep[Manohar, Sachrajda (2008)]{Manohar08}.

\begin{figure}[!]
\resizebox{0.9\columnwidth}{!}
{\includegraphics{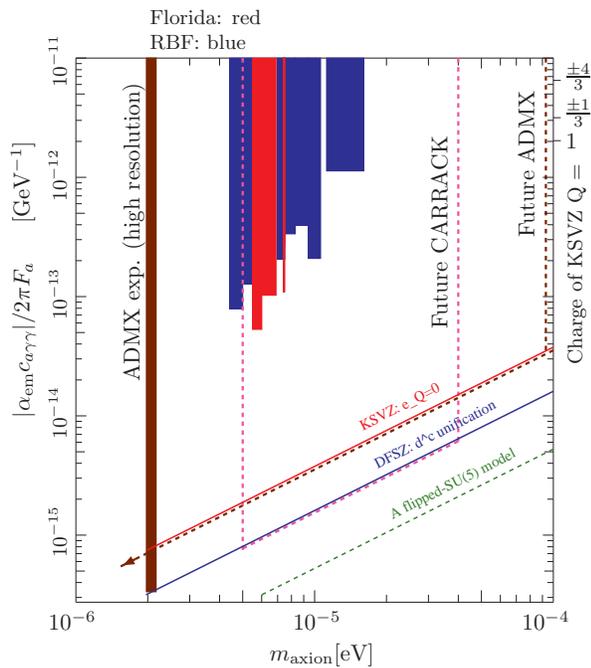}}
\caption{The bounds on cosmic axion searches with some theoretical predictions.  The coupling on the vertical axis is the coefficient of $\bf E\cdot B$. The future CARRACK and ADMX experiments are from \cite{Imai08,vanBibber08}.
}\label{fig:Cosmic}
\end{figure}

If the axion is the CDM component of the universe, then they can be detected even though it may be very difficult. The feeble axion coupling can be compensated by the huge number of axions, since the number density is $\sim F_a^2$ and the cross section is $\sim 1/F_a^2$. So, there is a hope to detect cosmic axions, which has been realized by Sikivie's cavity detector \citep[Sikivie (1983)]{SikDet}. But the Sikivie detector has the technical limitation for the interesting large and median regions of the $F_a$ window. For example, the $F_a$ region $F_a>10^{13}$ GeV advocated in anthropic arguments needs too large cavity size and the supergravity mediation preferred region $F_a\sim 5\times 10^{10}$ GeV requires 1.6 mm order cavities. For technically preferred axion masses in the region $10^{-6}$ eV, one needs a low temperature cavity with dimension $O(>10^4~\rm cm^3)$ and a magnetic field strength of $O(10~\rm Tesla)$. The current status of cosmic axion search is shown in Fig. \ref{fig:Cosmic}.

\subsection{Axion cosmology beyond the window}\label{subsec:Anthropic}

If $F_a\gg 10^{12}$ GeV, an ${\cal O}(1)$ misalignment angle $\theta_1$ is ruled out from the cosmic energy density argument. However, if $\theta_1\ll 1$, the axion energy density can be within the closure density. Rather than fine-tuning $\theta_1$ to order $10^{-3}$ for a Planck scale $F_a$ \citep[Pi (1984)]{Pi}, the anthropic argument of Weinberg \citep[Weinberg (1987), Linde (1988)]{Weinanth87,Linde88}, that life forms can evolve in a universe with a sufficiently long lifetime, can be used for an allowable $\theta_1$.

The homogeneous axion field value (with $a\to -a$ symmetry) right after inflation can take any value between 0 and $\pi F_a$ or $\theta_1=[0,\pi]$ because the height of the axion potential is negligible compared to the total energy density right after inflation. So, in the axion context with only the misalignment production of axions, the CDM density is chosen as a random number by the spontaneous symmetry breaking of the U(1)$_{\rm PQ}$. Even in the multi component CDM models, including axion, the axion misalignment angle can play as the random number. This singles out axion physics, as stressed in \citep[Tegmark, Aguirre, Rees, Wilczek (2006)]{Tegmark06}, from all other anthropic arguments without axion in that selecting an axion vacuum has an unavoidable random process that fixes the key cosmological parameter. This also distinguishes axions from WIMPs, super-WIMPs etc. where the abundance is fixed by particle physics parameters and not by a primordial random process. So $\Omega_a$ may be at the required value by an appropriate initial misalignment angle in models with axions with $F_a> 10^{12}$ GeV. Tegmark {\it et al.} studied the landscape scenario for 31 dimensionless parameters and some dimensionful parameters with which habitable planets are considered for the assumed nuclear physics parameters \citep[Barr, Seckel (1992)]{Barr92}. For example, Fig. 12 of \citep[Tegmark, Aguirre, Rees, Wilczek (2006)]{Tegmark06} presents the scalar fluctuation $Q\simeq\delta\rho/\rho$ vs the matter density per CMB photon $\xi$, in which the anthropically chosen point is shown as the star. In models with axion, this point results from a random number after inflation.
If a WIMP is the sole candidate for CDM, one obtains just one number for $\delta\rho/\rho$ from particle physics parameters, which may not fit to the observed point of that figure. So, we may need the CDM favored WIMP and in addition the axion with $F_a>10^{12}$ GeV, with the axion CDM fraction $R_a=\Omega_a/\Omega_{\rm CDM}$. But this large $F_a$ anthropic region has a potential conflict with the WMAP 5 year data, as presented in the $F_a$ vs. $E_I$(= the inflation energy scale) plane of Fig. 2 of \citep[Herzburg, Tegmark, Wilczek (2008)]{Tegmark08}. For $R_a=1$, for example, $F_a\ge 10^{14}$ GeV is inconsistent with the WMAP 5 year data on the upper bound on the isocurvature fluctuation $\alpha_a<0.072$ \citep[Komatsu {\it et al.} (2008)]{WMAP08}.

\begin{widetext}

\begin{figure}[!h]
\resizebox{1\columnwidth}{!}
{\includegraphics{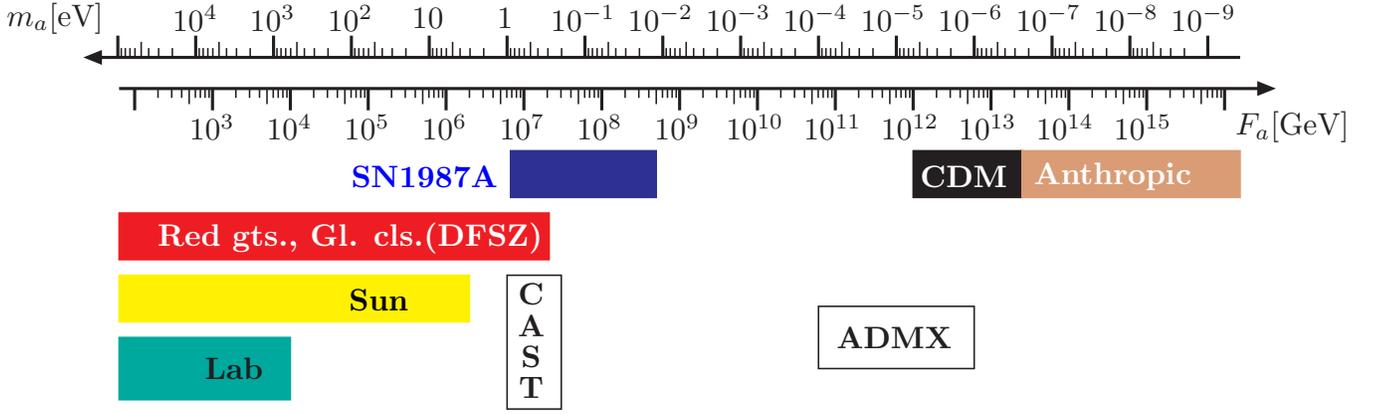}}
\caption{A cartoon for the $F_a$ bounds.}\label{fig:FaCartoon}
\end{figure}
\end{widetext}

From the study of outer space axions, we present a cartoon for the $F_a$ bound in Fig. \ref{fig:FaCartoon} where the future CAST and ADMX experimental regions are also marked.

\subsection{Quintessential axion}
In light of SUSY breaking in supergravity, it is generally believed that at least a hidden confining force is needed at an intermediate scale. This hidden sector and the observable sector couple extremely weakly in most phenomenological models. This scheme fits very well in the heterotic string framework and in heterotic M-theory. In cosmology, on the other hand, we have the important dark energy problem already for a decade \citep[Permutter {\it et al.} (1998), Riess {\it et al.} (1998)]{Perlmutter98,Riess98},  which has led to a lot of interest in quintessence models since the late 1980s \citep[Wetterich (1988)]{Wetterich88}. The quintessence related to axion physics is called ``quintessential axion"(QA) which was suggested in \citep[Kim, Nilles (2003)]{KimNilles03}. There were attempts to identify one of the MD axions as the quintessential axion \citep[Kim (2000), Choi (2000)]{Kim00,Choi00}.

To explain the dark energy in terms of a QA, one requires the VEV of QA not to roll down until recently. Of course, it is required for the current vacuum energy density of the classical QA to be of order $\lambda^4\approx (0.003~ \rm eV)^4$. These two conditions restrict the QA decay constant $f_q$ and the QA mass $m_q$. We can parametrize the QA ($\phi$) potential as
\begin{equation}
V[\phi]=\lambda^4 U(\xi),\quad \xi=\frac{\phi}{f_q}.
\end{equation}
For $\omega=p/\rho<-1+\delta$, we require $f_q>\sqrt{(2-\delta)/6\delta}~M_P|U'|$  where $U'=dU/d\xi$
\citep[Kim, Nilles (2003)]{KimNilles03}.
Generically, one needs a Planckian scale quintessential axion decay constant $f_q$. So, the QA mass is extremely small, $\lesssim 10^{-32}$ eV. As a result, there are two problems to be resolved to achieve the QA idea: a large decay constant and an extremely shallow QA potential.

It has long been believed that the MI axion has rather a robust model independent prediction of its decay constant \citep[Choi, Kim (1985), Svrcek, Witten (2006)]{Choiharm85,Svrcek06}. Recently, however, it was shown that the MI axion may not be model independent since the decay constant may depend on the compactification scheme in warped internal space, $ds^2=h_w^2\eta_{\mu\nu} dx^\mu dx^\nu +g_{mn}(y)dy^m dy^n$ \citep[Dasgupta, Firouzjahi, Gwyn (2008)]{Dasgupta08},
\begin{equation}
F_a=\sqrt{\frac{2}{\beta}}~\frac{m_s^2}{M_P}
\end{equation}
where $\beta$ depends on the warping in the compact space $y\in K$,
\begin{equation}
\beta=\frac{\int d^6y \sqrt{g_{(6)}}e^{-\phi}h_w^{-2}}{ \int d^6y \sqrt{g_{(6)}}h_w^{2}}.
\end{equation}
Thus, the MI axion with a small $\beta$ can be a QA if the QCD axion decay constant can be in the intermediate scale. This possibility may be realizable in some composite axion models as recently suggested in \citep[Kim, Nilles (2009)]{KimNill09}.


\section{AXION DETECTION EXPERIMENTS}\label{sec:AxionDet}

There are currently a variety of experiments searching for axions, whether they are left over from
the big bang or produced in stars or the laboratory. Though these experiments search for axions
at a variety of mass and coupling scales they all rely on the Primakoff process for which
the following coupling, $c_{a\gamma\gamma}$ is given in Eq. (\ref{aggcoupling}),
\begin{equation}
{\cal L}=c_{a\gamma\gamma}\frac{a}{F_a}\{F_{\rm em}\tilde F_{\rm em}\},\ \ c_{a\gamma\gamma} \simeq \bar c_{a\gamma\gamma}-1.98\label{caggexppart}
\end{equation}
where $\bar c_{a\gamma\gamma}={\rm Tr} Q^2_{\rm em}|_{E\gg M_Z}$.

\subsection{Solar axion search}\label{subsec:SolAxiondet}

\subsubsection{Axion Helioscopes}

Axions produced in the nuclear core of the sun will free-stream out and can possibly be detected
on Earth via an axion helioscope, first described in 1983 \citep[Sikivie (1983, 1985)]{SikDet,Sikivie85} and developed into a practical laboratory detector in 1988 \citep[van Bibber, McIntyre, Morris, Raffelt (1989)]{vanBibber89}. The technique relies on solar axions converting into low energy X-rays as they pass through a strong magnetic field. The flux of axions produced in the sun is expected to follow a thermal distribution with a mean energy of $\langle E \rangle = 4.2$ keV. The integrated flux at Earth is expected to be $\Phi_a = g^2_{10} 3.67\times 10^{11}\;{\rm cm^{-2}s^{-1}}$ with $g_{10} = (\alpha_{\rm em}/2\pi F_a)c_{a\gamma\gamma} 10^{10}$ GeV \citep[Ziotas {\it et al.} (2005)]{Ziotas05}. The probability of a solar axion converting into a photon as it passes through a magnet with field strength $\textbf{B}$ and length $\textbf{L}$
is given as:
\begin{equation}
P = \left(\frac{\alpha_{\rm em}c_{a\gamma\gamma}BL}{4\pi F_a}\right)^2 2L^2 \frac{1-\cos(qL)}{(qL)^2}.
\label{eq:Solar_Conversion}
\end{equation}
Here $c_{a\gamma\gamma}$ is defined as the coupling of the axion to two photons as given in Eq. (\ref{caggexppart}), while $q$ is the momentum difference between the axion and the photon, defined as $q = m_a^2/2E$ where $E$ is the photon energy. To maintain maximum conversion probability the axion and photon fields need to remain in phase over the
length of the magnet, thus requiring  $qL < \pi$ \citep[van Bibber, McIntyre, Morris, Raffelt (1989)]{vanBibber89}. For low mass axions $q \rightarrow 0$ leading to a maximum conversion probability. More massive axions will begin to move out of phase with the photon waves though this can be compensated for by the additon of a buffer gas to the magnet volume, thus imparting an effective mass to the conversion photon \citep[van Bibber, McIntyre, Morris, Raffelt (1989)]{vanBibber89} and bringing the conversion probability back to the maximum. Various axion masses can be tuned to by varying the gas pressure.

An initial axion helioscope was built at Brookhaven in 1992 and used a 2.2 ton iron core dipole magnet
oriented at Sun with a proportional chamber for X-ray detection \citep[Lazarus {\it et al.} (RBF Collaboration, 1992)]{Lazarus92}. It was followed by a 4 tesla superconducting helioscope, developed by the University of Tokyo, which ran for 1 week with
an evacuated bore in 1997 \citep[Moriyama {\it et al.} (1998), Ootani {\it et al.} (1999)]{Moriyama98,Tokyo99} and for 1 month with a helium filled bore in 2000 \citep[Inoue {\it et al.} (2002)]{Inoue02}. Though
both managed to set limits over a wide mass range their sensitivities were still well above even the
most optimistic KSVZ axion couplings. Recently though, the University of Tokyo group released data taken between December 2007 and April 2008, which was able to set a limit of $g_{a\gamma\gamma} < 5.6 - 13.4 \times
10^{-10}\;{\rm GeV}^{-1}$ for the axion in the mass range $0.84 < m_a < 1.00\;{\rm eV}$ \citep[Inoue {\it et. al.} (Tokyo Axion Helioscope Collaboration, 2008)]{Inoue08}.

In order to push into proposed axion model space third generation axion helioscopes have been developed at CERN (the CERN Axion Solar Telescope (CAST)) and at the University of Tokyo. Utilizing a prototype Large Hadron Collider (LHC) magnet with $L=9.3$~m and $B=9$ tesla CAST began taking data in 2003. It utilizes a rail system to track Sun for 90 minutes a day at sunrise and sunset and its dual magnet bore allows it to employ up to four different X-ray detectors (one on each end of each magnet bore). Currently a time-projection chamber (TPC), a Micromegas (micromesh gaseous structure) detector and an X-ray reflective telescope with a charge coupled device (CCD) detector are all used to detect converted X-rays. Results from the combined 2003 and 2004 runs yield limits on axion-photon-photon couplings to $c_{a\gamma\gamma}/F_a < 7.6 \times 10^{-8}\rm GeV^{-1}$ \citep[Andriamonje {\it et al.} (CAST Collaboration, 2007)]{CAST07}. The experiments second phase utilizing $^4$He and $^3$He buffer gases is currently underway with the latter gas allowing for axion searches in proposed model space up to mass $\sim$ 1 eV.

\subsubsection{Bragg Diffraction Scattering}

\begin{figure}[!h]
\centering
\epsfig{figure=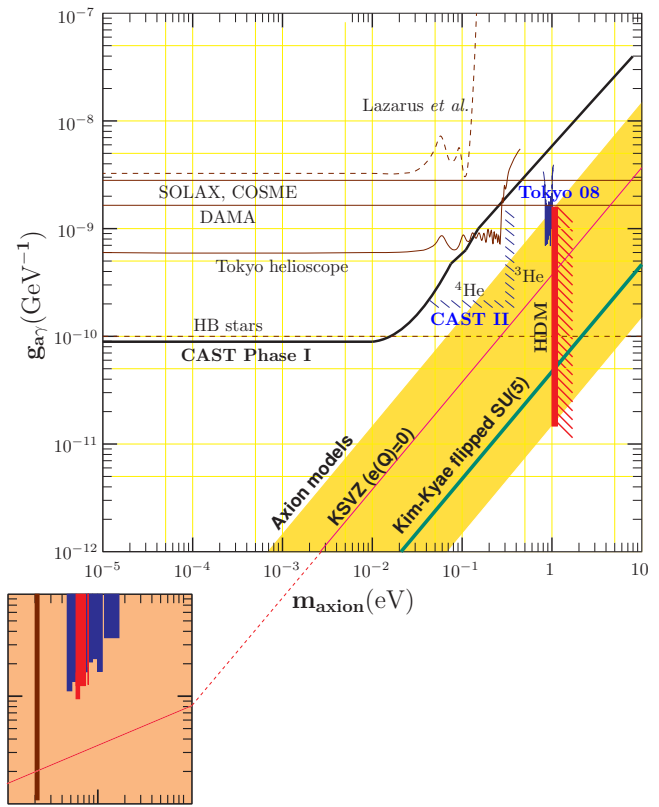, width=8.6cm}
\caption{Exclusion plot of axion-photon coupling versus axion mass \cite{Carosi08}. The black bold line limit is for the phase one of the CAST experiment and results with inclusion of buffer gas are expected to push up in mass reach to plausible axion models. The field theoretic expectations are shown together with the string theory $Z_{12-I}$ model of \cite{ChoiKS07}.  In the lower left apricot box, Fig. \ref{fig:Cosmic} is located.}\label{CASTexpTh}
\end{figure}

An alternative to axion helioscopes was proposed in 1994, using crystal detectors to search for X-rays generated by coherent axion-to-photon conversion which
meet the Bragg conditons \citep[Paschos, Zioutas (1994)]{Paschos94}. Various dark matter WIMP search collaborations were able to look through their data sets and set limits on possible interactions from solar axions. These included germanium experiments such as COSME \citep[Morales {\it et al.} (2002)]{COSME02} and SOLAX \citep[ Avignone {\it et al.} (1998)]{SOLAX98}, the reactor germanium experiment TEXONO \citep[Chang {\it et al.}]{TEXONO07}, as well as the DAMA experiment \citep[Bernabei {\it et al.} (2001, 2003)]{DAMA01,Dama03} which utilized NaI crystals. The limits from these searches can all be seen in Fig. \ref{CASTexpTh}. One advantage of this technique is that their sensitivity is independent of axion mass, as long as one can neglect any nuclear recoils \citep[Carosi, van Bibber (2007)]{CarosiLNP07}.

Current results for solar axion searches can be seen in Fig. \ref{CASTexpTh}

\subsubsection{Geomagnetic Conversion}

It has recently been pointed out \citep[Davoudiasl, Huber (2006)]{Geomagnetic06} that solar axions might pass through Earth and convert to X-rays on the other side as they pass through the earth's magnetic field. They could then be detected by X-ray telescopes and the solar X-ray background could be effectively shielded by Earth.

\subsection{Search for cosmic axions}\label{subsec:CAxiondet}

Cosmic axions left over from the Big Bang may be detected utilizing microwave cavity haloscopes
\citep[Sikivie (1983, 1985)]{SikDet,Sikivie85}.
The strategy relies on primordial axions drifting
through a microwave cavity immersed in a strong static magnetic field in which they can
resonantly convert to microwave photons. The cosmic axions feeble interactions can be in part compensated by
their large numbers; since the number density goes as $\sim F_a^2$ while their cross section goes
as $\sim 1/F_a^2$.  If the axion makes up the majority of CDM in the universe, its local density is expected to be roughly 0.45 GeV/cm$^3$ \citep[Gates, Gyuk, Turner (1995)]{Gates95}, which yields a number density
of $\sim 10^{14}$ axions/cm$^3$ if one assumes a 4.5 $\mu$eV axion. The expected microwave signal will be
a quasi-monochromatic line beginning at the microwave frequency corresponding
to the axion mass and slightly broadened upward due to the axion virial distribution, with expected
velocities of order $10^{-3}c$ implying a spread in energies of $\delta E/E \sim 10^{-6}$.

There could also be an additional signal from non-thermalized axions
falling into the galaxy's gravitational well which would yield very sharp signals due to their low
predicted velocity dispersion ($< 10^{-7}c$) \citep[Sikivie (2003)]{Sikivie03}.

\begin{figure}[!]
\resizebox{0.9\columnwidth}{!}
{\includegraphics{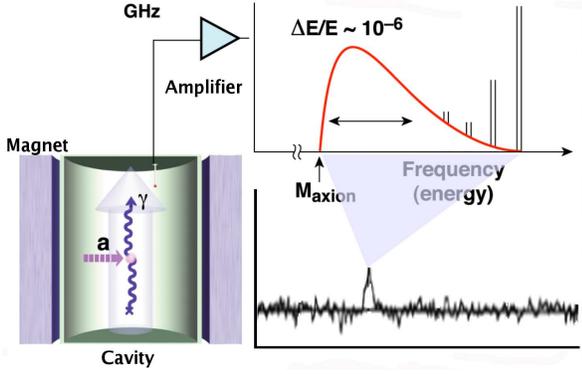}}
\caption{Outline of the general configuration of a resonant microwave cavity detector along with the
associated singal expected from axion-photon conversions. This diagram includes both the virial component
as well as possible lines from coherent axions.
}\label{fig:Cavity_diagram}
\end{figure}

\subsubsection{General Detector Properties}

Since the Lagrangian for axions coupling to a magnetic field goes as
\begin{equation}
{\cal L}_{a\gamma\gamma} = \left(\frac{\alpha_{\rm em} c_{a\gamma\gamma}}{2\pi F_a}\right)a\textbf{E}\cdot\textbf{B}\ ,
\end{equation}
the only resonant modes which can couple to axions are those that provide an axial electric field
component (transverse magnetic or TM modes).
The expected power generated from axion-to-photon conversions in the cavity is given by \citep[Sikivie (1985)]{Sikivie85}
\begin{eqnarray}
  P_a & = & \left(\frac{\alpha_{\rm em} c_{a\gamma\gamma}}{2\pi F_a}\right)^2 V
  B^2_0 \rho_a C_{lmn} \frac{1}{m_{a}} \text{min}\left(Q_L,Q_a\right)  \nonumber\\
  & = & 0.5\times 10^{-26}\,\text{W}
  \left(\frac{V}{500\,{\ell}}\right)
  \left(\frac{B_0}{7\,\text{T}}\right)^2 C_{lmn}
  \left(\frac{c_{a\gamma\gamma}}{0.72}\right)^2 \nonumber\\
  & & \times \left(\frac{\rho_a}{0.5\times 10^{-24}\,\text{g}\,\text{cm}^{-3}}\right)
  \nonumber\\
  & & \times \left(\frac{m_{a}}{2\pi(\text{GHz})}\right)
  \text{min}\left(Q_L,Q_a\right)\;,
  \label{eq:carosi-Power}
\end{eqnarray}
where $V$ is the cavity volume, $B_0$ is the magnetic field strength, $\rho$ is the local axion
mass density, $m_a$ is the axion mass, $C_{lmn}$ is a form factor which describes the overlap of the
axial electric and magnetic fields of a particular $TM_{lmn}$ mode, $Q_L$ is the microwave cavities loaded
quality factor (defined as center frequency over bandwidth) and $Q_a$ is axion quality factor defined as
the axion mass over the axion's kinetic energy spread.
The mode dependent cavity form factor is defined as
\begin{equation}
C_{lmn} = \frac{|\int_V d^3x \vec{E}_{\omega}\cdot\vec{B}_0|^2}{B_0^2V\int_V d^3x \epsilon |\vec{E}_{\omega}|^2}
\label{eq:form_factor}
\end{equation}
where $\vec{E}_{\omega}(\vec{x})e^{i\omega t}$ is the oscillating electric field of the $TM_{lmn}$ mode,
$\vec{B}_0(\vec{x})$ is the static magnetic field and $\epsilon$ is the dielectric constant of the
cavity space.
For a cylindrical cavity with a homogeneous longitudinal $\vec{B}$-field the $T_{010}$ mode yields the
largest form factor with $C_{010} \approx 0.69$ \citep[Bradley (2003)]{Bradley03}.

The mass range of cosmological axions is currently constrained between $\mu$eV and meV scales which
corresponds to converted photon frequencies between several hundred MHz and several hundred GHz.
Since larger microwave cavities correspond to lower resonant frequencies and lighter axions are more
likely to contribute to the dark matter density experiments have been designed to start searching
at the low end of the frequency range. At these frequencies cavities can only scan a few kHz at a time in order
to maintain the maximum quality factor. Axial metallic and dielectric tuning rods are utilized to
tune the cavities resonant frequency as it scans over the possible axion mass range. The scan rate is
determined by the amount of time it takes for a possible axion signal to be detected over the microwave
cavity's intrinsic noise and is governed by the Dicke radiometer equation \citep[Dicke (1946)]{Dicke46},
\begin{equation}
  \text{SNR} = \frac{P_a}{\bar{P}_N} \sqrt{Bt} = \frac{P_a}{k_B T_S} \sqrt{\frac{t}{B}}\ .
  \label{eq:carosi-Radiometer}
\end{equation}
Here $P_a$ is the power generated by axion-photon conversions (Eq. (\ref{eq:carosi-Power})),
$P_N=k_B B T_S$ is the cavity noise power, $B$ is the signal bandwidth, $t$ is the integration time,
$k_B$ is Boltzmann's constant and $T_S$ is the system temperature (electronic plus physical temperature).
The scan rate for a given signal to noise is given by
\begin{eqnarray}
  \frac{df}{dt} & = & \frac{12\,\text{GHz}}{\text{yr}}
  \left(\frac{4}{\text{SNR}}\right)^2
  \left(\frac{V}{500\,\text{l}}\right)
  \left(\frac{B_0}{7\;T}\right)^4
  \label{eq:carosi-Rate}\\
  & &
  \times\;C^2\left(\frac{c_{a\gamma\gamma}}{0.72}
  \right)^4\left(\frac{\rho_a}{0.45\;{\rm GeV/cm}^3}\right)^2 \nonumber\\
  & & \times
  \left(\frac{3K}{T_{\text{S}}}\right)^2
  \left(\frac{f}{\text{GHz}}\right)^2
  \frac{Q_{\text{L}}}{Q_{\text{a}}}
  \nonumber
\end{eqnarray}

One can see from Eq. (\ref{eq:carosi-Radiometer}) that even a tiny expected signal power can be made
detectable by either increasing the signal power ($P_a \propto VB_0^2$), increasing the integration time
$t$ or minimizing the system noise temperature $T_S$. Technology and costs limit the size and strength of
the external magnets and cavities and integration times are usually $t \sim 100$ seconds in order to scan an
appreciable bandwidth in a reasonable amount of time. As a result the majority of development has
focused on lowering the intrinsic noise of the first stage cyrogenic amplifiers.

\subsubsection{Microwave Receiver Detectors}

Initial experiments were undertaken at Brookhaven National Laboratory \citep[DePanfilis {\it et al.} (Rochester-Brookhaven Collaboration, 1987)]{DePanfilis87} and the
University of Florida \citep[Hagmann {\it et al.} (U. of Florida, 1990)]{Hagmann90}, but their modest sized cavities and magnet fields meant they
were still factors of 10-100 times away from plausible axion model space. There are currenly two active second generation experiments underway, the Axion Dark Matter
eXpereiment (ADMX) at Lawrence Livermore National Laboratory (LLNL), USA and the Cosmic Axion Research with Rydberg Atoms in Cavities at Kyoto (CARRACK) experiment in Japan. Both experiments utilize large microwave cavities immersed in a strong static magnetic field to
resonantly convert axions to photons but they go about detecting these photons in two different ways.
ADMX uses ultra-sensitive microwave receivers while CARRACK uses Rydberg atoms to detect single photons.

\begin{figure}[!]
\resizebox{0.9\columnwidth}{!}
{\includegraphics{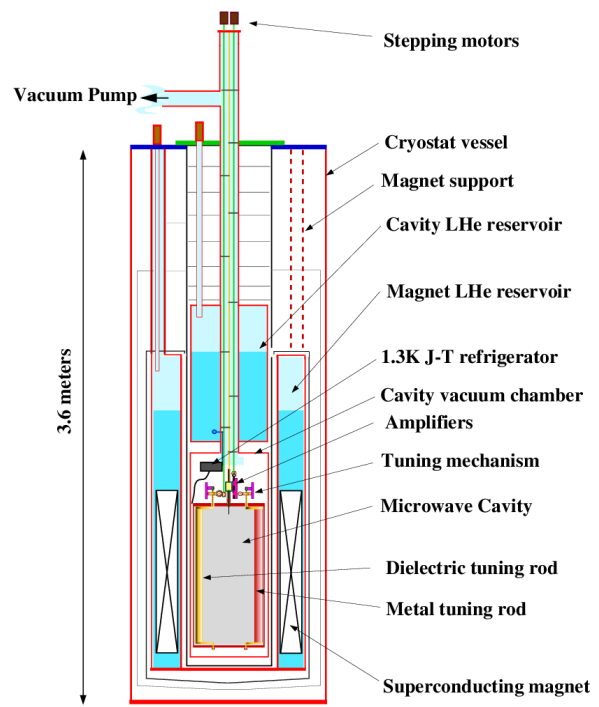}}
\caption{Schematic of the ADMX experiment \cite{CarosiLNP07}.
}\label{fig:ADMX_schem}
\end{figure}

The ADMX experiment is a collaboration of LLNL, MIT, the University of Florida, Lawrence Berkeley National
Laboratory (LBNL), U.C. Berkeley, U. of Chicago and Fermilab, and has been operating in various modes
since February, 1996. A diagram of the experiment can be seen in Fig. \ref{fig:ADMX_schem}.
ADMX consists of an 8.5 Tesla superconducting magnet, 110 cm in length with a 60 cm clear bore. A 200 liter
stainless steel microwave cavity plated in ultra-pure copper is suspended below a cryogenic stage in the center
of the B-field. Power generated in the cavity is coupled to an adjustable antenna vertically
input through the top cavity plate. Any signal is then boosted by extremely low noise cryogenic amplifiers
before being sent through a double-heterodyne mixing stage. Here the GHz range signal is mixed down
to an intermediate 10.7 MHz, sent through a crystal bandpass filter, and then mixed down to audio
frequencies at 35 kHz. This audio signal is then analyzed by fast-Fourier-transform (FFT) electronics
which measure over a 50 kHz bandwidth centered at 35 kHz. There is also a ``high resolution'' channel in
which the signal is mixed down to 5 kHz and sent through a 6 kHz wide bandpass filter. Time traces
of the voltage output, consisting of $2^{20}$ data points, taken with a sampling frequency of 20 kHz is then
taken, resulting in a 52.4 second sample with 0.019 Hz resolution \citep[Duffy {\it et al.} (2006)]{Duffy06}.

Since the system noise is dominated by the first stage of amplification great care was taken in choosing
the cryogenic amplifiers. The initial ADMX data runs utilized Heterojunction Field Effect
Transistor (HFET) amplifiers developed by the National Radio Astronomy Observatory (NRAO) \citep[Daw, Bradley (1997)]{Daw97}. Even though they had noise temperatures of only 2~K, the quantum noise limit at a GHz (defined as $T_q = h\nu/k_B$) is only 50~mK. As a result a great deal of development went into replacing the HFETs
with more sensitive Superconducting Quantum Interference Devices (SQuIDs) which had noise temperatures
of only 15\% the quantum limit \citep[Bradley (2003)]{Bradley03}. As of this writing data is being taken using the SQuIDs for the first stage of amplification.

Results from the initial run using HFET amplifiers have already probed plausible axion model space in the
axion mass range between 2.3 - 3.4 $\mu$ eV \citep[Bradley (2003)]{Bradley03}. Results from a high resolution search have probed further into coupling space over a smaller mass range, 1.98-2.17 $\mu$eV \citep[Duffy {\it et al.} (2006)]{Duffy06}. As of this review ADMX is scanning over the mass range corresponding to 800-900 MHz using SQuID amplifiers.

\begin{figure}[!]
\resizebox{1.0\columnwidth}{!}
{\includegraphics{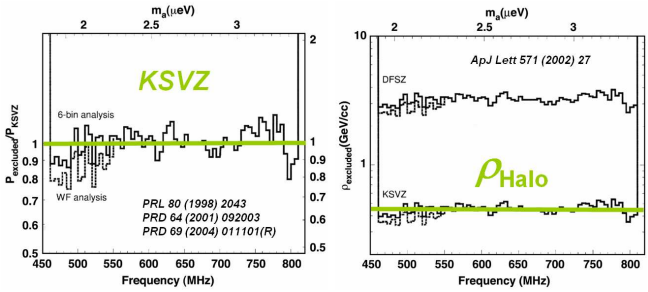}}
\caption{Medium resolution limits from the ADMX experiment. The left-side plot shows
limits to axion coupling assuming a dark matter halo density of $\rho = 0.45~\rm GeV/cm^3$ (upper region excluded). The right-side plot shows limits on the axion contribution to the dark matter halo density assuming the axion has either the KSVZ or the DFSZ couplings \cite{Bradley03}.
}\label{fig:med_res}
\end{figure}

\begin{figure}[!]
\resizebox{0.9\columnwidth}{!}
{\includegraphics{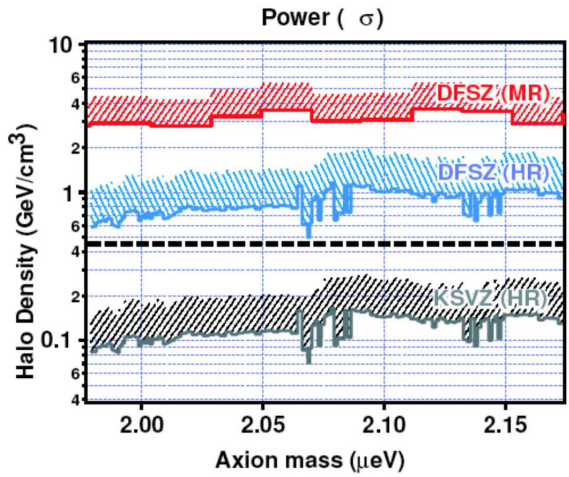}}
\caption{High resolution limits from the ADMX experiment. Limits given in terms of axion contribution to the dark matter halo density assuming the axion has either the KSVZ or the DFSZ coupling strength. The medium resolution plot for that mass range for the DFSZ coupling is also given as red \cite{Duffy06}. The KSVZ axion with $e_Q=0$ shown above gives $\rho_{a}$ at Earth less than $0.16~{\rm GeV/cm^3}$ which corresponds to $\Omega_a\le 0.36$. So the ADMX line of Fig. \ref{fig:Cosmic} using Eq. (\ref{axionenergyfrac}) crosses the $e_Q=0$ KSVZ line and goes down to the DFSZ line.
}\label{fig:high_res}
\end{figure}

\subsubsection{Rydberg Atom Detectors}

\begin{figure}[!]
\resizebox{0.9\columnwidth}{!}
{\includegraphics{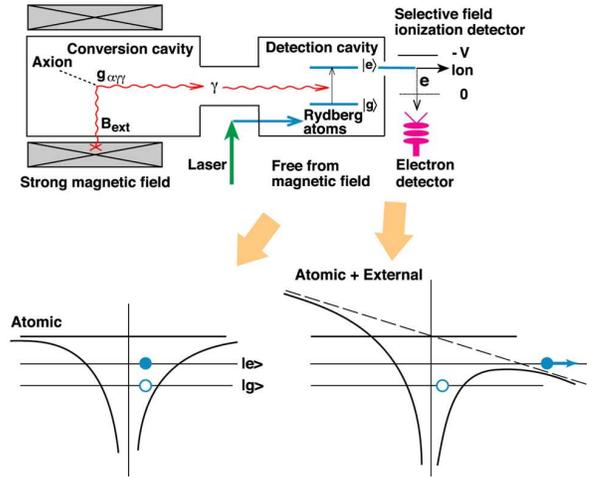}}
\caption{General schematic of CARRACK experiment utilizing Rydberg atoms to recover single photons
generated in the microwave cavity \cite{Tada06, CarosiLNP07}.
}\label{fig:CARRACK_diagram}
\end{figure}

The CARRACK experiment has published proof of concept papers for their detection technique using Rydberg atoms as opposed to low noise amplifiers \citep[Tada {\it et al.} (2006)]{Tada06}. The experimental setup can be seen in Fig. \ref{fig:CARRACK_diagram}. In it rubidium atoms are excited into a Rydberg state ($|0>\rightarrow |n>$), and move through a detection cavity coupled to an axion conversion cavity. The spacing between energy levels is tuned to the appropriate frequency utilizing the Stark
effect and the Rydberg atoms large dipole transition moment ensures efficient photon detection (one
photon per atom, $|ns>\rightarrow|np>$). The atoms are then subjected to a selective field ionization
allowing the atoms with the higher energy state ($|np>$) to be detected \citep[Carosi, van Bibber (2007)]{CarosiLNP07}. The advantage of this system is that Rydberg atoms act as single photon detectors and thus do not suffer from quantum noise limitations.

Though still in the development phase CARRACK has already gone through two iterations, CARRACK 1 and
CARRACK 2, and it has measured cavity emission at 2527 MHz down to a temperature of 67~mK,
which is a factor of two below the quantum noise floor for that frequency. Their eventual goal is 10~mK
\citep[Tada {\it et al.} (2006)]{Tada06}. One disadvantage of this technique is that one cannot detect signals finer than the bandpass of the cavity, of order $\sim 10^{-5}$, which negates searches for late infall coherent
axions.

\subsection{Laser searches}

In addition to cosmological and solar axion searches there is also a class of laboratory axion searches which utilize laser photons ($\gamma_{\rm laser}$) traversing a magnetic field. Here the polarized laser photons can scatter off virtual photons ($\gamma_{\rm v}$) provided by the magnetic field and convert into axions
$\gamma_{\rm laser}+\gamma_{\rm v} \rightarrow a$. Currently laser axion searches have fallen into two
general categories. The first technique looks
for magneto-optical effects of the vacuum due to polarized laser photons disappearing from the beam as
they are converted into axions. The second looks for photons converting into axions in the presence of a magnetic field, which are then transmitted through a wall and converted back into photons by a magnetic field
on the other side, so called ``light shining through walls'' experiments.

\subsubsection{Polarization shift of laser beams}

\begin{figure}[!h]
\resizebox{0.9\columnwidth}{!}
{\includegraphics{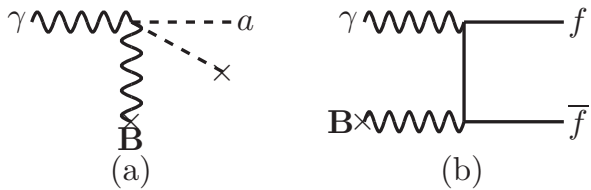}}
\caption{Possible processes leading to a vacuum
dichroism.}\label{fig:axionlike}
\end{figure}

There can be the axion-photon-photon anomalous coupling of the form $a{\bf E} \cdot{\bf B}$. The laser induced axion-like particle search employing this coupling has been performed since early 1990s by the Rochester-Brookhaven-Fermilab-Trieste (RBFT) group \citep[Cameron {\it et al.} (1991)]{BFRT91}. A few years ago, the same type experiment by the PVLAS collaboration was performed with an initial positive signal with $F_a\sim 10^6$ GeV \citep[Zavattini {\it et al.} (2006)]{PVLAS1} we discussed earlier.
This has led to some exotic models where a vacuum dichroism is achieved by producing axion-like particles as shown in Fig. \ref{fig:axionlike}(a). Because of the nonrenormalizable interaction implied in Fig. \ref{fig:axionlike}(a), one may reconclie this model with the astrophysical bound \citep[Mohapatra, Nasri (2007)]{Mohapatra07}. Or if light milli-charged particles are produced in the strong magnetic field as shown in Fig. \ref{fig:axionlike}(b) a vacuum dichroism is achieved as discussed in \citep[Gies, Jaeckel, Ringwald (2006), Masso, Redondo (2006), Kim (2007)]{Gies06,Masso06,Kim07milli}.
Here, the polarization of the laser beam is looked for. With more data accumulation, there is no convincing evidence for an axionlike particle with  $F_a\sim 10^6$ GeV at present, contrary to an earlier confusion \citep[Zavattini {\it et al.} (2007), Chou {\it et al.} (2008), Yoo (2007), Chen {\it et al.} (2006), Dupays {\it et al.} (2005)]{PVLAS2,GammeV08,Yoo08,Chen06,Dupeys05}. But, this incident has lead to the current search of axionlike particles at DESY \citep[Ringwald (for the 4th Petras Workshop) (2008)]{Ringwald08}.

\subsubsection{Light shining through walls}

\begin{figure}[!]
\resizebox{1.0\columnwidth}{!}
{\includegraphics{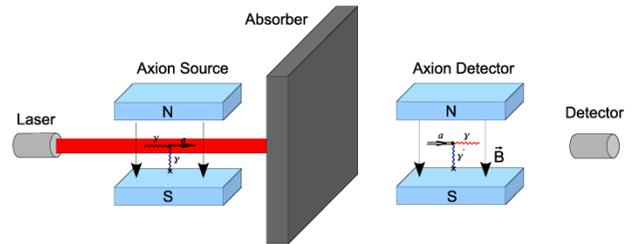}}
\caption{Schemtic of ``light shining through walls'' experiment \cite{Battesti07}.
}\label{fig:Shining}
\end{figure}

The ``Light shinging through walls'' technique for searching for axions was first proposed in 1987
by \citep[van Bibber {\it et al.} (1987)]{vanBibber87} and recently a model study has been presented \citep[Adler, Gamboa, Mend\'ez, L\'opez-Sarri\'on (2008)]{Adler08}. The general experimental layout can be seen in Fig. \ref{fig:Shining} where
polarized laser photons pass through the magnetic field with $\vec{E} || \vec{B}$ and any converted
axions (or other psuedoscalar particles) can continue through an absorber to be re-converted to photons on the
other side.

The probability for a photon to convert into an axion as it traverses the ``Axion Source'' region
is given by
\begin{equation}
P_{\gamma \rightarrow a} \propto \frac{1}{4}\left(\frac{\alpha_{\rm em} c_{a\gamma\gamma}}{2\pi F_a}
BL\right)^2\frac{1-\cos(qL)}{(qL)^2}.
\label{equ:laser_convert}
\end{equation}
This is the same probability for an axion to convert back into a detectable photon in the
``Axion Detector'' region on the otherside of an absorbor, which leaves the total probability for detecting a photon-axion-photon conversion as $P_{\gamma \rightarrow a \rightarrow \gamma} = P_{\gamma \rightarrow a}^2$ (ignoring photon detection efficiencies of course) \citep[Battesti {\it et al.} (2007)]{Battesti07}. There is a maximum detectable axion mass for these laser experiments as a result of the oscillation length becoming shorter than the magnetic field length causing a degradation of the form factor $F(q) = 1-\cos(qL)/(qL)^2$ but this can be compensated for by multiple discrete dipoles.

The first experiment using this technique was performed by a collaboration consisting of
Rochester-Brookhaven-Fermilab-Trieste (RBFT) in early 1990s \citep[Cameron {\it et. al.} (1993)]{BFRT93}. Using two superconducting dipole magnets ($L = 4.4$~m and $B = 3.7$~T)  and a laser ($\lambda = 514$ nm and $P = 3$ W)
with an optical cavity providing $\sim 200$ reflections in the axion generating region they were able to set upper limits on axion couplings of $g_{a\gamma\gamma} < 6.7\times 10^{-7}$ GeV (95\% C.L.) for pseudoscalars with a maximum mass of $m_a < 10^{-3}$ eV \citep[Cameron {\it et al.} (1993)]{BFRT93}.

\begin{figure}[!]
\resizebox{1.0\columnwidth}{!}
{\includegraphics{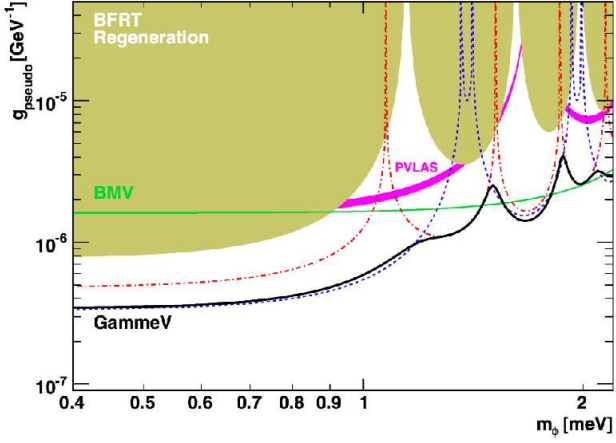}}
\caption{Current limits on axion coupling from the GammeV collaboration \cite{Yoo08,GammeV08}.
}\label{fig:GammeV}
\end{figure}

Recent photon regeneration experiments include the BMV collaboration at LULI \citep[Robilliard {\it et al.} (2007)]{BMV07} which uses a short pulsed-field magnet and the GammeV collaboration at Fermilab \citep[Chou {\it et al.} (2008)]{GammeV08} which uses a Tevatron dipole magnet ($L=6$~m and $B=5$~T) with an optical barrier in the middle. Both of which have ruled out the signal reported by the PVLAS (see next section). Fig. \ref{fig:GammeV} illustrates the current bounds from these latest regeneration experiments.
Recently it has been shown that photon regeneration experiments can be resonantly enhanced by
encompassing both the production and reconversion magnets in matched Fabry-Perot optical resonators
\citep[Sikivie, Tanner, van Bibber (2007)]{Sikivie07}.

\subsubsection{Magneto-Optical Vacuum Effects}

An alternative to the ``shining light through walls'' technique is to look for the indirect effect
of photons in polarized laser light converting into axions as the beam traverses a magnetic field.
\begin{figure}[!]
\resizebox{1.0\columnwidth}{!}
{\includegraphics{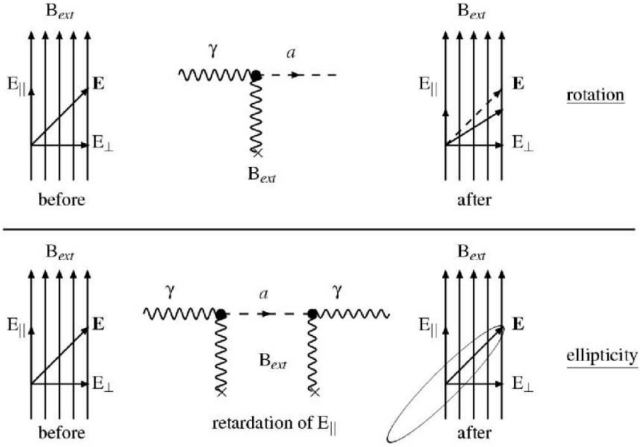}}
\caption{The upper figure shows the effect of dichroism as photons converting into axions cause
a rotation of the linear beams polarization vector by an amount $\epsilon$. The lower figure
shows virtual axions inducing birefringence in which the linear beam acquires ellipticity $\Psi$  \cite{Battesti07}.
}\label{fig:Mag_Opt}
\end{figure}
Fig. \ref{fig:Mag_Opt} illustrates the two different ways in which axion interactions can modify a
polarized laser beam, induced dichroism and vacuum birefringence. Vacuum dichroism occurs when a polarized
laser beam passes through a dipole magnet with the electric field component $\vec{E}$ at a non-zero
angle $\phi$ relative to the $\vec{B}$. The photon component parallel to the $\vec{B}$ will have a small
probability to convert into axions causing the polarization vector to rotate by an angle $\epsilon$.
Vacuum birefringence is due to the induced ellipticity of the beam ($\Psi$) as a result of virtual axions.
It should be noted that higher order QED diagrams, or ``light-by-light scattering'' diagrams,  are
expected to contribute to vacuum birefringence as well. Each of these effects can be estimated as
\begin{equation}
\Psi \approx N \frac{B^2L^3m^2_a}{384
\omega}\left(\frac{c_{a\gamma\gamma}}{F_a}
\right)^2\sin(2\theta),
\label{equ:Vac_effects_1}
\end{equation}
\begin{equation}
\epsilon \approx N\frac{B^2L^2}{64}
\left(\frac{c_{a\gamma\gamma}}{F_a}
\right)^2\sin(2\theta),
\label{equ:Vac_effects_2}
\end{equation}
in the limit that $m_a^2L/4\omega \ll 1$. Here $L$ is the effective path lengths, $N$ is the number of
paths the light travels in the magnetic field, $m_a$ is the axion mass, $\omega$ is the photon energy and
$\theta$ is the photon polarization relative to the magnetic field \citep[Battesti {\it et al.} (2007)]{Battesti07}.

The initial experiment looking for magneto-optical vacuum effects was carried out by the RBFT
Collaboration in the early 1990s \citep[Semertzidis {\it et. al.} (1990)]{BFRT90Sem}. This experiment used a single-pass length 8.8~m magnet with a magnetic field of $B \sim 2.1$~T and $N=500$. It set a limit on the polarization rotation of $\epsilon < 3.5\times 10^{-10}$ which was still 3 orders of magnitude higher than that
expected by light-by-light scattering and almost 15 orders of magnitude greater than a $m_a \sim 10^{-3}$~eV axion.

Recently the early PVLAS collaboration reported the positive detection of vacuum dichroism. This experiment consists of a 1~m long 5~T superconducting magnet with a angular frequency $\Omega_{\rm mag}$ of the magnet rotation and a 6.4~m long Fabry-Perot cavity giving the pass number $N = 2\Omega_{\rm mag}/\pi \sim 44,000$. It registered a polarization shift of
\begin{equation}
\epsilon = (3.9 \pm 0.5) \times 10^{-12} {\rm rad\;pass^{-1}}
\label{equ:PLVAS_pol}
\end{equation}
which translates to an allowed mass range of a neutral pseudoscalar boson of 1~meV$ \leq m_b \leq$1.5~meV and a coupling strength of
$1.5\times 10^{-3}{\rm GeV}^{-1} \leq c_{a\gamma\gamma} /F_a \leq  8.6\times 10^{-3}{\rm GeV}^{-1}$ \citep[Zavattini {\it et al.} (PVLAS Collaboration) (2006)]{PVLAS1}. Though the report of this positive signal has been retracted \citep[Zavattini {\it et al.} (PVLAS Collaboration) (2007, 2008)]{PVLASE,PVLAS2}, the interest it raised has lead to a number of more advanced experimental searches such as some of the new laser regeneration experiments mentioned previously.

\begin{widetext}\vskip 0.2cm
\section{Theories for Very Light Axions}\label{sec:AxionModels} \hskip 1cm
\begin{table}[!h]
\begin{center}
\begin{tabular}{l|c}
   \hline\\
   [-0.95em]
Axions from &  Order of $F_a$\\[0.2em]
  \hline\\
   [-0.95em]
 String theory& String scale or Planck scale\\[0.3em]
 $M$-theory  & String or the scale of the 11th dimension\\[0.3em]
 Large extra ($n$) dimension& Combination of the fundamental mass $M_D$ and extra dimension radius $R$ \\
Composite models & Compositeness scale\\[0.3em]
Renormalizable  theories& The U(1)$_{\rm PQ}$ global symmetry breaking scale\\[0.2em]
 \hline
\end{tabular}
\end{center}
\caption{Natural scales of $F_a$. For $n$ large extra dimension, the Planck mass is $M_P\simeq M_D(R/M_D)^{n/2}$.
}\label{Models}
\end{table}

\end{widetext}

The axion couplings come in three types: the PQ symmetry preserving derivative coupling $c_1$ term, the PQ symmetric $c_2$ term, and the anomalous $c_3$ term. The PQ symmetry gives a gluon anomaly and $c_2+c_3$ must be nonzero. Generally, we can therefore define $a$ as a
pseudoscalar field without potential terms except the one arising from the gluon anomaly under certain basis (for example in the $c_2=0$ basis),
\begin{equation}
\frac{a}{F_a}\frac{1}{32\pi^2}G^a_{\mu\nu}
\tilde G^{a\mu\nu} .\label{gluonanom}
\end{equation}
Then, we note that this kind of nonrenormalizable anomalous term can arise in several ways. The natural scales of $F_a$ are shown in Table \ref{Models}. In any case, the essence of the axion solution (wherever it originates in Table \ref{Models}) is that the axion VEV $\langle a\rangle$ seeks $\theta=0$ whatever happened before. In this sense it is a cosmological solution. The potential arising from the anomaly term after integrating out the gluon field is the axion potential (with $c_2+c_3$) sketched in  Fig. \ref{fig:axionDW}.

\subsection{SM singlets without SUSY}

A complex SM singlet carrying the PQ charge can appear in many extensions of the SM: in grand unified theories(GUTs) \citep[Wise, Georgi, Glashow (1981)]{Wise81}, in composite models \citep[Kim (1985), Choi, Kim (1985), Babu, Choi, Pati, Zhang (1994)]{Kimcomp,Choicomp,Babu94}, and in models with extra dimensions \citep[Di Lella, Pilaftsis, Raffelt, Zioutas (2000)]{DiLella00}.

In the SU(5) GUT, the axion can be embedded in a complex ${\bf 24}=\Sigma$ \citep[Wise, Georgi, Glashow (1981)]{Wise81}, in which case the VEV $\langle\Sigma\rangle$ breaking SU(5) down to the SM and hence the axion decay constant is the GUT scale and is outside the axion window. On the other hand, a complex GUT singlet, whose VEV is not related to the GUT scale, can house the axion within the axion window. The SUSY generalization of the SU(5) GUT axion has been shown to be possible
\citep[Nilles, Raby (1982)]{NillesRaby82}. Recently, the electrophilic axion in view of the white dwarf evolution \citep[Isern, Garc\'ia-Berro, Torres, Catal\'an (2008)]{Isern08} with the two dark matter scenario \citep[Huh, Kim, Kyae (2008)]{HuhKK08} has been suggested in a SUSY flipped SU(5) \citep[Bae, Huh, Kim, Kyae, Viollier (2008)]{BaeHKKV08}.

\subsection{Composite axions}\label{Subsec:Compaxion}

A SM singlet for the very light axion can arise as a composite meson with an extra confining force whose scale is much larger than the electroweak scale. This confining force can be the hidden sector gauge group in supergravity or just an extra gauge group. Let us call this extra confining gauge group `axi-color' SU($N$). To create the QCD axion below the axi-color scale, there must be two classically conserved axial global symmetries \citep[Kim (1985), Choi, Kim (1985)]{Kimcomp,Choicomp}. With only one axial symmetry, there would not result a massless meson, as in the case of  one flavor QCD there is no massless meson since the only meson $\eta'$ becomes heavy by the instanton solution of the so-called U(1) problem \citep['t Hooft (1976)]{tHooftU1}. For two axial symmetries, we can consider two kinds of axi-quarks, $Q^{A\alpha}, \overline{Q}_{A\alpha}, q^A,$ and $\overline{q}_A$ where $A$ is the SU($N$) index and $\alpha$ is the SU(3)$_c$ index. For these vectorlike representations, $\bf (N,3)+(\overline{N},\overline{3})+(N,1)+(\overline{N},1)$ under SU($N$)$\times$SU(3)$_c$, mass terms are not introduced. The axi-color vacuum angle problem is solved basically by the massless axi-quarks, $Q$ and $q$. Even though $Q$ looks like a massless QCD quark, it cannot be considered as the massless quark solution of the QCD $\theta$ problem. After integrating out the axi-color degrees, we obtain an effective Lagrangian resulting from $Q$ and $q$. The axi-baryons are expected to be removed at the axi-color scale. Out of two kinds of mesons, one (the axi-color $\eta'$) is removed at the axi-color scale and the other remains exactly massless. However, this massless axi-color meson couples to the QCD anomaly and becomes a QCD axion through the $c_3$ term, becoming the so-called hadronic axion. Out of the two currents,
\begin{eqnarray}
\begin{array}{l}
\overline{J}_{\mu}^5= \overline{Q}\gamma_\mu\gamma_5 Q+
\overline{q}\gamma_\mu \gamma_5q\\
J_{\mu}^5= \overline{Q}\gamma_\mu\gamma_5 Q -
3\overline{q}\gamma_\mu \gamma_5q,
\end{array}\label{compositeaxion}
\end{eqnarray}
the divergence of $J^5_\mu$ corresponds to the massless meson $a$ below the axi-color scale,
\begin{equation}
\partial^\mu J_\mu^5=\frac{2N}{32\pi^2}G^{\alpha}_{\mu\nu}\tilde G^{\alpha\mu\nu},
\end{equation}
and hence we obtain the effective interaction (\ref{gluonanom}). In this minimal model, the domain wall number is $N$ \citep[Choi, Kim (1985)]{Choicomp,Choi85DW}. In a supergravity model of preons, a similar mechanism was used to realize a composite axion \citep[Babu, Choi, Pati, Zhang (1994)]{Babu94}, where the role of $q$ type matter is replaced by the meta-color gluino $\lambda'$ where meta-color is the binding force of preons. Even if the meta-color gluino obtains a mass of ${\cal O}(100\ \rm GeV)$, the QCD $\theta$ can be made within the experimental bound if $F_a$ is greater than $10^{11}$ GeV.

The composite axion of \citep[Chun, Kim, Nilles (1992)]{ChunKNmu92b} is a composite made of hidden-color scalars whose bilinears develop VEVs and break the PQ symmetry. This idea has been made more concrete in \citep[Kim, Nilles (2009)]{KimNill09}.

In the gauge mediated SUSY breaking scenario {\it a la} Intrilligator, Seiberg and Shih (ISS) \citep[Intrilligator, Seiberg, Shih (2006)]{ISS06}, for example, an SU($N_c$) confining group with $N_f$ flavors satisfying $N_c+1\le N_f<\frac32 N_c$ allows a SUSY breaking local minimum. If $N_f-N_c\ge 3$ (for example $N_c=7$ and $N_f=10$) with one type $Q^{A\alpha}+\overline{Q}_{A\alpha}$ and $N_f-3$ flavors of the type $q^A+\overline{q}_A$, then there can exists a suitable local minimum where the composite axion envisioned in Eq. (\ref{compositeaxion}) can be realized. In this case, the SUSY breaking scale and the composite axion scale are related as first tried in \citep[Kim (1984)]{Kim84}.

\subsection{Axions with extra dimensions}

With large extra dimensions, the axion identification involves a few parameters: the fundamental scale mass $M_{\rm F}$, the Kaluza-Klein (KK) radius $R$ and the number of extra dimensions $n$. In addition, there are several ways to allocate the field(s) containing the axion in the bulk and/or branes.

The possibility of large extra dimensions have been considered for the flat and warped extra dimensions. The TeV scale for $M_{\rm F}$ was the main motivation to look for the next level of the current experimental limit on the mm scale gravity \citep[Arkani-Hamed, Dimopoulos, Dvali (1998), Antoniadis, Arkani-Hamed, Dimopoulos, Dvali (1998)]{Arkani98A,Antoniadis98}. Because the axion scale is considered to be at the intermediate scale, a string theory at the intermediate scale  $M_{\rm F}$ has been considered also \cite[Burgess, Ibanez, Quevedo (1999)]{Burgess99}. With the Randall-Sundrum type warp factor \citep[Randall, Sundrum (1999)]{Randall99A}, it is possible to introduce the intermediate scale with a Planck scale $M_{\rm F}$ via the Giddings-Kachru-Polchinski stabilization mechanism \cite[Giddings, Kachru, Polchinski (2002)]{Giddings02}.

Here, we look only at the possibility of TeV scale $M_{\rm F}$. Since the Planck mass is given by $M_P\approx M_{\rm F} (R M_{\rm F})^{n/2}$, we obtain $n\ge 2$ for $M_{\rm F}\simeq 10$ TeV \citep[Hannestad, Raffelt (2002), Kanti (2008)]{Hannestad02,Kanti08}. The Lagrangian in $(4+n)-$dimensions ($(4+n)$D) with a bulk field axion can be written as \citep[Chang, Tezawa, Yamaguchi (2000)]{Chang00},
\begin{eqnarray}
&{\cal L}_{\rm eff}= \int d^ny \Big\{\frac12 M_{\rm F}^n
[\partial_\mu a\partial^\mu a+\partial_y a\partial^y a ]\nonumber\\
 &+\frac{\xi \alpha_{\rm em}}{\pi}\frac{a}{\overline{v}_{PQ}}
F^{\rm em}_{\mu\nu}\tilde F^{\rm em,\mu\nu}
\Big\}
\end{eqnarray}
where $\overline{v}_{PQ}$ is the PQ symmetry breaking scale at the fundamental scale order $M_{\rm F}$, and
\begin{equation}
a(x^\mu,{\bf y})=\sum_{{\bf n}={\bf 0}}^\infty a_{\bf n}(x^\mu) \cos\left(\frac{{\bf n}\cdot{\bf y}}{R}\right).
\label{axionKK}
\end{equation}
The 4-dimensional PQ symmetry breaking scale is $F_a\approx
(M_P/ M_{\rm F})^{A/n}\overline{v}_{PQ}$ where $A=|{\bf n}|=\sqrt{n_1^2 +\cdots+n_n^2}<n$ and $F_a$ falls between $\overline{v}_{PQ}$ and $M_P$. The very light axion is
the ${\bf n}=0$ component in Eq. (\ref{axionKK}), and the rest are the KK axions. The mass splitting of the KK axions are of order $1/R$, and the phenomenology of these KK axions for $a_{KK}\to 2\gamma$ has been nicely studied in \citep[Di Lella, Pilaftsis, Raffelt, Zioutas (2000)]{DiLella00}, from which we have $1/R\sim 1(10)$ eV for $n=2(3)$ for $M_{\rm F}\approx 1$ TeV.

The possibility of a $Z_2$ odd 5D gauge field in a warped fifth dimension has been suggested for a QCD axion under the assumption that all unwanted PQ symmetry breaking effects are suppressed \citep[Choi (2004)]{Choi04}. One such constraint is that the bulk fields carry the vanishing PQ charge.  

\subsection{SUSY breaking scale, axion and axino}\label{subsec:axinoth}

The 4D supergravity interactions with the vanishing cosmological constant was written in 1983 \citep[Cremmer, Ferrara, Girardello, van Pr\"oyen (1983)]{Cremmer83}. The PQ symmetry can be embedded in the supergravity framework  \citep[Kim (1984)]{Kim84},
\begin{equation}
\begin{array}{c}
W_{PQ}=(f_\delta A_1A_2-F_1^2) Z+(f_\epsilon A_1A_2-F_2^2) Z'\\
+f_Q A_1\overline{Q}_1Q_2
\end{array}\label{WAAZZ}
\end{equation}
where $Z,Z',A_1$ and $A_2$ are gauge singlet chiral fields, $\overline{Q}_1$ and $Q_2$ are chiral quark superfields, $f_\delta,f_\epsilon, F_1^2$, and $F_2^2$ are parameters. The superpotential (\ref{WAAZZ}) leads to the F-term SUSY breaking and the PQ symmetry breaking at a common scale at order ${\cal O}(F_1^2,F_2^2)$ if $f_\delta/f_\epsilon\ne F_1^2/F_2^2$. The $f_Q$ term defines the PQ charge of the heavy quark and the resulting axion is the KSVZ type. The PQ symmetry breaking scale is given by nonzero $\langle A_1A_2\rangle\simeq (F^2/\lambda)\cos(\beta-\alpha)$ and the SUSY breaking scale is given by nonzero $Z_F=-F^2\sin\alpha
\sin(\beta-\alpha)$ and $Z'_F=F^2\sin\alpha
\sin(\beta-\alpha)$ where $\lambda= \sqrt{f_\delta^2 +f_\epsilon^2}, F^2=\sqrt{F_1^4+F_2^4}, \tan\alpha=f_\epsilon/f_\delta$ and $\tan\beta=F_2^2/F_1^2$. The axino does not obtain mass at this level, but obtains a mass at order the soft SUSY breaking scale \citep[Chun, Kim, Nilles (1992), Chun, Lukas (1995)]{Chunaxino92a,ChunLukas95}. Note that $\langle Z'/Z\rangle\simeq -\cot\alpha$ with $\langle Z, Z'\rangle={\cal O}(F^2/M)$. The early discussions on axino can be found in \citep[Frere, Gerard (1983)]{Frere83}.

\subsection{The $\mu$ problem}

If Higgs doublets $H_{u,d}$ carry vanishing U(1) charges beyond the MSSM gauge charges, then the superpotential can contain $W_\mu=-\mu H_uH_d$ term where $\mu$ can be of order the fundamental scale since it is a supersymmetric term. This is problematic for the TeV scale electroweak symmetry breaking, which is the so-called $\mu$ problem \citep[Kim, Nilles (1984)]{KimNilles84}. This $\mu$ term is a supersymmetric Higgsino mass term and can be forbidden in $W$ by introducing some symmetry, continuous or discrete. The widely discussed ones are the PQ and R symmetries. In the supergravity framework, if the Higgs doublets carry one unit of PQ charge then the nonrenormalizable interactions of the form $S^2H_uH_d/M_P$ can be present in $W$ if $S^2$  and $H_uH_d$ carry the opposite PQ charges. Then, the resulting $\mu$ is of order $F_a^2/M_P$ which can be of order of the TeV scale \citep[Kim, Nilles (1984)]{KimNilles84}. With an intermediate hidden sector with hidden sector squarks  $Q_1$ and $\overline{Q}_2$, one may have a nonrenormalizable interaction of the form $Q_1\overline{Q}_2H_uH_d/M_P$. In this case also, the  hidden sector squark condensation at the intermediate mass scale can generate a TeV scale $\mu$ \citep[Chun, Kim, Nilles (1992)]{ChunKNmu92b}. [For the $B_\mu$ term, one may consider $(Q_1 Q_2^*/M_P^2)H_uH_d$ in the K\"ahler potential.] The superpotential is better to possess this kind of PQ symmetry and/or R symmetry \citep[Dine, MacIntire (1992), Hall, Randall (1991), Casas, Munoz (1993),
Kim, Nilles (1994), Kim(1999a)]{DineM92,Hall91,Casas93,
KimNilles94,Kim99plb}. If so, even if the nonrenormalizable interactions are not considered, the gravity mediation scenario can generate a TeV scale $\mu$ via the Giudice-Masiero mechanism \citep[Giudice, Masiero (1988)]{Giudice88}. In supergravity, the Higgsino mass term is present in the chiral fermion mass matrix given by \citep[Cremmer, Ferrara, Girardello, van Pr\"oyen (1983), Nilles (1984)]{Cremmer83,Nilles84}
\begin{equation}
e^{-G}[G^{ij}-G^iG^j-G^l(G^{-1})^k_lG^{ij}_k]\chi_{{\rm L}i}\chi_{{\rm L}j}
\end{equation}
where $G=K(\phi,\phi^*)-\ln|W|^2$. The term $e^{-G}G^{ij}$ gives $\mu\sim m_{3/2}$ if $K$ contains $H_uH_d$  \citep[Giudice, Masiero (1988)]{Giudice88} and $\mu\sim S^2/M_P$ if $W$ contains $S^2H_uH_d/M_P$ \citep[Kim, Nilles (1984)]{KimNilles84}.

In the next-to-MSSM(NMMSM) models with $W=SH_uH_d$, the $\mu$ term can be generated by the singlet VEV at the electroweak scale \citep[Cerde\~no, Hugonie, L\'opez-Fogliani, Mu\~noz, Teixeira  (2004), L\'opez-Fogliani, Mu\~noz (2006)]{Munoz04,Munoz06}. In a $Z'$-added MSSM($Z'$MSSM), the $\mu$ term can be successfully generated also \citep[Langacker, Paz, Wang, Yavin  (2008)]{Langacker08}.

Extending the MSSM gauge group which can be broken down to the MSSM at a high energy scale, one can generate a reasonable $\mu$. For example, there exists an interesting solution to the problem, $\lq\lq$Why is there only one pair of Higgsino doublets at low energy?", in the extended electroweak model \citep[Lee, Weinberg (1977)]{LeeWein77a}. This is dictated by the extended gauge symmetry. This one pair problem is elegantly solved in the SUSY Lee-Weinberg type model due to the antisymmetric Higgsino mass matrix \citep[Kim (2007)]{Kim07st}, reminiscent of the $\lq\lq$color" introduction to put the low-lying baryons in the completely symmetric representation {\bf 56} in the old flavor-spin SU(6) \citep[Han, Nambu (1965)]{Han65}.

Thus, explicit models toward a successful $\mu$ in the gravity mediation scenario can be constructed in extra singlet models, in SUSY-GUT models, through the superpotential, through the K\"ahler potential, and in composite models.

The loop effects are the important sources of SUSY breaking in the gauge mediated SUSY breaking scenario (GMSB) \citep[Dine, Nelson (1993), Dine, Nelson, Shirman (1995), Dine, Fischler, Srednicki (1981), Dimopoulos, Raby (1981), Dine, Fischler (1983)]{DineNelson93,DineShirman95, DineFiGMSB81,DimoRaby81,DineFish83}, in anomaly mediation SUSY breaking (AMSB) \citep[Randall, Sundrum (1999), Giudice, Luty, Murayama, Rattazzi (1999)]{Randall99Anom,Giudice99Anom}, and even in the mirage mediation scenario \citep[Choi, Jeong, Okumura (2005), Loaiza-Brito, Martin, Nilles, Ratz (2005)]{Choi05MM,mirage2}. The GMSB has been suggested to solve the flavor problem \citep[Gabbiani, Gabrielli,  Masiero, Silvestrini (1996)]{Masiero96}, present in the gravity mediation scenario. In this GMSB or any other loop generated SUSY breaking scenarios, the soft terms generated by supergravity effect are required to be subdominant compared to those arising from the loops or at best comparable to it. If it is sub-dominant as in the GMSB or AMSB, then there are some problems.

Firstly, the generation of $\mu$ is difficult because $\mu$ term generation via the Giudice-Masiero mechanism is subdominant at the TeV scale. One has to generate $\mu$ employing the PQ and/or R symmetries, which  however does not belong to the grand design of generating all TeV scale parameters dynamically. In this regard, another confining group around TeV scale has been proposed \cite[Choi, Kim (2000)]{{ChoiHD00}}, and the model presented there belongs to a composite SU(2)$_W$ axion discussed in Subsec. \ref{Subsec:Compaxion}, which was saved by introducing singlets and relevant couplings \citep[Luty, Terning, Grant (2001)]{Luty01}. Then, again it does not succeed in generating all TeV scale parameters dynamically.

Second, in the loop SUSY breaking scenarios for generating all TeV scale electroweak parameters by loops, there exists the $B_\mu/\mu$ problem \citep[Dvali, Giudice, Pomarol (1996)]{Giudice96}. Since it occurs at loop orders, we consider $\int d^4\theta H_u H_d X^\dagger$ for $\mu$ and $\int d^4\theta H_u H_d X X^\dagger$ for $B_\mu$ where the auxiliary component of $X$ develops a VEV. From this observation, one generically obtains $B_\mu\sim \mu\Lambda$ where $\Lambda\sim \mu/f^2$ can be greater than $\mu$ which was remedied by making $B_\mu$ appear at a two loop order \citep[Dvali, Giudice, Pomarol (1996)]{Giudice96}. This $B_\mu/\mu$ problem occurs basically from the difference of the engineering dimensions of the $B_\mu$ and $\mu$ terms. Both generically appear at one loop order with the coefficient $g^2/16\pi^2$, and hence in describing the electroweak scale $B_\mu$ term lacks one power of  $g^2/16\pi^2$.  Recently, a better solution employing a K\"ahler potential $H_uH_d(\ln X+\ln X^\dagger )$ has been suggested \citep[Giudice, Kim, Rattazzi  (2008)]{Giudice08}, which can be compared to the original Giudice-Masiero  K\"ahler potential $H_uH_dX^\dagger+\cdots$. There exist several more ideas toward the $B_\mu/\mu$ problem \citep[Cohen, Roy, Schmaltz (2007), Murayama, Nomura, Poland (2008), Roy, Schmaltz (2008), Cho (2008)]{Cohen07,Murayama08,Roy08,Cho08}

\begin{figure}[!h]
\resizebox{0.7\columnwidth}{!}
{\includegraphics{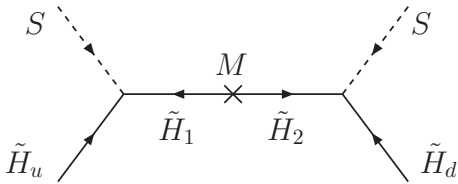}}
\caption{The generation of the $\mu$ term by a seesaw mechanism.}\label{fig:muterm}
\end{figure}

Maybe, the nonrenormalizable interactions are the easy solution of the $\mu$ and $B_\mu/\mu$ problems even in the GMSB also. Here however one introduces another scale. Without a detail knowledge on the ultra-violet completion of the MSSM, the nonrenormalizable interactions are usually assumed to be suppressed by the Planck mass $M_P$. But, there can be some heavy mass scale $M$, which can be somewhat smaller than the Planck mass $M_P$, for the seesaw mass of the required nonrenormalizable interactions. In string compactifications, it is known explicitly that $M$ can be different from $M_P$ \citep[Choi, Kim (2006), Kim, Kim, Kyae (2007), Choi, Kobayashi (2008)]{ChoiKSorb06,KimJHKyae07,ChoiKS08}. A simple diagram giving an $M$ dependence is shown in Fig. \ref{fig:muterm} where SU(2) doublets $H_1$ and $H_2$ forms a vectorlike superheavy pair.
This is a kind of the seesaw mechanism of Higgsino doublet  pairs. For this scenario, a superpotential possessing the PQ symmetry can be constructed
\begin{eqnarray}
&W=\frac12 mX^2+f_XX^3+\frac12 S^2T+XH_1H_2\nonumber\\
&-f_1SH_1H_u-f_2SH_2H_d+\cdots
\end{eqnarray}
where $\langle X\rangle=M$ and $H_uH_d$ term is forbidden by the PQ symmetry. Then, the $\mu$ term for $H_u$ and $H_d$ (which give mass to up and down type quarks, respectively) is given by $\mu=\langle f_1f_2S^2/M\rangle$
\citep[Kim, Nilles (1984)]{KimNilles84}. If $\langle S\rangle$ is lowered to the hidden sector confining scale of order $\sim 10^{10-12}$ GeV in the GMSB, the Higgsino mass can be made around the TeV scale by adjusting $f_1f_2/M$. One may construct models with appropriate F-terms such that $B_\mu$ and $m^2_{\rm soft}$ are of the same order in the GMSB, e.g. through the PQ symmetry preserving term in the K\"ahler potential
\begin{equation}
\int d^4\vartheta \frac{f_1f_2T^*}{X}H_uH_d +{\rm h.c.}
\end{equation}
which also gives a $\mu$ term.

In the so-called $\lq\lq$mixed mediation(M-mediation) scenario", with comparable moduli, anomaly and gauge mediations, that includes in its parameter space  the GMSB, the AMSB, the mirage mediation, and the deflected mirage mediation \citep[Everett, Kim, Ouyang, Zurek (2008)]{Everett081}, the loop generated $\mu$ term in general has a severe $B_\mu/\mu$ problem. It seems that the model presented in an AMSB scenario \citep[Pomarol, Rattazzi (1999), Rattazzi, Strumia, Wells (2000)]{Pomarol99,Rattazzi00} has the basic ingredient for the solution of $B_\mu/\mu$ problem according to a PQ symmetry as stressed in \citep[Giudice, Kim, Rattazzi (2008)]{Giudice08}. This can be gleaned also from the axion shift symmetry in the mirage mediation scenario \citep[Nakamura, Okumura, Yamaguchi (2008)]{Yamaguchi08}.

\subsection{Axions from superstring} \label{subsec:superstring}

The most interesting theory housing axions is superstring theory. Axions from string are described by effective field theory below the compactification scale. If it arises from the spontaneous symmetry breaking of a tree-level global symmetry as we discussed so far, the answer is simple: There is no such axion since string theory would not allow any global symmetry. If the compactification process leads to the SM, the renormalizable terms in this effective theory respect the gauge symmetry SU(3)$_c\times$SU(2)$\times$U(1)$_Y$ and the global symmetries: the baryion number U(1)$_B$ and the separate lepton numbers U(1)$_{Li}$. On the other hand, if the nonrenormalizable terms are allowed, one can write for example $q_L l_L u_R d_R$, breaking both the baryon number and the lepton number. Including the nonrenormalizable terms, thus the SM does not respect the baryon and lepton numbers. Similarly, there is no PQ global symmetry if we are allowed to write all nonrenormalizable terms. For the PQ symmetry, the situation is more severe. Suppose the singlet carrying the PQ charge is $\sigma$. Then, $\sigma^*\sigma$ respects the PQ symmetry but $\sigma^2$ and $\sigma^{*2}$ do not respect the PQ symmetry, which has led to the discussion on the gravitational effects on the axion \citep[Kamionkowski, March-Russell (1992), Barr, Seckel (1992), Holman, Hsu, Kephart, Kolb, Watkins, Widrow (1992), Ghigna, Lusignoli, Roncadelli (1992), Dobrescu (1997)]{Barr92,Holman92,Kamionkowski92,Ghigna92,Dobrescu96}. Therefore, the PQ symmetry cannot be discussed in general terms in terms of matter fields only, when we include gravity in the discussion such as in string theory.

Thus, in string compactification one must consider the gravity multiplet also. Here, the gauge singlet bosonic degrees in the gravity multiplet are graviton $g_{MN}(M,N=0,1,\cdots,9)$, antisymmetric tensor $B_{MN}$ and dilaton $\Phi$. In 10D, $g_{MN}$ and $B_{MN}$ are gauge fields.  A 4D action on Minkowski space $x^\mu(\mu=0,1,2,3)$ is obtained by compactifying six internal spaces $y^i(i=4,\cdots,9)$ with the compact volume $V_Z$. Some bosonic degrees from the 10D antisymmetric tensor field behave like pseudoscalars in the 4D effective theory. Thus, the axion candidates, if they are not arising from the matter multiplets, must be in $B_{MN}$. The pseudoscalar fields in $B_{MN}$ are like phase fields in axion models in field theory. Because there is no global symmetry in string theory, there must be no massless $B_{MN}$, otherwise the shift symmetry of $B_{MN}$ must have worked as a global symmetry.
From the tree-level equations of motion all pseudoscalar $B_{MN}$ fields are not massless. For example, if a shift symmetry of $B_{MN}$ is related to an anomaly as in the PQ current case, we consider that the shift symmetry is already broken. In other words, there is no shift symmetry of pseudoscalar $B_{MN}$ if it is not anomalous.

One must deal with these bosonic degrees in string compactification to see whether these components do not lead to terms in the potential (or the superpotential in SUSY models), which is very technical and model-dependent procedure. Here, we briefly discuss the story on axions from string and comment on its phenomenological viability. Some relevant recent reviews describing details can be found in \citep[Svrcek, Witten (2006), Conlon (2006)]{Svrcek06,Conlon06}. The M-theory discussion was presented in \citep[Choi (1997), Svrcek, Witten (2006)]{Choi97,Svrcek06}.

The pseudoscalar fields in $B_{MN}$ come in two categories, one the tangential component $B_{\mu\nu}$ and the other $B_{ij}$. $B_{\mu\nu}$ can be discussed in any string compactification and hence is called $\lq$model-independent'(MI) while $B_{ij}$ depends on the compactification scheme as its internal coordinates $i$ and $j$ imply and is called $\lq$model-dependent'(MD). After presenting the string formulae containing $B_{MN}$, we will discuss the MI axion in Subsubsec. \ref{subsubsec:MIa} and then the MD axions present in much more speculative models in \ref{subsubsec:MDa}. One can go directly to Subsubsec. \ref{subsubsec:MIa} if not interested in the details on the origin of the couplings.

Now, there exists the standard formulae for the string action \citep[Polchinski (1988), Eq. (13.3.22)]{PochinskiBk}, which was lacking in the early days of string axions \citep[Witten (1984), Choi, Kim (1985)]{Witten85a,Choiharm85}. The Type II dilaton $\phi_{II}$ and coupling $g_{II}=e^{\phi_{II}}$ are related to the 10D gravitational coupling $\kappa_{10}$ by $M_{10}^8=1/\kappa_{10}^2=4\pi/g_{II}^2\ell_s^8$  where $\alpha'=\ell_s^2/(2\pi)^2$. For the Type II string, there are NS-NS and R-R fluxes which can give anomalous couplings. These complicated system housing pseudoscalars is reviewed in \citep[Conlon (2006)]{Conlon06} with a possible conclusion of the difficulty of realizing a QCD axion in string models with a workable moduli stabilization \cite[Kachru, Kallosh, Linde, Trivedi (2003)]{Kachru03}. In heterotic string models, there does not exist a reasonable moduli stabilization mechanism, even though an ambitious attempt has been proposed \citep[Becker, Becker, Fu, Tseng, Yau (2006)]{Becker06}. We however discuss heterotic string axions below because string axions were found first in heterotic string models and key couplings on axion phenomenology might be similarly discussed also for the Type II string. As in the Type II string, the heterotic string coupling is related to the dilaton $\Phi$ as $g_{h}=e^{\Phi}$. The kinetic energy terms of $g_{MN}, B_{MN}$ and $A_M$ are \citep[Polchinski (1988), Eqs. (12.1.39) and (12.3.36)]{PochinskiBk},
\begin{widetext}
\begin{eqnarray}
&&{\cal L}_{KE}=\sqrt{-g_{10}}e^{-2\Phi}\left\{
\frac{M_{10}^8}{2}  R
-\frac{M_{10}^8}{4} \left|dB_{\it 2}-\frac{\omega_{\it 3}}{M_{10}^2 g_h^2}\right|^2
-\frac{M_{10}^8\alpha'}{8 g_h^2} {\rm Tr_v}|F_{\it 2}|^2
\right\}\nonumber \\
&&\quad\quad=\sqrt{-g_{10}}\left\{
\frac{2\pi}{g_h^2\ell^8_s}  R
-\frac{\pi}{g_h^2\ell^8_s}  \left|dB_{\it 2}-\frac{\omega_{\it 3}}{M_{10}^2 g_h^2}\right|^2
-\frac{1}{4(2\pi) g_h^2\ell^6_s } {\rm Tr_v}|F_{\it 2}|^2
\right\}\label{BMNKE}
\end{eqnarray}
\end{widetext}
where ${\rm Tr_v}$ is the trace over vector representation and the Chern-Simons 3-form is
\begin{equation}
\omega_{\it 3}={\rm Tr_v}\left( A_{\it 1}\wedge d A_{\it 1}+\frac23  A_{\it 1}\wedge  A_{\it 1}\wedge  A_{\it 1}\right).
\end{equation}
For E$_{8}\times $E$_{8}$, there is the adjoint representation and we use $\frac{1}{30}{\rm Tr_a}$ in place of  ${\rm Tr_v}$. For the compact internal volume $V_Z$, the Planck mass is $M_P=4\pi V_Z/g_s^2 \ell_s^8$ and the 4D gauge coupling constant is $g^2_{YM}= 4\pi g_s^2 \ell_s^6/V_Z$ or  $\alpha_{YM}= g_s^2 \ell_s^6/V_Z$. In most compactifications, the SM gauge fields arise from the level $k=1$ embedding and the coupling $\alpha_{YM}$ is the coupling strength at the compactification scale. If the SM gauge fields are embedded in the level $k$, the SM gauge coupling at the compactification scale will be smaller by the factor $k$.
For interactions of $B_{MN}$, we consider
the Bianchi identity, the gauge invariant couplings of gaugino $\chi$ \citep[Derendinger, Ibanez, Nilles (1985), Dine, Rohm, Seiberg, Witten (1985)]{Derendinger85,DineRohm85}, and the Green-Schwarz terms \citep[Green, Schwarz (1984)]{Green84}
\begin{eqnarray}
&dH=\frac{1}{16\pi^2}({\rm tr} R\wedge R-{\rm tr} F\wedge F),\ H_{MNP}\overline{\chi} \Gamma^{MNP}\chi  ,\nonumber\\ &B\wedge {\rm tr}F\wedge F\wedge F\wedge F+\cdots
\label{BMNcoupling}
\end{eqnarray}
where $H_{MNP}$ is the field strength of $B_{MN}$, $F$ is the field strength of the gauge field $A$, the gauge invariant fermion coupling is the SUSY counterpart of the relevant terms of Eq. (\ref{BMNKE}), and $\cdots$ denotes more Green-Schwarz terms. It was argued that the $H_{MNP}$ coupling to the gaugino must be a perfect square \citep[Dine, Rohm, Seiberg, Witten (1985)]{DineRohm85}, which gives a vanishing cosmological constant even for a nonvanishing gaugino condensation with nonzero $\langle H_{MNP}\rangle$ \citep[Derendinger, Ibanez, Nilles (1985), Dine, Rohm, Seiberg, Witten (1985)]{Derendinger85,DineRohm85}.

\subsubsection{Model-independent axion} \label{subsubsec:MIa}

$B_{\mu\nu}$ with $\mu$ and $\nu$ tangent to the 4D Minkowski spacetime is the MI axion present in all string compactifications \citep[Witten (1984)]{Witten84}. Because it is a 4D gauge boson, one cannot write potential terms in terms of $B_{\mu\nu}$ and it is massless if one neglects the anomaly term.
The number of transverse degrees in $B_{\mu\nu}$ is one and can be expressed as a pseudoscalar $a$ by dualizing it, $H_{\mu\nu\rho}\propto F_a\epsilon_{\mu\nu\rho\sigma}
\partial^\sigma a$. Even though it is massless at this level, the Bianchi identity of (\ref{BMNcoupling}) gives an equation of motion of $a$ as $\partial^2 a=(1/32\pi^2 F_a)G^a_{\mu\nu}\tilde G^{a\mu\nu}$ which hints that $a$ might be an axion. For it to be really a QCD axion, $c_2+c_3$ should be nonzero as discussed in Subsec. \ref{subsec:axionmass}. It is known that $c_3=1$ \citep[Witten (1985)]{Witten85b} with $c_3$ defined in Eq. (\ref{Axionint}). The other possible couplings are given by the second term of (\ref{BMNcoupling})
\begin{equation}
\frac{F_a}{M^2_{10}}\epsilon_{\mu\nu\rho\sigma}
\partial^\sigma a_{MI} \overline{\chi} \Gamma^{\mu\nu\rho}\chi = \frac{F_a}{M^2_{10}}\overline{\chi} \gamma_\sigma\gamma_5  \chi\partial^\sigma a_{MI}\label{conestring}
\end{equation}
which is the $c_1$ term defined in Eq. (\ref{Axionint}). There is no $c_2$ term, and $c_2+c_3=1$ and hence $H_{\mu\nu\rho}$ is really an axion and is model-independent.\footnote{Nevertheless, its property may depend on models in warped space \citep[Dasgupta, Firouzjahi, Gwyn (2008)]{Dasgupta08}.} This is a hadronic axion. This MI hadronic axion can have a nonvanishing $c_1$ and hence its phenomenology can be different from that of the KSVZ hadronic axion. So, in Eqs. (\ref{coneu},\ref{coned}) for the MI hadronic axion, one has to add the relevant $c_1$ term from Eq. (\ref{conestring}). The domain wall number of the MI axion has been shown to be $N_{DW}=1$ by considering the coupling of the MI axion to string $X^M(\sigma,\tau)$ on the world sheet $\int d^2\sigma \epsilon^{\alpha\beta}B_{\mu\nu}\partial_\alpha X^\mu\partial_\beta X^\nu$ \citep[Witten (1985)]{Witten85a}. $F_a$ is about $10^{-3}$ times the Planck mass \citep[Choi, Kim (1985a)]{Choiharm85}, and the correct relation, obtained from (\ref{BMNKE}), is $F_a/k=\alpha_c M_P/2^{3/2}\pi\sim 10^{16}$ GeV where $k$ is the level of the SM embedding and $\alpha_c$ is the QCD coupling constant \citep[Svrcek, Witten (2006)]{Svrcek06}. But the value $F_a\sim 10^{16}$ GeV most probably overcloses the universe.

An idea for lowering the MI axion decay constant may be the following. In some compactification schemes, an anomalous U(1)$_{\rm an}$ gauge symmetry results, where the U(1)$_{\rm an}$ gauge boson eats the MI axion so that the U(1)$_{\rm an}$ gauge boson becomes heavy. This applies to the MI axion since the coupling $\partial^\mu a_{MI}A_\mu^{\rm an}$ is present by the Green-Schwarz term \citep[Witten (1984), Chun, Kim, Nilles (1992)]{Witten84,ChunKNmu92b}. In fact, even before considering this anomalous U(1)$_{\rm an}$ gauge boson, the possibility was pointed out by Barr \citep[Barr (1985)]{Barr85Anom}, which became a consistent theory after discovering the anomalous U(1)$_{\rm an}$ from string compactification \citep[Dine, Seiberg, Witten (1987), Atick, Sen (1987), Dine, Ichinose, Seiberg (1987)]{Dine87an,Atick87,Dine87antwo}. Then, a global symmetry survives down the anomalous U(1)$_{\rm an}$ gauge boson scale. A detailed scenario is the following. The way anomalous U(1)$_{\rm an}$ with the gauge transformation, $\theta_{\rm an}\to\theta_{\rm an}+{\rm constant}$, is obtained is by calculating U(1)$_{\rm an}$ charges of fermions. Thus, we have a non-vanishing $c_2$ in Eq. (\ref{Axionint}) as $m\overline{\psi}_L\psi_R \exp(i c_2 \theta_{\rm an})$ and $c_3$ of the MI axion as $c_3\theta_{MI}\{F\tilde F\}$. For all gauge group factors, the anomaly units are calculated and they are shown to be identical \citep[Casas, Kathehou, Munoz (1988), Kim (1988)]{Casas88,Kim88}. For the MI axion to be the part of a gauge boson, it must be a true Goldstone boson without an anomaly, i.e. it should be exactly massless; so let us transform away the $c_3$ term by a phase redefinition of fermions such that $\bar c_2=c_2-c_3\langle\theta_{MI}\rangle/\theta_{\rm an}$ and $\bar c_3=0$ can be made for all gauge fields, i.e. $a_{MI}$ coupling to anomalies vanishes for all gauge groups. Because the longitudinal gauge boson $a_{MI}$ is removed away, we are left with the $\bar c_2$ term only, $m\overline{\psi}_L\psi_R \exp(i\bar c_2 \theta_{\rm an})+{\rm h.c}$, without the need to consider the gauge symmetry  U(1)$_{\rm an}$. At low energy, however, the term $m\overline{\psi}_L\psi_R \exp(i\bar c_2 \theta_{\rm an})$ has a global symmetry, $\theta_{\rm an}\to\theta_{\rm an}+{\rm constant}$, with $\theta_{\rm an}$ not depending on $x^\mu$. Thus, the interaction $m\overline{\psi}_L\psi_R \exp(i\bar c_2 \theta_{\rm an})+{\rm h.c.}$ explicitly shows a global U(1) axial symmetry or the PQ symmetry below the U(1)$_{\rm an}$ gauge boson mass scale: $\psi_L\to\psi_L e^{-i\theta_{\rm an}/2}$ and $\psi_R\to\psi_R e^{i\theta_{\rm an}/2}$. This global PQ symmetry can be broken in the axion window as in the field theoretic axion models. However, this idea on the decay constant does not work necessarily because most fields, including those removed at the GUT scale, carry the U(1)$_{\rm an}$ charge.

\subsubsection{Model-dependent axion} \label{subsubsec:MDa}

In 4D, $B_{MN}$ contains more pseudoscalars $B_{ij}$ with $i$ and $j$ tangent to the compact space $V_Z$. If they are axions, these are MD axions. The number of massless $B_{ij}$ modes at the KK mass level is the second Betti number of the compact space \citep[Green, Schwarz, Witten (1987), Eq. (14.3.10)]{GreenBk}, which was discussed in the early days in \citep[Witten (1984,1985), Choi, Kim (1985)]{Witten84,Witten85a,Choi85MD}. The string propagation on $M_4\times V_Z$ can be described by a suitable nonlinear sigma model. In this sigma model description, when a closed string wraps $V_Z$ topologically nontrivially then there are world-sheet instantons due to the map $S_1\to U(1)$. It was known that the world-sheet instantons are present precisely if the second Betti number is nonzero \citep[Green, Schwarz, Witten (1987)]{GreenBk}, and hence the MD axions are expected to receive non-negligible masses non-perturbatively \citep[Wen, Witten (1986), Dine, Seiberg, Wen, Witten (1986,1987)]{Wen86,Wen86A,Wen87A}, but it may be a model dependent statement \citep[Polchinski (2006)]{Polchinski06}. If a MD axion is known to have no potential term except the anomaly terms, then one should check the $c_2$ and $c_3$ couplings to confirm that it is really an axion. There has been no example presented yet in this way for a MD axion. If a MD axion is present, its decay constant is expected to be near the string scale as explicitly given as $F_{MD}=\alpha_C^{1/3}M_P/2^{3/2}\pi k^{1/3}g_s^{2/3}$ from the anomaly term alone in \citep[Svrcek, Witten (2006)]{Svrcek06}. The Green-Schwarz term integrated over $V_Z$ leads to this kind of the decay constant for the MD axion \citep[Choi, Kim (1985)]{Choi85MD}. However, as commented above, one has to calculate the corresponding $c_2$ term also to pinpoint the MD axion decay constant $F_{MD}$.

\subsubsection{Toward a plausible QCD axion from string} \label{subsubsec:stringQCDa}

A key problem in string axion models is to find a method obtaining a QCD axion at the axion window ($10^{9}\ {\rm GeV}\le F_a\le 10^{12}\ {\rm GeV}$) but an attractive model toward this direction is still lacking. Thus, the most pressing issues is the problem of introducing a detectable QCD axion from superstring. It includes the search for an approximate PQ symmetry toward a detectable QCD axion.

The conditions for compactified manifolds in warped space toward lowering the MI axion decay constant has been discussed in
\citep[Dasgupta, Firouzjahi, Gwyn]{Dasgupta08}, but its realization seems nontrivial.

The idea of localizing MD axions at fixed points toward lowering the decay constant has been proposed in \citep[I. W. Kim, J. E. Kim (2006)]{IWKim06}. It uses the warp factor idea and one needs a so-called Giddings-Kachru-Polchinski throat \citep[Giddings, Kachru, Polchinski (2002)]{Giddings02} in the Type-II string, but in the heterotic string a non-K\"ahler $V_Z$ is needed \citep[Becker, Becker, Fu, Tseng, Yau (2006)]{Becker06}. Indeed the warp factor is obtained in this way, which however has the power law behavior.

The intermediate scale string models can introduce the axion window as the ultra-violet completion scale \citep[Burgess, Iba\~nez, Quevedo (1999)]{Burgess99}. On the other hand, in this case the large radius to generate the Planck mass is the scale needing explanation.

We note that an attractive method of obtaining $F_a$ in the axion window is through the composite axion from superstring as discussed in Subsec. \ref{Subsec:Compaxion}. However, from string construction the composite axion has not been obtained so far.

Even if one lowered some $F_a$, we must consider the hidden sector also in estimating the axion masses and decay constants as discussed below.

\subsubsection{Hidden sector confining forces, axion mixing, and approximate PQ symmetry} \label{subsubsec:ApproxPQ}

\begin{figure}[!h]
\resizebox{0.7\columnwidth}{!}
{\includegraphics{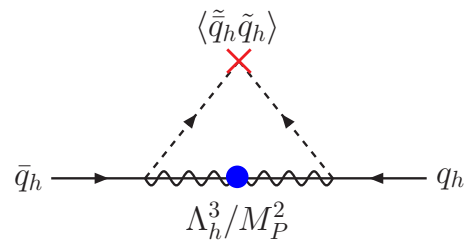}}
\caption{The hidden-sector squark condensation breaks chiral symmetry and generate hidden-sector quark masses.}\label{fig:hquarkmass}
\end{figure}

With the hidden-sector confining forces, we need at least two (QCD and one hidden-sector) $\theta$s which have to be settled to zero and hence we need at least two axions. For definiteness, let us consider only one more confining force at an intermediate scale, which may be the source of gravity mediation or GMSB. In this case, at least one MD axion is assumed to be present, and the axion mixing must be considered. We assume that one decay constant is in the intermediate scale. Here, there is an important (almost) theorem: the {\it cross theorem} on decay constants and condensation scales. Suppose that there are two axions $a_1$ with $F_1$ and $a_2$ with $F_2\ (F_1\ll F_2)$ which couple to  axion potentials with scales $\Lambda_1$ and $\Lambda_2\ (\Lambda_1\ll \Lambda_2)$. The theorem states that \citep[Kim (1999, 2000), I. W. Kim, J. E. Kim (2006)]{Kim99,Kim00,IWKim06}: according to the diagonalization process in most cases with generic couplings, the larger potential scale $\Lambda_2$ chooses the smaller decay constant $F_1$, and the smaller potential scale $\Lambda_1$ chooses the larger decay constant $F_2$. So, just obtaining a small decay constant is not enough. The hidden sector may steal the smaller decay constant. Probably, the QCD axion is left with the larger decay constant. We can turn this around such that the hidden sector instanton potential is shallower than the QCD instanton potential since the instanton potential is proportional to the light quark mass as discussed in Subsec. \ref{subsec:axionmass}. If a hidden-sector quark mass is extremely small, then the QCD axion can win the smaller decay constant and the other axion is an extremely light axion which can be used to fit the observed dark energy \citep[Perlmutter {\it et. al.} (1998), Riess {\it et. al.} (1998), Komatsu {\it et. al} (WMAP Collaboration, 2008)]{Perlmutter98,Riess98,WMAP08} which is named as the quintessential axion \citep[Kim, Nilles (2003)]{KimNilles03}. This can be easily realized if some hidden-sector squark condensations are very small as Fig. \ref{fig:hquarkmass} can generate hidden-sector quark masses \citep[I.-W. Kim, J. E. Kim (2006)]{IWKim06}.

Since it is difficult to obtain a reasonable light MD axion, it has been tried to look for an approximate PQ symmetry from string compactification. Only one reference exists from a realistic string compactification because of the difficulty of calculating all approximate PQ charges of quarks \citep[Choi, I.-W. Kim, J. E. Kim (2007)]{ChoiKS07}. After all, the topologically attractive $B_{ij}$ may not be the QCD axion we want. In this regard, we note that there already exists a field theoretic work regarding an approximate PQ symmetry, starting with a discrete $Z_9$ symmetry \citep[Lazarides, Panagiotakopoulos, Shafi (1986)]{Lazarides86}. Later, the gravitational nonperturbative effects such as wormholes and black holes were phenomenologically studied in view of any global symmetries \citep[Giddings, Strominger (1988), Lee (1988), Gilbert (1989)]{Giddings88,LeeK88,Gilbert89}. It is known that the PQ symmetry breaking operators in the superpotential must be forbidden up to dimension 8 \citep[Kamionkowski, March-Russell (1992), Barr, Seckel (1992), Holman, Hsu, Kephart, Kolb, Watkins, Widrow (1992), Ghigna, Lusignoli, Roncadelli (1992), Dobrescu (1997)]{Barr92,Holman92,Kamionkowski92,Ghigna92,Dobrescu96}. If we introduce an {\it approximate} PQ symmetry, it is better to forbid the PQ symmetry breaking operators up to dimension 8 in the superpotential, possibly up to dimension 7 with reasonably small couplings somewhere.

In this spirit, it is worthwhile to check approximate PQ symmetries in string derived models. The MSSMs presented in \citep[Kim, Kyae (2007), J. E. Kim, J.-H. Kim, Kyae (2007), Kim (2007a, 2007b)]{KimKyae06,KimJHKyae07, Kim07un,Kim07st} satisfy most phenomenological constraints and one can check approximate global symmetries. But it is a tedious work, and so far only for the model of \citep[Kim, Kyae (2007)]{KimKyae06} an approximate PQ symmetry has been checked out. In searching for an approximate global symmetry in a string derived model, there are so many Yukawa couplings to be considered that a complete study up to all orders is almost impossible. For example, in \citep[Choi, I.-W. Kim, J. E. Kim (2007)]{ChoiKS07} there appear $O(10^4)$~  $d=7$ superpotential terms and it is not a trivial task to find an approximate PQ symmetry direction, considering all these terms. Up to dimension 7 terms, there exists an approximate PQ symmetry which is spontaneously broken. The resulting axion coupling with photon has been calculated in \citep[Choi, I.-W. Kim, J. E. Kim (2007)]{ChoiKS07} which is presented in Fig. \ref{CASTexpTh} together with the CAST and Tokyo axion search bounds \citep[Andriamonje {\it et. al.} (2007), Inoue {\it et. al.} (2008)]{CAST07,Inoue08}. But the axion decay constant is not lowered. It is because the needed singlet VEVs, leading to the low energy MSSM, carry PQ charges. This is a generic problem for observable axions from superstring.
Comparing this to the MI axion case with the anomalous U(1)$_{\rm an}$, it may be easier to realize the observable axion with an approximate PQ symmetry.

\section{Axino cosmology }\label{sec:Axinocosmology}
Supersymmetrization of axion models includes the fermionic superpartner axino $\tilde a$ and the scalar superpartner saxion as discussed in Subsec. \ref{subsec:axinoth}. Both saxion and axino masses are split from the almost vanishing axion mass if SUSY is broken. The precise value of the axino mass depends on the model, specified by the SUSY breaking sector and the mediation sector to the axion supermultiplet~\citep[Nilles (1984)]{Nilles84}. Most probably, the saxion mass is around the soft mass scale $M_{\rm SUSY}$. The axino mass should also be near this scale as well. But the axino mass can also be much smaller than $M_{\rm SUSY}$ \citep[Frere, Gerard (1983), Kim, Masiero, Nanopoulos (1984), Chun, Kim, Nilles (1992a)]{Frere83,KimMas84,Chunaxino92a} or much larger than $M_{\rm SUSY}$ \citep[Chun, Lukas (1995)]{ChunLukas95}. Therefore, we take the axino mass as a free parameter here.

The decoupling temperature of the axino supermultiplet is of order \citep[Rajagopal, Turner, Wilczek (1991)]{Raja91},
\begin{equation}
\tadec =10^{11}\gev \left(\frac{F_a}{10^{12}\gev} \right)^2\left(\frac{0.1}{\alpha_c} \right)^3\label{Tdec}
\end{equation}
where $\alpha_c$ is the QCD coupling constant.

The saxion cosmology is a simple extension of the standard cosmology with saxion mass around the SUSY breaking scale \citep[J. E. Kim (1991), Chang, H. B. Kim (1996), Asaka, Yamaguchi (1999))]{Kimsaxion91,Chang96,AsakaYama99}, but its effect is not so dramatic as the effect of the axino. Therefore, here we focus on the axino cosmology \citep[Rajagopal, Turner, Wilczek (1991), Covi, J. E. Kim, Roszkowski (1999), Covi, H. B. Kim, J. E. Kim, Roszkowski (2001), Choi, J. E. Kim, Lee, Seto (2008)]{Raja91,Kimsaxion91,CKKR01,ChoiKY08}. In the moduli stabilization scenario of Ref. \cite{Kachru03}, the saxion VEV has been estimated in \citep[Choi, Jeong (2007)]{Choi07}.

The axino cosmology depends crucially on the nature of R-parity. If R-parity is conserved and the axino is lighter than the neutralino, then most probably the axino or gravitino (in case of GMSB) is the LSP.
If R-parity is not conserved, the neutralino can decay to ordinary SM particles, which has been discussed extensively in \citep[Allanach, Dedes, Dreiner (2004)]{Allanach04} and references cited there.

Now we focus on R-parity conservation. The neutralino, if it is the LSP, is a natural candidate for dark matter. Due to TeV scale sparticle interactions, the thermal history of neutralinos allows them to be dark matter. But, imposing a solution of the strong \Ca\Pa~ problem via the axion, the thermal history involves contributions from the axion sector, notably by the axino. Since axino cosmology depends on neutralino and gravitino number densities, let us comment on the neutralino and gravitino cosmologies  before discussing the effect of the axino. The neutralino cosmology depends on the neutralino freeze-out temterature \citep[Lee, Weinberg (1977), Drees, Nojiri (1993)]{LeeWein77b,Drees93} and the gravitino cosmology depends on the reheating temperature after inflation \citep[Weinberg (1982)]{Wein82}. Here we list several relevant temperatures in the axino cosmology
\dis{
&\tadec={\rm axino\ decoupling\ temperature}\\
&\treh={\rm reheating\ temperature\ after\ inflation}\\
&T_{fr}={\rm neutralino\ freeze-out\ temperature}\\
&\tarad={\rm axino-radiation\ equality\ temperature}\\
&\td = {\rm radiation\ temperature\ right\ after\ \tilde {\it a}\ decay}
\label{Temps}
}

Here, we are interested in the axino domination of the dark matter density. The cold axino dark matter possibility in the evolution history occurs either a heavy axino decayed already or axino has not decayed yet. If the axino has not decayed yet, the current axino CDM can be estimated using $\tadec$ or $\treh$.  If it has decayed already, the past cold axino dark matter requires the existence of $\treh^{\rm min}$ at some earlier time
\begin{equation}
 \frac43 m_{\tilde{a}} Y_{\tilde{a}}(\treh^{\rm min})= \td
\end{equation}
so that $ Y_{\tilde{a}}(\treh)=n_{\tilde a}/s\ge Y_{\tilde{a}}(\treh^{\rm min})$ at the time of the reheating after inflation, where $\treh^{\rm min}$ is the temperature above which axinos dominate the universe before they decay.

\subsection{Neutralino and gravitino}\label{subsec:gravitino}

The neutralino LSP seems the most attractive candidate for CDM simply because the TeV order SUSY breaking scale introduces the LSP as a WIMP \citep[Goldberg (1983), Ellis, Hagelin Nanopoulos, Olive, Srednicki (1984)]{Goldberg83,Ellis84DM}.
The neutralino, which was in the thermal equilibrium in the early universe, decouples and freezes out when the annihilation rate becomes smaller than the Hubble parameter. The freeze-out temperature $T_{fr}$ is normally given by $m_\chi/25$  \citep[Lee, Weinberg (1977b), Kolb, Turner (1990)]{LeeWein77b,KolbTur90}, e.g. $4 \gev$ for $100 \gev $ neutralino. Obviously, the neutralino relic density is not affected by the axino for $\td > T_{fr}$ since neutralinos were in thermal equilibrium after the axino decay. This is the standard neutralino dark matter. With the axino introduction, therefore, we study the case $\td < T_{fr}$.

Gravitinos in the universe are important if they dominate the dark matter fraction now or they had affected the result of nucleosynthesis. Thermal gravitinos produced at the Planckian time are important if $m_{3/2}\sim 1$ keV \citep[Pagels, Primack (1982)]{Pagels82}. However, in the inflationary scenario these Planckian time  gravitinos are not important now. For a heavy gravitino,  it was observed that the gravitino decay affects the nucleosynthesis \citep[Weinberg (1982)]{Wein82}, which was suggested to be solved by inflation \citep[Krauss (1983), Khlopov, Linde (1984)]{KraussInos83, Khlopov84}. Then, the gravitino number density is roughly estimated in terms of the reheating temperature after inflation $n_{3/2}\propto \treh$. To estimate the cosmological bound on $\treh$ rather accurately, a full supergravity interaction \citep[Cremmer, Ferrara, Girardello, van Pr\"oyen (1983)]{Cremmer83} has been used and applied to the dissociation problem of rare light elements such as deuterium, etc., resulting in $\treh<10^9~\gev$ \citep[Ellis, Kim, Nanopoulos (1984)]{Ellis84}. A recent calculation of $\treh$ has been performed using the nucleosynthesis code to look for $^7$Li destruction  and/or $^6$Li overproduction \citep[Kawasaki, Kohri, Moroi (2005, 2008)]{Kawasaki05,Kawasaki08}, following the earlier work of \citep[Cyburt, Ellis, Fields, Olive (2003)]{Cyburt03}, which led to a stronger bound $\treh<10^8~\gev$ if the gravitino is lighter than the gluino and $\treh<10^7~\gev$ if the gravitino is heavier than the gluino. This gravitino problem is absent if the gravitino is the next LSP (NLSP), $m_{\tilde a}<m_{3/2}<m_{\chi}$, since a thermally produced gravitino would decay into an axino and an axion which would not affect the BBN produced light elements \citep[Asaka, Yanagida (2000)]{Asaka00}.

If the gravitino is the LSP with the stau or neutralino as the NLSP, the gravitino can be the CDM even in the constrained MSSM (or mSUGRA) for some parameter space, avoiding the BBN and $b\to s\gamma$ constraints \citep[Boehm, Djouadi, Drees (2000), Ellis, Olive, Santoso, Spanos (2004), Roszkowski, Ruiz de Austri, Choi (2005), Cerdeno, Choi, Jedamzik, Roszkowski, Ruiz de Austri (2006)]{Boehm00,Ellis04,Roszkowski05,Cerdeno06}.

\subsection{Axino}

Thus, in SUSY theories  we must consider a relatively small reheating temperature $10^{7-8}\gev$. The axino cosmology must also be considered with this low reheating temperature.

In principle, the axion supermultiplet is independent from the observable sector in which case we may take the axino mass as a free parameter of from keV scale to a value much larger than the gravitino mass \citep[Chun, Kim, Nilles (1992a), Chun, Lukas (1995)]{Chunaxino92a, ChunLukas95}. Light axinos ($m_{\tilde a}\lesssim 100~\gev$) can be a dark matter candidate, and they have been studied extensively as a warm dark matter candidate \citep[Rajagopal, Turner, Wilczek (1991)]{Raja91} with the reheating temperature given in \citep[Brandenburg, Steffen (2004)]{Brandenburg04}, or a CDM candidate \citep[Covi, J. E. Kim, Roszkowski (1999), Covi, J. E. Kim, H. B. Kim, Roszkowski (2001), Roszkowski, Seto (2007), Seto, Yamaguchi (2007), Asaka, Yanagida (2000)]{CKRosz99,CKKR01, Roszkowski07,Seto07,Asaka00}. Heavy axinos, however, cannot be the LSP and can decay to the LSP plus light particles. This heavy axino decay to neutralinos has been already considered  \citep[Chun, Lukas (1995)]{ChunLukas95}.  The heavy axino possibility was considered briefly in studying cosmological effects of the saxion by \citep[Kawasaki, Nakayama, Senami (2007), Kawasaki, Nakayama (2008)]{Kawasaki07,Kawasaki08}. A more complete cosmological analysis of the heavy axino has been discussed in \citep[Choi, Kim, Lee, Seto (2008)]{ChoiKY08prd}.

\begin{figure}[!]
\resizebox{0.95\columnwidth}{!}
{\includegraphics{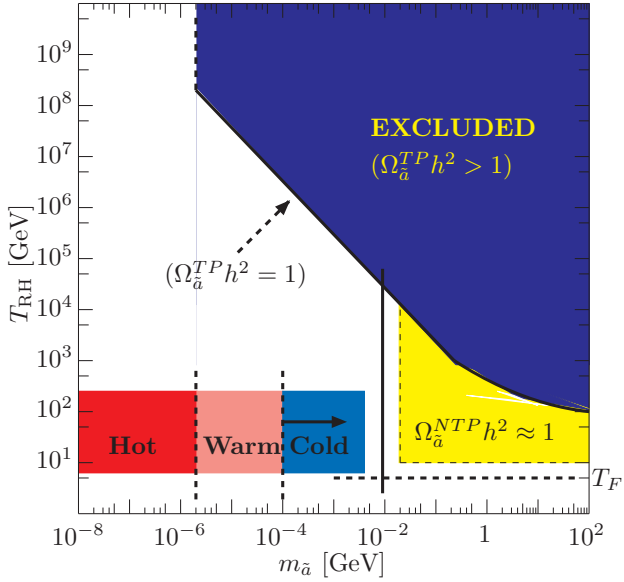}}
\caption{The solid line is the upper bound from TP. The yellow region is the region where NTP can give cosmologically interesting results ($\Omega_{\tilde a}^{\rm
NTP} h^2\simeq1$). The freezeout temperature is
$T_F\approx\frac{m_\chi}{20}$.}\label{fig:axinocdm}
\end{figure}

Since the CDM fraction of the universe is roughly 0.23 \citep[Komatsu {\it et al.} (WMAP Collaboration, 2008)]{WMAP08}, let us focus on the CDM possibility of the axino or axino related neutralino.

For the axino to be the LSP, it must be lighter than the lightest neutralino and gravitino. In this case, we do not have $T_{\rm D}$ of Eq. (\ref{Temps}).
If the lightest neutralino is the NLSP,  $m_{\tilde a}<m_{\chi}<m_{3/2}$, the thermal production (TP) mechanism gives the afore-mentioned bound on the reheating temperature after inflation. At the high reheating temperature, the TP is dominant in the axino production \citep[Covi, J. E. Kim, Roszkowski (1999)]{CKRosz99}. If the reheating temperature is below the critical energy density line, there exists another axino CDM possibility from the nonthermally produced (NTP) axinos which result from neutralino decay \citep[Covi, J. E. Kim, H. B. Kim, Roszkowski (2001)]{CKKR01}. This situation is sketched in Fig. \ref{fig:axinocdm}. We note that with R-parity conservation, the double production of low mass axinos are negligible in supernovae, and hence from SN1987A there does not result a useful exclusion region in the low mass region.

Since the final axino energy fraction is reduced by the mass ratio, $\Omega_{\tilde a}h^2=(m_{\tilde a}/m_{\chi})\Omega_{\chi}h^2$ for $m_{\tilde a}<m_{\chi}<m_{3/2}$, the stringent cosmologically constrained MSSM parameter space for $m_\chi$ can be expanded. As shown in this figure, the NTP axinos can be CDM for relatively low reheating temperature ($<10$ TeV) for 10 MeV $<m_{\tilde a}<m_{\chi}$. In Fig. \ref{fig:axinocdm} the thin dash line corner on the RHS corresponds to the MSSM models with $\Omega_\chi h^2<10^4$, and a small axino mass renders the possibility of the axino forming 23\% of the closure density. If all SUSY mass parameters are below 1 TeV, then probably $\Omega_\chi h^2<100$ (the thick solid-dash line corner on the RHS) but a sufficient axino energy density requires $m_{\tilde a}>1$ GeV. Thus, if the LHC does not detect the neutralino needed for its closing of the universe, the axino closing is a possibility \citep[Choi, Roszkowski, Ruiz de Austri (2008), Covi, Roszkowski, Small (2002), Covi, Roszkowski, Ruiz de Austri, Small (2004), Choi, Roszkowski (2006)]{ChoiKY08,CoviRS02,
CoviRAu04,ChoiRosz06}. If the NLSP is stau with the axino or gravitino LSP, the previously forbidden stau LSP region is erased. In this case, the CDM axino is similar to the bino LSP case but it is easier to detect the stau signal due to the charged stau at the LHC \citep[Brandenburg, Covi, Hamaguchi, Roszkowski, Steffen (2004)]{Brandenburg04}. However, the efforts to detect axinos may be difficult \citep[H. B. Kim, J. E. Kim (2002)]{HBKim02}.

\begin{figure}[!]
 \resizebox{0.95\columnwidth}{!}
{\includegraphics{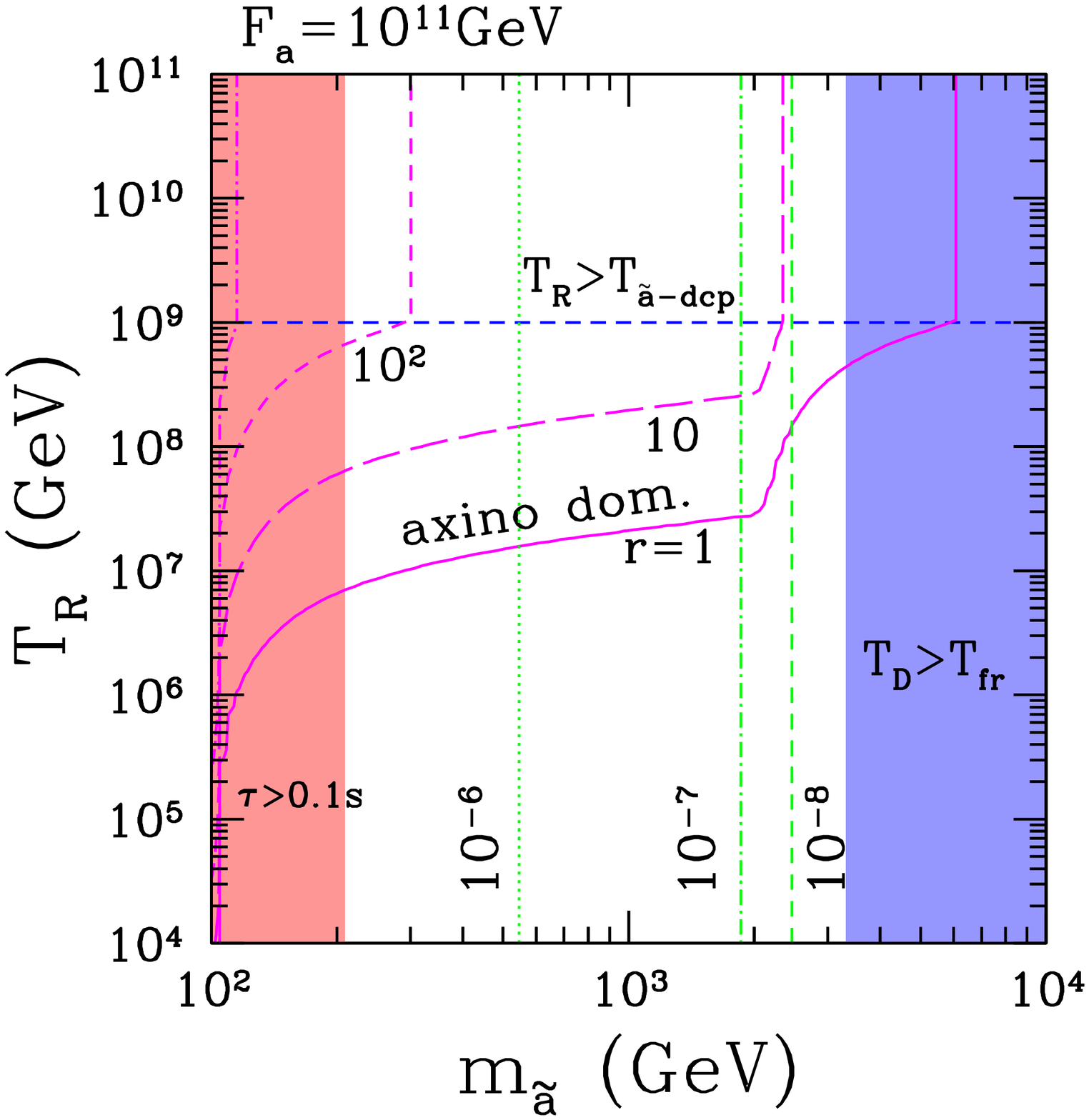}}
  \caption{The $\treh$ vs. $m_\axino$ plot for $m_\chi=100~\gev$ and $F_a=10^{11} \gev$. }\label{fig:axinoChoi}
\end{figure}

In the GMSB scenario, the gravitino mass is generally smaller than the neutralino mass and possibly smaller than the axino mass, for which case the cosmological effect has been studied by \citep[Chun, H. B. Kim, J. E. Kim (1993), H. B. Kim, J. E. Kim (1995)]{AxinoGravitino93,HBKim95}.

For a heavy axino decaying to a neutralino, we present a $\treh$ vs. $m_\axino$ plot for $F_a=10^{11} \gev$ in Fig. \ref{fig:axinoChoi}. The region $\treh>\tadec$ is above the dashed blue line. The axino lifetime greater than $0.1$ s is denoted by the red shaded region in the LHS. The blue shaded region in the RHS is where the axino decays before the neutralino decouples ($\td > T_{fr}$). The magenta lines (horizontal) are the contours of the entropy increase due to the axino decay, $r\equiv S_f/S_0$. Above $r=1$ lines axinos dominate the universe before they decay. The green lines (vertical) denote the $\langle \sigma_{ann} v_{rel}\rangle$, where $\sigma_{ann}$ is the neutralino annihilation cross section, in units of ${\rm GeV}^{-2}$ which are used to give the right amount of neutralino relic density. In Fig. \ref{fig:axinoChoi} we use neutralino and gluino masses as $m_\chi=100\gev$ and $m_\gluino = 2 \tev$, respectively. For a larger $F_a$ and a heavier neutralino mass, the green lines move to the right \citep[Choi, Kim, Lee, Seto (2008)]{ChoiKY08prd}.



\noindent {\bf Acknowledgments\ :} We have benefitted from the discussions with K.-J. Bae, S. M. Barr, Kiwoon Choi, Ki-Young Choi, D. K. Hong, J.-H. Huh, K. Imai, H. D. Kim, I.-W. Kim, A. Melissinos, C. Mu\~noz, H. P. Nilles, S. Park, S. Raby, G. G. Raffelt, A. Ringwald, K. van Bibber, and K. Zioutas. This work is supported in part by the Korea Research Foundation, Grant No. KRF-2005-084-C00001.

\bibliography{Exp_Sec}

\end{document}